\documentclass[twocolumn,superscriptaddress,10pt,pre,floatfix,aps]{revtex4-1}
\pdfoutput=1
\usepackage{hyperref}
\usepackage{graphicx}
\usepackage{mathtools}
\usepackage{bm}
\usepackage{color}
\usepackage{multirow}
\usepackage{commath}
\usepackage{amsmath}
\usepackage{amsfonts}
\usepackage{amssymb}
\usepackage{bbm}
\usepackage{widetable}
\usepackage{enumerate}
\usepackage{multirow}

\usepackage{algpseudocode}
\usepackage{algorithm}
\newcommand{\argmin}{\operatornamewithlimits{argmin}}

\newcommand{\emptycirc}{${\color{white}\bullet}\mathllap{\circ}$}
\newcommand{\filledcirc}{${\color{black}\bullet}\mathllap{\circ}$}
\begin{document}

\title{ Prediction of allosteric sites and mediating interactions
  through bond-to-bond propensities }

\author{B.R.C. Amor} \affiliation{Department of Chemistry, Imperial
  College London, London SW7 2AZ, United Kingdom}
\affiliation{Institute of Chemical Biology, Imperial College London,
  London SW7 2AZ, United Kingdom} \author{M.T. Schaub}
\affiliation{Department of Mathematics, Imperial College London,
  London SW7 2AZ, United Kingdom}
\thanks{present address: Université catholique de Louvain, Belgium}
\author{S.N. Yaliraki} \affiliation{Department of Chemistry, Imperial
  College London, London SW7 2AZ, United Kingdom}
\affiliation{Institute of Chemical Biology, Imperial College London,
  London SW7 2AZ, United Kingdom} \author{M. Barahona}
\affiliation{Department of Mathematics, Imperial College London,
  London SW7 2AZ, United Kingdom} \affiliation{Institute of Chemical
  Biology, Imperial College London, London SW7 2AZ, United Kingdom}

\begin{abstract}
  Allosteric regulation is central to many biochemical
  processes. Allosteric sites provide a target to fine-tune protein
  activity, yet we lack computational methods to predict
  them. Here, we present an efficient graph-theoretical approach for
  identifying allosteric sites and the mediating interactions that connect
  them to the active site. Using an atomistic graph with edges weighted
  by covalent and non-covalent bond energies, we obtain a bond-to-bond 
  propensity that quantifies the effect of instantaneous bond fluctuations
  propagating through the protein. 
  We use this propensity to detect the sites and
  communication pathways most strongly linked to the active site,
  assessing their significance through quantile regression and 
  comparison against a reference set of 100 generic proteins. We exemplify
  our method in detail with three well-studied allosteric proteins: caspase-1,
  CheY, and h-Ras, correctly predicting the location of the allosteric
  site and identifying key allosteric interactions.  
  Consistent prediction of allosteric sites is then attained in 
  a further set of 17 proteins known to exhibit allostery. 
  Because our propensity measure runs in almost linear time, it offers a scalable
  approach to high-throughput searches for candidate allosteric sites.
\end{abstract}

\maketitle

\section{Introduction}
Allostery is a key molecular mechanism underpinning control 
and modulation in a variety of cellular 
processes~\cite{monod1963allosteric, perutz1989mechanisms}.  
Allosteric effects are those induced on the main functional site of a
biomolecule by the binding of an effector at a distant site, e.g.,
the binding of a co-factor modulating the catalytic rate
of an enzyme~\cite{nussinov2013allostery}.  Despite the importance of
such processes, there is still a lack of understanding as to how the
interactions at the allosteric site propagate across the protein and
affect the active site.  In this paper, we present a graph-theoretic
approach that uses atomistic structural data to identify allosteric
sites in proteins, as well as bonds and residues involved in this
propagation.  By defining an edge-to-edge transfer function, which can
be understood as a Green's function in the edge space of the protein
graph, we compute a bond-to-bond propensity that captures
the effect induced on any bond of the molecule by the propagation of
perturbations stemming from bonds at the active site.  This propensity can be
computed efficiently to predict allosteric sites and key bonds which
are prominently involved in mediating the allosteric propagation.

The growing realisation that all proteins exhibit innate dynamic behaviour
\cite{frauenfelder1991energy, henzlerwildman2007} and the discovery of
allosteric effects in single domain proteins~\cite{volkman2001two} have reaffirmed
the ubiquitousness of this form of regulation; potentially, any
protein could be allosteric \cite{Gunasekaran2004}.  This fact opens
up important experimental directions: drugs targeted at
allosteric sites could offer improved specificity and control compared
to traditional drugs that bind at the active site~\cite{nussinov2013allostery}.  
Efficient methods able to identify putative allosteric sites are therefore of great current
interest~\cite{hardy2004searching}.  To date, computational approaches
to finding allosteric sites have involved statistical coupling
analysis \cite{lockless1999evolutionarily}, molecular
dynamics~\cite{grant2011novel, weinkam2012structure,
  ota2005intramolecular}, machine
learning~\cite{demerdash2009structure}, and normal mode
analysis~\cite{panjkovich2012exploiting}.  For a comprehensive review
see Ref.~\cite{collier2013emerging}.

Classic thermodynamic models of allostery (such as the
Monod-Wyman-Changeux~(MWC)~\cite{monod1965nature} and
Koshland-N\'emethy-Filmer (KNF) models~\cite{koshland1966comparison})
were formulated to explain cooperativity in multimeric proteins in
terms of conformational transitions in a protein
landscape~\cite{hilser2012structural,weinkam2012structure}.  Although
such models reproduce broad experimental features (e.g. the sigmoidal
binding curves), they offer little insight into the molecular
mechanisms driving and defining the underlying conformational
transitions.  Attempts to identify the specific residues involved in
allosteric transitions have led to the idea of allosteric pathways,
which aim to describe the routes through which an excitation
propagates through the protein~\cite{lockless1999evolutionarily,
  del2009origin, zhuravlev2010protein}.  Indeed, recent
experimental~\cite{muller2014donor,li2014anisotropic} and
computational
\cite{martinez2011mapping,fujii2014observing,nguyen2009energy,gnanasekaran2011communication}
work has shown that energy flow in globular proteins is anisotropic.
Some of these studies have connected this anisotropy to the allosteric
properties of the
protein~\cite{li2014anisotropic,gnanasekaran2011communication}.  Our
work builds on this line of research and aims at finding allosteric
sites by using graph-theoretical techniques to quantify efficiently
the propagation of perturbations through a protein structure described
in atomistic detail.  In~\cite{li2014anisotropic} the authors find
that internal energy flow in albumin is anisotropic, and that this
flow is altered by binding of an allosteric ligand. Here, we also find
that the propagation of perturbations internally is
anisotropic. However, we use the term `allosteric' in a more specific
way, to describe locations distant from the active site where a
perturbation can have a functional effect on the active site.  The
identification of such distant sites 
and the pathways connecting them to the active-site,
has become an area of considerable interest~\cite{gerek2011change,
  ota2005intramolecular, kaya2013mcpath}.

The connection between the behaviour of a diffusion process (e.g., a
random walk) on a network and the vibrational dynamics of that network
is well established in the biophysical literature
\cite{nakayama1994dynamical, leitner2008energy}.  Previous
network-based methods for protein structure analysis have made use of
shortest path calculations~\cite{del2006residues}, community detection
algorithms~\cite{del2007modular}, and random walks on
networks~\cite{chennubhotla2007signal}.  However, such methods almost
universally used \textit{coarse-grained} protein descriptions at the
level of residues, i.e., they are based on residue-residue interaction
networks (RRINs)~\cite{amitai2004network} that neglect atomistic
detail.  Although methods that use molecular dynamics simulations to
derive edge weights for RRINs from the cross-correlations of residue
fluctuations have yielded interesting
results~\cite{ghosh2007study,sethi2009dynamical}, such approaches are
computationally costly.  Furthermore, Ribeiro and Ortiz have recently
shown that RRINs are critically dependent on the chosen cut-off
distance, and that using \textit{energy-weighted} networks that
include the covalent interactions of the backbone is crucial for 
correctly identifying signal 
propagation pathways~\cite{ribeiro2014determination,ribeiro2015energy}.  
Our findings below show that efficient methodologies which can exploit
the physico-chemical detail of atomistic, energy-weighted protein networks 
can lead to enhanced identification of allosteric sites and relevant individual
mediating interactions in a number of important cases.

Our analysis starts by building an atomistic graph model of the
protein: nodes are atoms and (weighted) edges represent individual
bonds, with weights given by energies from interatomic potentials. 
The graph includes both covalent bonds and 
weak, non-covalent bonds (hydrogen bonds, salt bridges, 
hydrophobic tethers and electrostatic interactions).  
Details of the construction of the graph are given in
Section~\ref{sec:ntwk_cnstr} and in Refs.~\cite{delmotte2011protein,
  amor2014uncovering}.  
The resulting all-atom graph is analysed using the edge-to-edge transfer matrix $M$, 
which is akin to a discrete Green's function in the \textit{edge space} of the graph and
has been recently introduced in Ref.~\cite{schaub2014structure} to
study non-local edge coupling in graphs.  In this paper, we derive a
new, alternative interpretation of the matrix $M$ and show that it
provides a means to extracting the level of influence that the
fluctuations of an edge have on any other edge of the graph 
(for detailed mathematical derivations, 
see Materials and Methods, Section~\ref{sec:M-matrix} and \textit{SI}).
We use this notion to calculate the \textit{propensity} of each bond, $\Pi_b$,
i.e., a measure of how strongly bond $b$ is coupled to the active site
through the atomistic graph.  Because allosteric effects are
reflected on induced changes in weak bonds, yet
mediated through the whole protein network,
our bond-to-bond formalism provides a natural way of uncovering how
the long-range correlations between bonds contribute to allosteric
signalling.  Crucially, recent algorithmic developments
\cite{spielman2004nearly, kelner2013simple} allow these computations
to be carried out in \textit{almost linear time} (in the number of
edges). Therefore, in contrast to most other computational approaches,
our method is easily scalable to large systems with tens of thousands
of atoms. 

To establish if a bond has a high propensity $\Pi_b$, and
to detect important bonds (and residues), we use quantile regression
to compare each bond to the ensemble of bonds within the protein 
at a similar geometric distance from the active-site
(described in Materials and Methods, Section~\ref{sec:qr}).  Quantile regression
(QR)~\cite{koenker2005quantile} is a robust statistical technique
previously employed in medicine~\cite{wei2006quantile},
ecology~\cite{knight2002variation} and
econometrics~\cite{buchinsky1994changes}.  We additionally confirm our
findings by computing the statistical significance of the bond
propensity against a reference set of 100 representative proteins
randomly drawn from the Structural Classification of Proteins (SCOP) database
(see Section~\ref{sec:scop}). This reference set provides us with a
pre-computed structural bootstrap against which any protein can be
tested to detect statistically significant bonds, further reducing the
computational cost of our method.

In Sections~\ref{sec:caspase-1}---\ref{sec:h-Ras}, 
we showcase our procedure through the detailed analysis of 
three important allosteric proteins: caspase-1, CheY, and h-Ras.  
In each case, given structural data and the location of the known active site, 
we correctly predict the location of the allosteric site and uncover
communication pathways between both sites. Each of the three examples
serves to highlight particular aspects of the method. In the case of caspase-1,
comparison of our results with those obtained using coarse-grained
residue-residue interactions networks (RRINs) shows that incorporating
atomistic physico-chemical detail can indeed be necessary for the reliable
identification of the allosteric site.  
In the case of CheY, we
illustrate how further information can be gained by incorporating dynamic data 
from ensembles of NMR structures: the variance of the propensity across the NMR ensemble 
reveals residues involved in allosteric signalling which cannot be identified from the
static X-ray crystal structure alone.  
In the case of h-Ras, our method shows that signal
propagation between the active and allosteric sites is crucially
dependent on the interaction between the protein and specific
structural water molecules.  
Having demonstrated the insight into
allosteric mechanisms offered by our method, we then evaluate it against a
test set with a further 17 allosteric proteins (see Section~\ref{sec:test_set}).  
We find that the bond-to-bond
propensity is a good predictor of a site's allosteric propensity,
suggesting it could be used to guide efforts in 
structure-based discovery of drugs as allosteric effectors.

\section{Results}

\subsection{Identification of the allosteric site and functional residues in caspase-1}
\label{sec:caspase-1}

Our first example is caspase-1, an allosteric protein of great
importance in apoptotic processes~\cite{amor2014uncovering}.
Caspase-1 is a tetramer composed of two asymmetric dimers, each 
containing one active site.  Using the PDB atomic structure
(PDB: 2HBQ), we constructed an atomistic, energy-weighted graph
representation of the protein based on interaction potentials, as described in
Section~\ref{sec:ntwk_cnstr}~\cite{delmotte2011protein,
  amor2014uncovering}.

\begin{figure*}[tb]
  \centering
  \includegraphics[width = \textwidth]{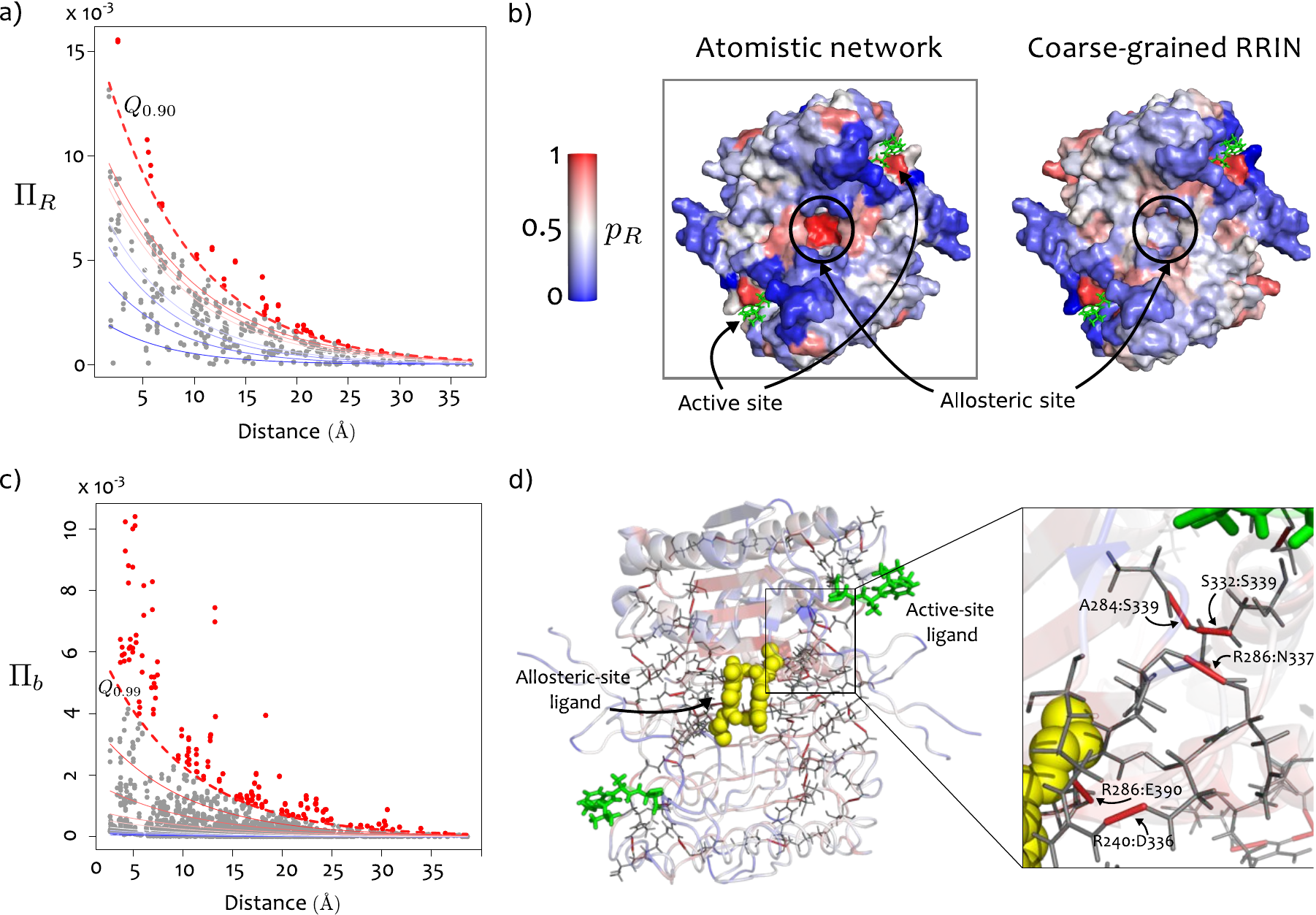}
  \caption{\textbf{Bond-to-bond propensities identify the allosteric
      site and atomistic pathway in caspase-1.} 
    (a) The propensities of all residues $\Pi_R$ are plotted against
    their distance from the active site.  The lines correspond to the
    quantile regression estimates for the $p$-th quantiles $Q_{p}$,
    with $p =0.1,0.2,\ldots,0.8,0.9$. The dashed red line
    indicates the $Q_{0.90}$ cut-off used for identifying important
    residues.  (b) The quantile scores $p_R$ for each residue are
    mapped onto the surface of caspase-1.  The active-site ligand is
    shown in green.  The allosteric binding site 
    is identified as a hot-spot of high propensity.  When a
    coarse-grained residue-residue interaction network with cut-off
    of 6\AA~is used (right), the allosteric binding site is not
    identified.  (c) The propensities of bonds $\Pi_b$ are
    plotted against their distance from the active site with the $Q_{0.99}$
    quantile indicated by the dashed line.  (d) High
    quantile score bonds ($p_b \geq 0.99$) are shown on the structure.
    Bonds between R286:E390, R240:D336, R286:N337, A284:S332, and
    S332:S339 have large quantile scores and form contiguous pathways
    between the active and allosteric sites.  The active site ligand is
    shown in green and the allosteric ligand is shown as yellow spheres.
}
  \label{fig:caspase}
\end{figure*}

In order to quantify how strongly each bond is coupled to the active site, 
we calculate the propensities $\Pi_b$ 
for all bonds in the protein, as given by Eq.~\eqref{eq:pp_bond}.  
We also aggregate the bond propensities for each residue to obtain 
the residue score $\Pi_R$, as given by Eq.~\eqref{eq:pp_residue}.  
To rank bonds and residues according to their significance, 
we compute the corresponding quantile
scores $p_b$ and $p_R$, respectively, obtained via quantile regression
as in Eq.~\eqref{eq:log_pb_fit}.  These quantile scores allow us
to establish which bonds (and residues) have high propensity values 
as compared to bonds (or residues) at the same distance from the
active site in the protein (Fig.~\ref{fig:caspase}a~and~\ref{fig:caspase}c).

Our method finds a hot spot of residues with high quantile scores in a
cavity at the dimer-dimer interface (Fig.~\ref{fig:caspase}b
left). This site has been previously identified by Scheer~\textit{et
  al.} as the binding site for a small molecule inhibitor of caspase-1
\cite{datta2008allosteric}.  Table~\ref{tbl:caspase} shows that the
allosteric residues, i.e., residues within 3.5\AA~of the allosteric
inhibitor, have significantly higher propensities than
non-allosteric residues (Wilcoxon rank sum, $p < 0.0005$). 
Residues E390, S332 and R286, 
which have been found to belong to a hydrogen bond 
network between the active and allosteric sites~\cite{datta2008allosteric}, 
have respectively the 3rd, 13th, and 15th highest
quantile scores of the 260 residues in each dimer of caspase-1.

Making use of the physico-chemical detail afforded by our atomistic description, 
we find the bonds with high propensity that lie on communication
pathways connecting the allosteric site to the active-site ligand.
Concentrating on the top quantile $p_b \geq 0.99$
(Fig.~\ref{fig:caspase}c), the two interactions in the salt bridges
between residues E390 and R286 have quantile scores of 0.996 and
0.990, and their combined propensity
gives this salt bridge the highest quantile score in the protein.  It
is known that these salt-bridges are directly disrupted by the
allosteric inhibitor~\cite{datta2008allosteric}.  In addition, our
method reveals other important bonds lying between the active and
allosteric sites (Fig.~\ref{fig:caspase}d), including hydrogen bonds
between Arg240:Asp336 ($p_b = 0.999$), S332:S339 ($p_b= 0.996$),
R286:N337 ($p_b = 0.992$), and A284:S332 ($p_b = 0.990$).  Bonds in
this pathway have previously been identified by Datta \textit{et al}
as being functionally important: the corresponding alanine mutations
cause 230-fold (R286A), 130-fold (E390A), 3.7-fold (S332A) and 
6.7-fold (S339A) reductions in
catalytic efficiency~\cite{datta2008allosteric}.

The atomistic detail is important for the outcome of the analysis.  
If instead of employing an all-atom graph description,
we carry out the same calculations on a coarse-grained residue-residue
interaction network (RRIN)~\cite{del2006residues,chennubhotla2007signal} 
with cut-off radius of 6 \AA, the allosteric site of caspase-1 is no longer identified
as a hot spot (Fig.~\ref{fig:caspase}b right) and the allosteric
residues do not have significantly higher propensity
compared to other residues (Wilcoxon rank sum, $p=0.5399$). 
The results obtained with RRINs are in general dependent on the cut-off radius used.
For caspase-1, the allosteric site is not detected in RRINs with cut-off radii
of 6 \AA, 7 \AA~and 8 \AA.  The allosteric site is found to be significant at 10\AA, 
but the signal is still considerably weaker than when using the atomistic network (Table S6).
These findings highlight that while an atomistic model of the protein
structure may not always be needed, it can indeed be important for the
detection of allosteric effects in proteins; in this case, 
the strength of the pair of salt bridges formed by E390 and E286, 
which is crucial for the allosteric communication in caspase-1,
is not captured by RRINs.  Other recent results have similarly demonstrated the
importance of both covalent bonds and hydrogen bonds to signal
transmission within proteins~\cite{ribeiro2015energy}.  
Yet in other cases (e.g., CheY in the following section), this level of physico-chemical detail
seems to be less important, and RRINs are able to capture allosteric communication. 
An extended, in-depth analysis of the results obtained with all-atom networks and
RRINs for a variety of proteins and cut-off radii can be found in the SI (Section~6).

\begin{table}
  \small
  \caption{ 
    Quantile scores for the propensities of residues within 3.5\AA~of the allosteric
    site of caspase-1 computed from the atomistic graph and from a residue-based network (RRIN) with cut-off radius of 6 \AA.  The average quantile scores of allosteric residues ($\overline{p_{R,\text{allo}}}$) and non-allosteric residues ($\overline{p_{R,\text{rest}}}$) are also presented.}
  \label{tbl:caspase}
  \medskip
  \begin{tabular*}{\columnwidth}{@{\extracolsep{\fill}}ccccc}
    \hline
    Residue & \multicolumn{2}{c}{$p_R$ (Atomistic network)}& \multicolumn{2}{c}{$p_R$ (RRIN)} \\
    \hline
    & Dimer 1 & Dimer 2 & Dimer 1 & Dimer 2 \\
    \cline{2-3} \cline{4-5}
    R240 & 0.772 & 0.734 & 0.562 & 0.562\\
    L258 & 0.394 & 0.408 & 0.168 & 0.168\\
    N259 & 0.828 & 0.832 & 0.324 & 0.324\\
    F262 & 0.654 & 0.652 & 0.464 & 0.464\\
    R286 & 0.938 & 0.928 & 0.838 & 0.838\\
    C331 & 0.634 & 0.646 & 0.724 & 0.724\\
    P335 & 0.206 & 0.196 & 0.450 & 0.450\\
    E390 & 0.990 & 0.992 & 0.318 & 0.318\\
    R391 & 0.982 & 0.984 &  0.258 & 0.258 \\
    \hline
    $\overline{p_R^\text{allo}}$
    & 0.711 & 0.708 & 0.4567 & 0.4567\\
    $\overline{p_R^\text{rest}}$
    & 0.481 & 0.492 & 0.4793 & 0.4789 \\
         \hline
  \end{tabular*}
\end{table}

\subsection{Uncovering allosteric communication pathways in CheY}
\label{sec:CheY}
\subsubsection{Identification of the phosphorylation site of CheY}

\begin{figure*}[tb]
  \centering
  \includegraphics[width = 0.9 \textwidth]{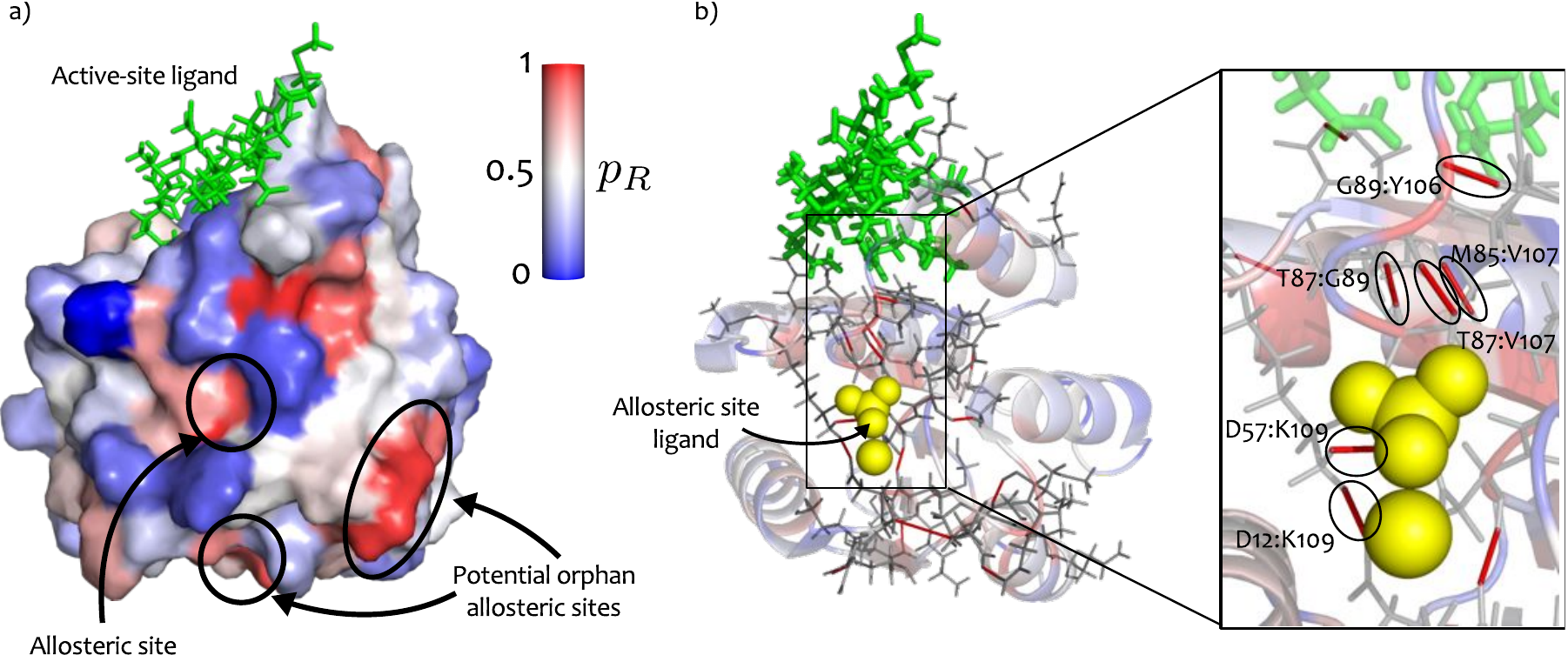}
  \caption{\textbf{Allosteric phosphorylation site in CheY is
      identified by its high propensity.} (a) Residue quantile
    scores $p_R$ are mapped onto the surface of CheY.  The allosteric
    phosphorylation residue D57 is identified as a hot-spot.  We identify two
    other distant sites, which could serve as potential orphan targets for 
    allosteric effectors.  (b) The top 3\% of bonds by quantile score (i.e., 
    $p_b \geq 0.97)$ are indicated on the structure.  The blow-up shows 
    high-quantile score non-covalent bonds that form propagation pathways between the
    allosteric ligand (yellow spheres) and the ligand binding site (green).}
  \label{fig:chey}
\end{figure*}

CheY is a key protein in bacterial chemotaxis. When CheY binds to the
flagellar motor switch protein (FliM), it causes a change in the
rotation direction of the flagellar motor, thus regulating the
tumbling rate of \textit{E. coli}.  This regulation is achieved
through a post-translational modification of CheY: phosphorylation of
CheY at the distant residue D57 increases its affinity for FliM,
making this an interesting example of a single-domain
allosteric protein.

Following the same procedure, we calculated the
propensity of each bond and residue (relative to the FliM binding site)
in fully activated CheY (PDB ID: 1F4V) bound to
Mg$^{2+}$, BeF$_3$ and FliM.  We identify a number of hot-spot surface
residues with high quantile scores (Fig.~\ref{fig:chey}a), including the
phosphorylation site, D57 ($p_R = 0.96$).
Again, residues in the allosteric site ($<3.5$~\AA~from the phosphorylation
site) have higher average quantile score than non-allosteric residues
($\overline{p_{R,\text{allo}}} = 0.61 > \overline{p_{R,\text{rest}}} =
0.43$), and four of the seven residues in the allosteric site have
high quantile scores, $p_R \geq 0.9$ (Table~\ref{tbl:chey}).  
In addition, we find a number of previously unidentified distant
surfaces with high quantile scores (Fig.~\ref{fig:chey}a), 
which could correspond to putative (orphan) allosteric sites.

\begin{table}
  \small
  \caption{Propensities of residues in CheY relative to the active site, ranked by quantile score 
  ($p_R \geq 0.90$).  Residues marked with a star are within 3.5~\AA~of the allosteric effector.}
  \label{tbl:chey}
  \medskip
  \begin{tabular}{ccc}
    \hline
    Residue & $\Pi_R^{\text{act}}$ & $p_R$ \\
    \hline
    D12 & 0.0076 & 1 \\
    E89* & 0.0370 & 0.984 \\
    N62 & 0.0017 & 0.984 \\
    D57* & 0.0094 & 0.968 \\    
    K45 & 0.0015 & 0.968 \\
    T87* & 0.0283 & 0.968 \\
    M85 & 0.0321 & 0.968 \\
    E35 & 0.0019 & 0.952  \\    
    L116 & 0.0189 & 0.952 \\
    W58* & 0.0247 & 0.936  \\
    L43 & 0.0030 & 0.921  \\
    F124 & 0.0120 & 0.905  \\
    L120 & 0.0189 & 0.905 \\    
    \hline
  \end{tabular}
\end{table}

In contrast to caspase-1 above, using a RRIN with cut-off radius of 6 \AA, 
we find that the phosphorylation site of CheY is identified as a hot-spot:
the average quantile score of allosteric residues is much higher for 
the rest of the residues ($\overline{p_{R,\text{allo}}} = 0.72 >
\overline{p_{R,\text{rest}}} = 0.46$).  The RRIN detection is robust over a
range of cut-off radii between 6\AA-16\AA~(Table~S6 and Fig.~S5).  This result suggests that sometimes
(as for CheY) it is the \textit{topology} of the protein
structure that is important for signal propagation, whereas in other cases (as
for caspase-1) the specific \textit{atomistic
structure} given by the \textit{chemistry} of the side-chain
interactions matters for allosteric propagation.  Our all-atom methodology
incorporates both aspects consistently. 

\subsubsection{Comparing propensities of active and inactive structures helps identify allosteric communication networks}
 
To get a more detailed picture of the pathways involved in allosteric
communication, we examined the specific bonds with high propensity in
the structure of fully activated CheY (1F4V). Considering
high quantile scores ($p_b \geq 0.97$), we find several bonds
connecting the allosteric phosphorylation site to the key binding site
residue Y106 (Fig.~\ref{fig:chey}b).  One pathway comprises bonds
between T87:E89 ($p_b=0.991$) and E89:Y106 ($p_b = 0.977$), whereas a
second pathway is formed by K109, which has high quantile score bonds
with D12 ($p_b=1$) and D57 ($p_b=0.993$).  These residues have been
discussed extensively in the biochemical literature and are known to
be crucial for allosteric signalling (see Discussion).

\begin{figure}[tb]
  \centering
  \includegraphics[width = \columnwidth]{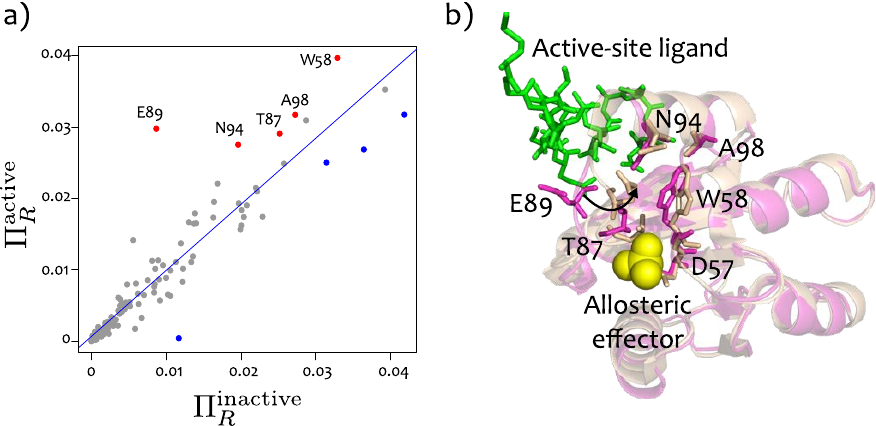}
  \caption{\textbf{Comparison of residue propensities between active
      and inactive conformations of CheY.}  (a) The
    propensities most increased in the active X-ray structure (1F4V) as compared to 
    to the inactive X-ray structure(3CHY), as identified by Cook's
    distance, are coloured red and labelled.  
    (b) Superposition of active (1F4V - beige) and inactive (3CHY
    - pink) conformations.  The residues found in (a) form a pathway between the
    allosteric site and the ligand binding surface.}
  \label{fig:chey_act_inact}
\end{figure}

In addition to fully activated CheY, we also studied four additional
structures corresponding to conformations of CheY
across a range of activation stages (details of the PDB files and
an in-depth comparison is given in SI Section 3).  Importantly, the
profiles of bond-to-bond propensities are similar across all
conformations (Fig.~S1), highlighting the robustness of the propensity scores to
local dynamical rearrangements across different conformations.
In particular, the propensities of residues in the active (1F4V) and inactive (3CHY)
conformations show a strong positive correlation ($r=0.94$, Fig.~\ref{fig:chey_act_inact}a).  
Using Cook's distance, a well-known method for detecting
influential points in linear regression~\cite{cook1979influential}, we
identified E89, N94, T87, A98, and W58 as 
the residues with highly increased propensity in the
active conformation as compared to the inactive conformation. 
Superposition of the active and inactive structures shows that the
large displacement of E89 causes the formation of a tighter network of
interactions involving N94, T87, and W58 in the active conformation 
(Fig.~\ref{fig:chey_act_inact}b).  
Interestingly, the propensity of the allosteric phosphorylation site D57 
is similar in the active and inactive conformations; 
in the inactive conformation, D57 forms a stronger hydrogen bond with K109 
than it does in the active conformation, yet the weakening of this bond in the active conformation is
compensated for by the formation of the network involving W58 and
E89. 
Hence activation induces a structural re-arrangement of the
network of bonds that connect the phosphorylation site to the active
site.

\subsubsection{Variability of bond-to-bond propensities in NMR ensembles uncovers transient effects in the allosteric network}

CheY exists in dynamic equilibrium between its active and inactive
conformations. Indeed, X-ray structures have revealed an intermediate
conformation with only the binding site adopting the active
conformation~\cite{dyer2006switched, lee2001crystal_2}.

To explore the effect of small structural changes on the propensities
of residues of CheY, we analysed 20 NMR
structures of the inactive conformation \textit{apo}-CheY (PDB: 1CYE)
and 27 NMR structures of the fully activated CheY bound to the
phosphate mimic BeF$_3$ (PDB: 1DJM).  We calculated the average
$\langle{\Pi_R}\rangle_{\text{NMR}}$ and the standard deviation
$\text{SD}(\Pi_R)_{\text{NMR}}$ of the propensity of each residue
over the ensemble of NMR structures.
We then compared these properties computed over the NMR ensemble against
those obtained from the X-ray structure.

The results of this comparison (NMR ensemble \textit{vs.}  X-ray structure) are
different for the inactive and active structures, suggesting that the dynamical reconfigurations
have a (consistent) effect on our measure.
For the inactive \textit{apo}-CheY, the average NMR propensity over the
ensemble ${\langle{\Pi_R^{\text{inact}}}\rangle}_{\text{NMR}}$ for
each residue is strongly correlated ($r^2=0.96$) with its X-ray
propensity $\Pi^{\text{inact}}_{\text{R, X-ray}}$ (Fig.~S2a).
For the active Che-Y, however, the correlation is weaker ($r^2=0.84$,
Fig.~S2b).
McDonald \textit{et al}~\cite{mcdonald2012segmental} have suggested
that phosphorylation causes a slight increase in the flexibility of CheY,
as signalled by increased B-factors and root mean square
fluctuations (RMSF) across the NMR ensemble for active CheY. 
This enhanced flexibility may account for the greater difference between the NMR ensemble and
the X-ray structures for the active conformation.

\begin{figure}[tb]
  \centering
  \includegraphics[width=8.5cm]{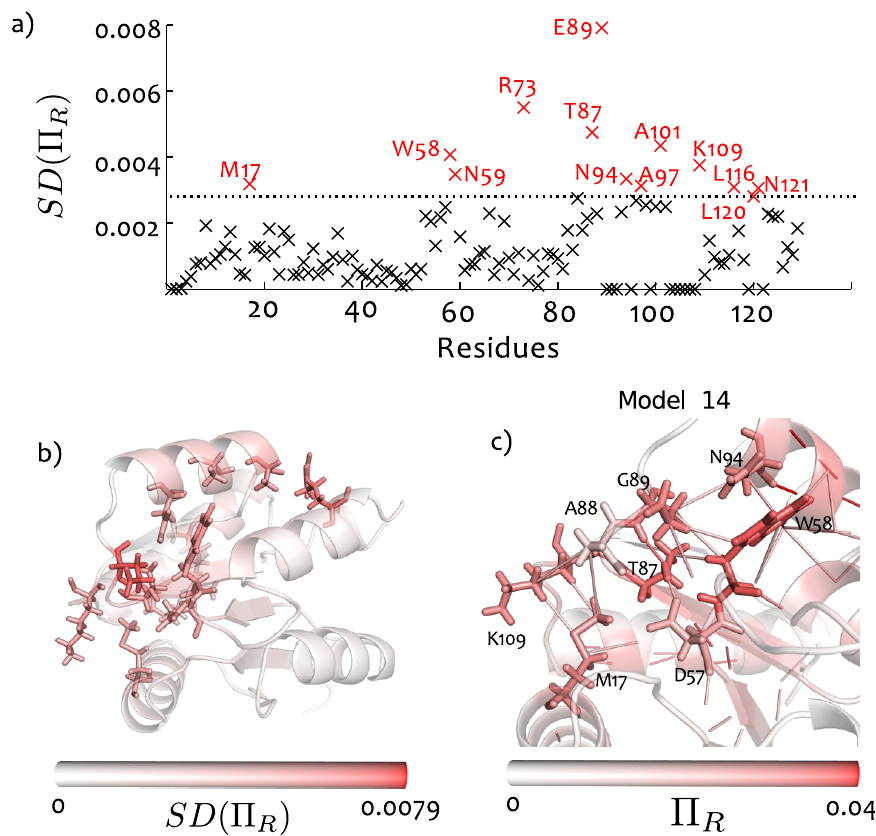}
  \caption{\textbf{Increased variability of the propensity in NMR
      structures of active CheY reveals additional relevant
      residues.}  (a) Standard deviation of the residue
    propensities recorded over the NMR ensemble of 27
    conformations corresponding to active CheY.  
    The dashed line separates the top 10\% of the residues
    by $SD(\Pi_R)$. Residue M17 has high NMR variability, 
    although it was not identified in the X-ray
    structure as having high $\Pi_b$.  
    (b) The residues with high standard deviation are
    indicated on the structure, coloured by their NMR
    standard deviation. (c) Interactions coupling
    M17 to Y106 and the active site is shown in one of
    NMR conformations (model 14) of the active CheY.  
    Residues coloured by their propensity 
    $\Pi_R$ in this particular conformation.}
  \label{fig:chey_nmr}
\end{figure}

The variability of the propensity of each residue, computed from the NMR active
ensemble, is shown in Fig.~\ref{fig:chey_nmr}a.  Among the residues with high (top 10\%) 
NMR standard deviation $\text{SD}(\Pi_R^{\text{act}})_{\text{NMR}}$, 
we find W58, T87, E89, and K109, which were also found to have
high propensities in the active X-ray structure.  These
residues are known to be functionally relevant, and recent NMR
relaxation-dispersion experiments have suggested that they form part
of an allosteric network undergoing asynchronous local
switching~\cite{mcdonald2012segmental}.  Other residues with high NMR
standard deviation are A101, R73, L116, K119, and N121.  Of these,
A101 lies in the alpha-helix forming the top half of the ligand
binding site, and the high variance of A101 and R73 can be explained
by an unstable hydrogen bond between the two residues, which is
transiently present across the active ensemble.  On the other hand,
L116 and N121 lie in the alpha-helix forming the other side of the
FliM binding site: L116 forms a transient alpha-helical hydrogen bond
with the ligand binding residue K119, and N121 forms fluctuating
hydrogen bonds with residues in, and adjacent to, the active site
(Fig.~\ref{fig:chey_nmr}b).

The large NMR variability of residue M17, which is 15\AA~away from the
active site, is of particular interest.  CheY is intolerant to
mutation of M17 \cite{bourret1993activation,smith2003investigation},
and it has been recently reported that this mutation causes chemical
shift changes at Y106~\cite{mcdonald2013colocalization}, a key residue
in the distant FliM binding site.  Our analysis shows that the
propensity of M17 is higher in the active structure (both
NMR and X-ray) than in the inactive structure: $ { \langle
  \Pi^{\text{act}}_{\text{M17}}\rangle}_{\text{NMR}}= 0.0173 >
\Pi^{\text{act}}_{\text{M17, X-ray}} = 0.0113 > \langle
\Pi^{\text{inact}}_{\text{M17}}\rangle_\text{NMR} = 0.0094 >
\Pi^{\text{inact}}_{\text{M17, X-ray}} = 0.0081$.  Furthermore, the
NMR standard deviation of the propensity is higher in the active than in
the inactive ensemble:
${SD(\Pi^{\text{act}}_{\text{M17}})}_{\text{NMR}} = 0.0032 >
{SD(\Pi^{\text{inact}}_{\text{M17}})}_{\text{NMR}}=0.0016$.  
All these results indicate that phosphorylation (i.e., activation) causes
transient pathways to form between M17 and the active site which are
not observed in the X-ray structure.  By examining bonds with high
propensity between M17 and Y106, we visually uncover a
communication pathway involving residue K109 and residues in the
flexible $\alpha4-\beta4$ loop: T87, A88, and E89.  Indeed, when we
examine the individual NMR conformation in which
M17 has the highest propensity, M17 bonds directly with A88 and is
indirectly connected to T87 through a hydrogen bond with K109
(Fig.~\ref{fig:chey_nmr}c).  This suggests that M17 is transiently coupled to Y106
through a network of hydrogen bonds and hydrophobic contacts not
captured in the active X-ray structure. In general,
the transient making-and-breaking of particular bonds in the NMR ensemble translates into 
highly variable propensities associated with functionally important allosteric residues.

\subsection{Structural water molecules are crucial to the allosteric communication network in h-Ras}
\label{sec:h-Ras}

The enzyme h-Ras is a GTPase involved in signal transduction pertaining to 
cell-cycle regulation~\cite{mccormick1995ras}.  Crystallographic evidence shows
that calcium acetate acts as an allosteric activator in this
process~\cite{buhrman2010allosteric}. By comparing the calcium
acetate-bound structure to the inactive structure, Buhrman~\textit{et
    al} have proposed a network of hydrogen bonds, involving structural water
molecules, linking the allosteric site to the catalytic residue
Q61~\cite{buhrman2010allosteric}.

\begin{table}[h!]
  \small
  \caption{Top bonds ranked by propensity quantile score for h-Ras ($p_b \geq 0.99$)}
  \label{tbl:hras}
  \medskip
  \begin{tabular*}{\columnwidth}{@{\extracolsep{\fill}}lccc}
    \hline
    Bond & $\Pi_b$ & Distance (\AA) & $p_b$ \\
    \hline
    Q99:HOH727 & 0.0051 & 14.8 & 0.9991 \\
    K117:G13 & 0.026 & 2.76 & 0.9983 \\    
    HOH727:S65 & 0.0067 & 12.2 & 0.9974 \\    
    R164:E49 & 0.0013 & 25.0 & 0.9974 \\
    I21:S17 & 0.019 & 4.83 & 0.9965 \\
    D47:R161 & 0.0015 & 21.6 & 0.9948 \\    
    H27:Q25 & 0.0075 & 10.8 & 0.9940 \\
    V8:L56 & 0.0010 & 9.05 & 0.9940 \\
    R161:D47 & 0.0013 & 21.6 & 0.9931 \\
    I24:K42 & 0.0035 & 14.8 & 0.9922 \\
    Q22:A146 & 0.017 & 5.09 & 0.9905 \\
    \hline
  \end{tabular*}
\end{table}

\begin{figure*}[htbt!]
  \centering
  \includegraphics[width = \textwidth]{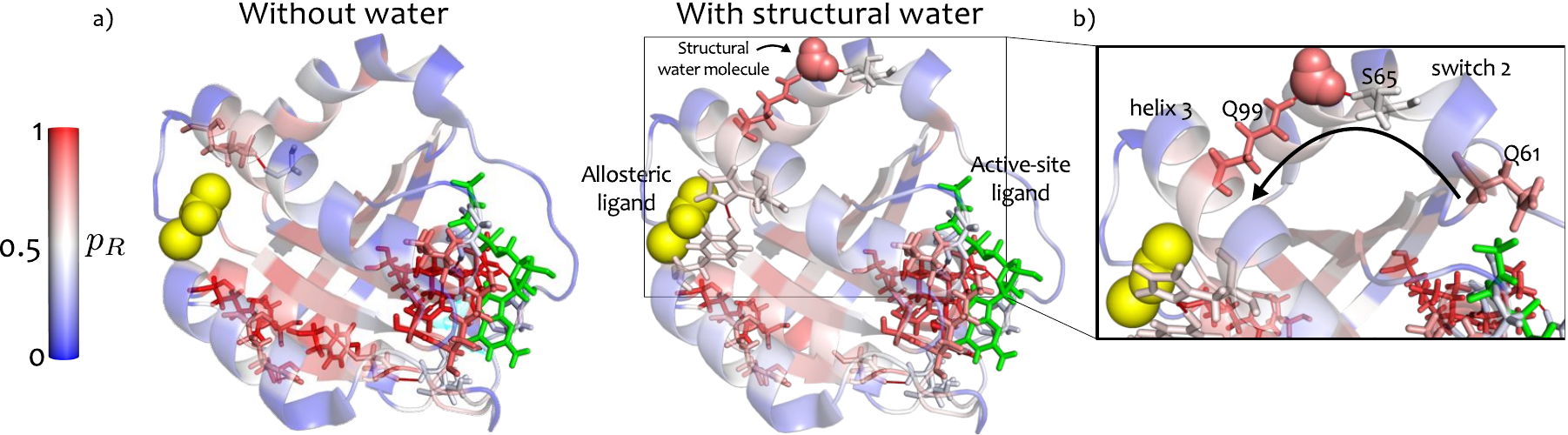}
  \caption{\textbf{Structural water molecules are essential for the
      allosteric pathway in hRas.} (a) Top percentile bonds by propensity quantile
    score ($p_b \geq 0.99$) are shown on the structure:
    the left panel shows pathways identified without the inclusion of
    water molecules, and the right panel when structural water
    molecules are included in the graph. The structural water allows the
    formation of a pathway between the bottom of the switch 2 region
    and the top of helix 3, where the allosteric binding site is
    situated. The crucial water molecule which connects Q99 and S65 is
    indicated.  (b) Blow-up indicating details of the pathway formed by
    Q99, a water molecule and S65, linking the allosteric pocket
    to the switch 2 region.  The catalytic residue Q61 is shown at the
    bottom of switch 2.}
  \label{fig:hras}
\end{figure*}

We have calculated the propensities and quantile scores 
of hRas (bound to substrate and allosteric activator, PDB code: 3K8Y) 
for two scenarios: with and without inclusion of structural water molecules in the graph. 
In the absence of water (Fig.~\ref{fig:hras}a left), we find no bonds or
residues with high quantile scores near the allosteric binding pocket.
When we include the 8 molecules of structural water present in the PDB file, 
we identify a high quantile bond between the allosteric site residue Y137 and H94,
and a pathway involving a structural water molecule that connects the
allosteric region to a catalytic residue (Fig.~\ref{fig:hras}b). In
Table~\ref{tbl:hras}, we show that the Q99-water and S65-water bonds
involved in this pathway have 1st and 3rd highest quantile scores out
of the 1159 weak interactions in the protein.

This water-mediated link between Q99 and S65 connects the allosteric
binding pocket on helix 3 with the helical structure known as the
switch 2 region, at the bottom of which lies Q61, which has been
identified as a key catalytic residue~\cite{buhrman2010allosteric}.
Our results thus suggest that structural water plays a crucial role in
coupling the allosteric effector to the catalytic residue Q61.

\subsection{Absolute bond propensities against a reference set from the SCOP protein database}
\label{sec:scop}

\begin{figure*}[tb]
  \centering
  \includegraphics[width = \textwidth]{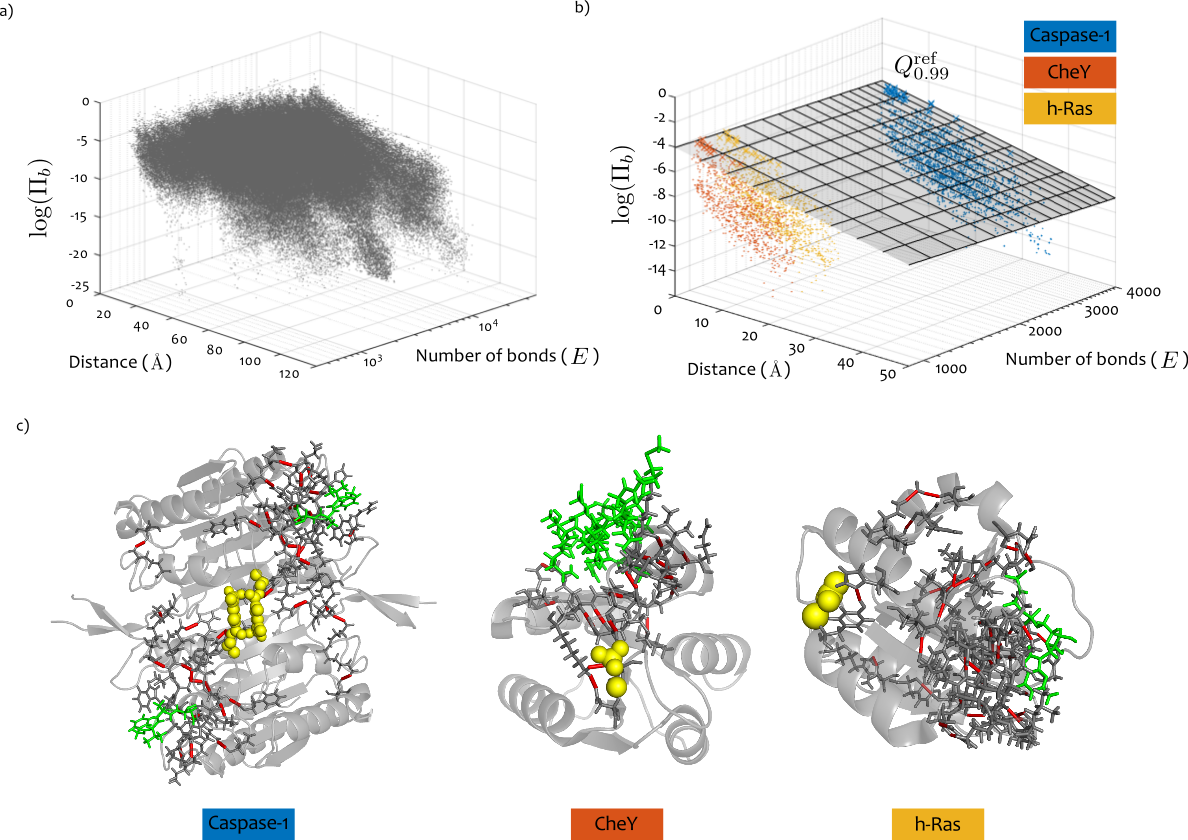}
  \caption{\textbf{Absolute propensities: calibration against the SCOP
      reference set.}  (a) The logarithm of the bond propensity
    $\log(\Pi_b)$ of all 465,409 weak bonds in the reference set
    (100 proteins from the SCOP database) plotted against $d$,
    the distance from their corresponding active site, and
    $E$, where $E$ is the number of weak bonds in the corresponding
    protein. 
      (b) The log propensities $\log(\Pi_b)$ for caspase-1
    (blue), CheY (orange), and h-Ras (yellow) are plotted together
    with the plane defining the 99th quantile fit obtained by solving
    the optimisation~Eq.~\eqref{eq:qr_multiple} against the SCOP set of
    bonds shown in (a).  For each of the three proteins, there are
    bonds lying above the 99th quantile plane. 
  (c) The bonds above the plane in (b) have $p^{\text{\tiny{ref}}}_b > 0.99$ and 
  are marked in red on the corresponding
  protein structures (active site ligand in green, allosteric ligand as yellow spheres).  
  The bonds thus identified play key allosteric roles, in agreement with the
  \textit{intrinsic} results in previous sections.  }
\label{fig:scop}
\end{figure*}

The quantile regression scores $p_b$ in the previous sections identify
bonds with high propensities as compared to other bonds
which are at a similar distance from the active site \textit{within the same protein}.  
To assess the \textit{absolute} significance of bond propensities, we
have assembled a reference set of 100 protein structures
from the SCOP database~\cite{murzin1995scop} (see SI, Section 4),  
and calculated the propensities with respect to the active site of all
465,409 weak bonds in this reference set (Fig.~\ref{fig:scop}a). 
Because the propensities are dependent on both the distance from the
active site, $d$, and the total number of weak interactions in the
protein, $E$, we apply quantile regression against both $d$ and $E$
(as given by Eq.~\eqref{eq:qr_multiple} in Materials and Methods) to obtain
fitted quantiles for the reference set.
The quantiles computed from this reference set can then be used to obtain absolute
bond propensity scores, denoted $p^{\text{\tiny{ref}}}_b$, for any given
protein without recomputing the regression. 

We have obtained the absolute quantiles 
$p^{\text{\tiny{ref}}}_b$ for the propensities of the three proteins (caspase-1,
CheY, and h-Ras) studied above (Fig.~\ref{fig:scop}b).
Reassuringly, the significant bonds are also found to be important according to the
absolute measure, 
with a strong correlation 
between quantile scores and absolute bond quantile scores
(Fig.~S3). Visualising the bonds with $p^{\text{\tiny{ref}}}_b \geq 0.99$ shows 
they form pathways between the active and
allosteric sites (Fig.~\ref{fig:scop}c).  These results confirm that these
bonds are important not only relative to other bonds and residues within each of 
the respective proteins, but also in absolute terms when compared to
the protein reference set.

\subsection{Validating the propensity measure: predicting allosteric sites in an extended set of proteins}
\label{sec:test_set}

To test the validity of our methodology, we have computed the
bond propensities for an additional 17 proteins known to exhibit allostery.  
Ten of these proteins were taken from a benchmark set collected by Daily~\textit{et
  al}~\cite{daily2009allosteric} and a further 7 were obtained through
an extensive literature search.  (Five proteins in~Ref.~\cite{daily2009allosteric} 
could not be used either due to the presence of non-standard amino-acids, to 
the absence of an allosteric ligand, or to a mismatch between the oligomeric state 
of the active and inactive structures.) 
The details and structures of all 20 proteins analysed in the paper 
are given in the SI (Table S2 and Figure~S4). 

\begin{figure*}[]
  \centering
  \includegraphics[width = \textwidth]{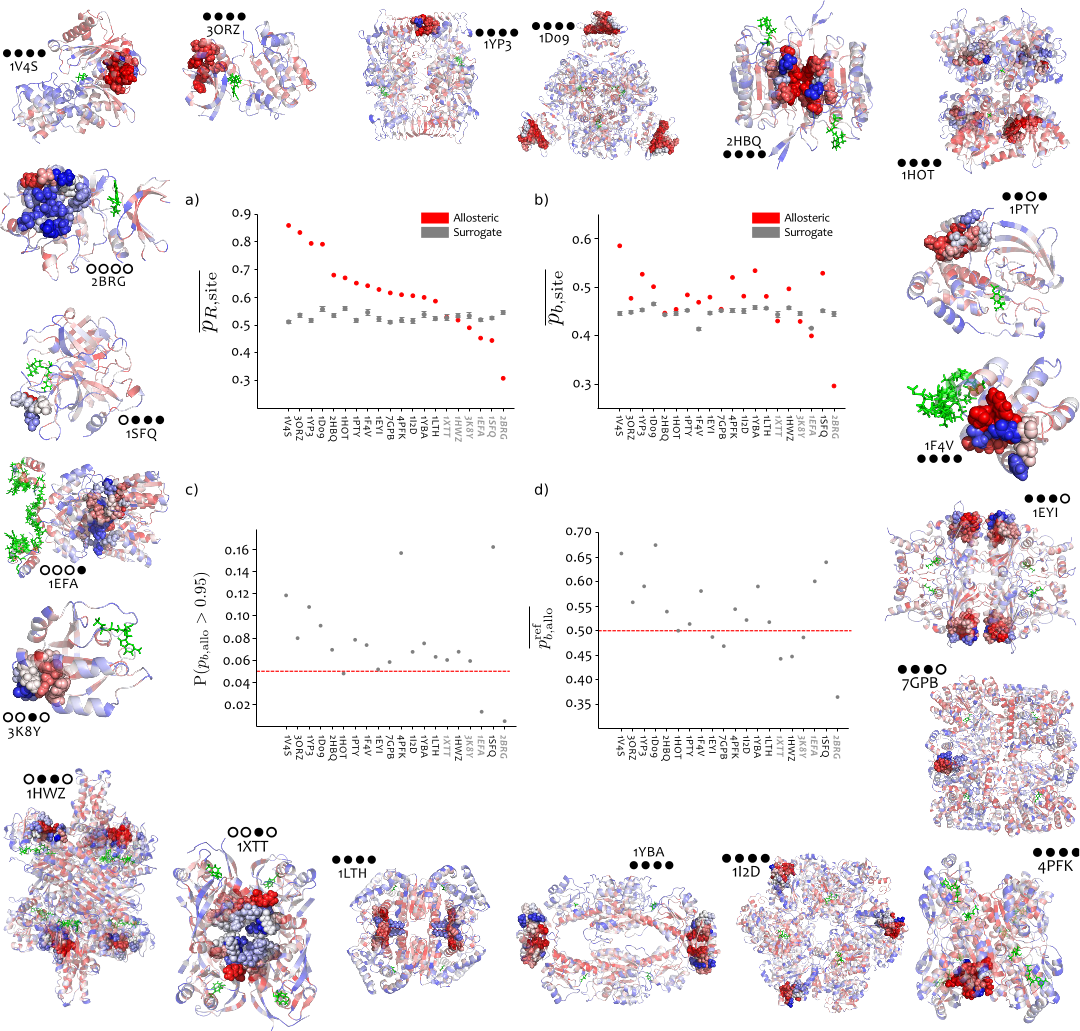}
  \caption{\textbf{Prediction of allosteric sites based on bond-to-bond
  propensity for a test set of 20 allosteric proteins.}
The structures of the 20 proteins in the test set (labelled by PDB code) 
have their residues coloured by their quantile score $p_R$, 
and the allosteric site is shown as spheres.  
For full details of these proteins, see Table S2 in the SI. 
The four statistics computed from our propensity are showed in the centre:
(a) average \textit{residue} quantile scores in the allosteric site
    $\overline{p_{R,\text{allo}}}$ (red) compared to the average
    score of 1000 surrogate sites $\langle\overline{p_{R,\text{site}}}\rangle_{\text{surr}}$ (grey), 
    with a 95\% confidence interval for the average from a bootstrap
    with 10000 resamples (see Section~\ref{sec:bootstrap});
(b) average \textit{bond} quantile scores in the allosteric site against
the equivalent bootstrap of 1000 surrogate sites;
 (c) tail of the distribution of bond propensities, i.e., 
 proportion of allosteric site bonds with quantile scores $p_{b,\text{allo}} > 0.95$.
 Proteins above the \textit{expected} proportion of 0.05 (red line)
 have a larger than expected number of bonds with high quantile scores; 
(d) average \textit{reference} bond quantile score in the allosteric site
    $\overline{p^{\text{ref}}_{b,\text{allo}}}$.
    The red dotted line indicates the expected value of 0.5, and proteins above
    this line have a higher than expected reference quantile score.
    For the numerical values of all measures see Table~S3 in the \textit{SI}.
The four circle code by each protein indicates whether the allosteric site
is identified (filled circle) or not identified (open circle) according to each of the four measures (a)--(d). 
19/20 allosteric sites are identified by at least one measure, and 15/20 sites are identified by 
at least three of four measures.}
  \label{fig:test_set}
\end{figure*}

For each protein, we calculated the propensity quantile scores 
of all its bonds and residues, both intrinsic ($p_b, p_R$) and absolute 
($p_{b}^{\text{ref}}$), with respect to their active site.
Again, no \textit{a priori} knowledge  about the allosteric site was used. 
Figure~\ref{fig:test_set} shows the structures of the 20 proteins coloured
according to the residue quantile score $p_R$, with the allosteric sites 
marked with spheres.  
To validate our findings on this test set, we used the location 
of the allosteric site \textit{a posteriori} and evaluated the significance 
of the computed allosteric quantile scores according to 
four statistical measures (Fig.~\ref{fig:test_set}a--d). 
See Section~\ref{sec:metrics} for a full description and definitions.

All combined, the allosteric site is detected significantly by at least one of the four measures 
in 19 out of 20 proteins in the test set, and is detected by three or more of our measures in 
15 out of 20 proteins in the test set. The full numerical values are given in the SI (Table S3). 
In practice, all statistical measures provide important and complementary information about the
distribution of bond propensities, 
and can be used in conjunction for the detection of allosteric sites. 

\section{Discussion}

Using a description of protein structural data in terms of 
an atomistic energy-weighted network with both covalent and non-covalent bonds, 
we have defined a graph-theoretic measure of bond propensity and 
used it to identify allosteric sites in proteins without prior
  information as to their location.  Our propensity measure
identifies bonds that are strongly coupled to the active site via
communication pathways on the protein graph, even if they might be
separated by large geometric distances. 
Allosteric sites correspond to hot spots, i.e., sites with high
propensity to perturbations generated at the active site, as measured
by their quantile score relative to other sites in the protein that
are at a similar distance from the active site. 
This finding suggests that the structural features embedded in the
architecture of the protein are exploited so as to enhance the
propagation of perturbations over long distances.

By using a representative reference set of 100 proteins randomly 
assembled from the SCOP database, we also computed absolute quantile scores 
to further confirm the significance of bond propensities. 
One advantage of this absolute measure is that
the quantile regression over the reference SCOP set does not need 
to be re-calculated, and the absolute bond quantile scores in any protein of
interest can be obtained directly against them, thus reduces the analysis time 
even further.

We have validated our method against a test set of 20 allosteric proteins
without using any \textit{a priori} information of their allosteric
sites.  We used our propensity quantile scores and a structural
bootstrap to define four statistical measures of significance 
based on the average and tail of the distribution of bond 
propensities in the allosteric site. The allosteric site is detected 
for 19/20 proteins, according to at least one statistical measure, 
and for 15/20, according to at least three of our four statistical measures.
These findings indicate the robustness of the bond-to-bond propensity 
as a predictor of allosteric sites, which could be used to guide structure-based drug 
discovery efforts, e.g., by ranking potential binding sites based on their
allosteric potential.  Our method also uncovers hot spots not previously
identified as allosteric sites (see our results for CheY in
Fig.~\ref{fig:chey}).  Hardy and Wells have discussed the existence of
`orphan' or `serendipitous' allosteric sites, i.e., sites targeted by
as-yet undiscovered natural effectors or open for exploitation by
novel small molecules \cite{hardy2004searching}.  The identified sites
could thus provide targets for mutational analysis or allosteric
small-molecule inhibition.

We have exemplified the use of atomistic propensities 
with the detailed analysis of three proteins (caspase-1, CheY, and h-Ras),  
focussing on the contribution of high propensity bonds to pathways (or networks) of 
weak bonds linking the active and allosteric sites.  The weak bond network we found 
in caspase-1 (E390/R286/S332/S339/N337) has been previously
tested experimentally and shown to be functionally
important~\cite{datta2008allosteric}.  In CheY, we found that bonds
between T87:E89 and E89:Y106, with very high quantile scores, are key to 
an important pathway for transmission of the signal
induced by phosphorylation, also consistent with experimental
evidence~\cite{dyer2006switched, mcdonald2012segmental,
  zhu1996tyrosine}.  
 We also found a second pathway in CheY involving
the bond K109:D57 (3rd highest quantile score). Interestingly,
mutation of K109 abolishes chemotactic activity
\cite{bourret1993activation} and has been proposed to form part of the
post-phosphorylation activation mechanism \cite{bellsolell1996three}.
Our analysis of bond propensities across active/inactive conformations and NMR data 
further confirmed that  K109 forms a central link in the communication
between the phosphorylation and binding sites in CheY.

Determination of protein structures from NMR solution experiments
results in multiple models, each consistent with
experimentally-derived distance restraints.  The resulting `ensemble'
of structures should be interpreted with caution, since variation
could be due to actual flexibility and thermal motion during the
experiment, or to inadequate (or under-constrained) interatomic distance
restraints. Hence the set of NMR structures is not a true thermodynamic
ensemble.  However, our analysis suggests that the variation within the
NMR structures can reveal functionally relevant
information.  For CheY, residues with highly variable
propensities across the NMR ensemble (E89/W58/T87/E89/K109) coincide with
those forming an asynchronously switching allosteric circuit after
phosphorylation, as revealed by NMR relaxation-dispersion
experiments~\cite{mcdonald2012segmental}.  We also identify residue
M17 as having high propensity in the NMR ensemble
due to the presence of a transient network of
interactions. This may explain experiments showing that mutation of
M17 has a functional effect and causes chemical shift changes at Y106
\cite{mcdonald2013colocalization}.  

Comparing the results across conformations indicates that propensities are
fairly robust to local dynamic fluctuations, as seen by the strong correlation
between active and inactive conformations and across 
NMR structures (Fig.~\ref{fig:chey_act_inact} and Figs.~S1~and~S2). 
As an additional confirmation of its robustness, we show in
SI (Section~6, Tables~S4~and~S5) that the propensities,
and the ensuing identification of significant residues and bonds,  
are generally robust to both randomness in the bond energies
and to the breakage of a large proportion of weak interactions.
On the other hand, as discussed above, our graph-theoretic analysis shows 
that further information about residues and bonds can be obtained by evaluating the highest variations 
induced by dynamical and structural variations.
A fuller investigation of the effect of dynamics on the calculated propensities 
using experimental data (NMR, conformational studies) and complemented 
with the analysis of molecular dynamics simulations  
would thus be an interesting area for future research.

The role of structural water molecules in mediating allosteric communication has
so far received limited attention. In a recent study of a PDZ domain, 
Buchli \textit{et al.} suggest that changes
in water structure could be responsible for mediating communication
with remote parts of the protein~\cite{buchli2013kinetic}.  Our analysis 
of h-Ras found that \textit{structural} water molecules
in the protein graph are necessary to reveal a pathway linking the
allosteric and active sites.  These results and the
findings of Buchli \textit{et al.}  suggest that novel methods to study
interaction networks between proteins and water are worth investigating.  
However, beyond including structural water when present 
in experimental structures (as in h-Ras here), the addition of bulk water 
would require the simulation of hydration, including energy minimisation and 
equilibration steps. This could constitute another direction of future research, since 
the computational efficiency of our method would make it possible
to analyse all-atom representations of such hydrated structures.

To what extent does the identification of the allosteric site require an
atomistic, chemically detailed construction of the graph?  %
To answer this question, we applied our propensity measure to 
residue-residue interaction networks (RRINs), the coarse-grained residue-level models 
used in almost all previous network analyses of proteins.
For caspase-1, we found that allosteric residues are not found significant 
in RRINs (across several different cut-off radii), whereas, on the other hand, 
the allosteric site of CheY was consistently detected in both the atomistic and 
residue-level descriptions.  This indicates that both coarser topological features, as well as more
detailed chemical communication pathways can be relevant depending on
the protein; e.g., in caspase-1, the binding of the allosteric
ligand perturbs a network of strong hydrogen bonds and salt-bridges as identified in our analysis.  
Therefore, the atomistic graph with detailed physico-chemical information can in some cases
provide important features underpinning the communication features of the protein. 
The analysis of coarse-grained models with a variety of cut-off radii for
all 20 proteins in our allosteric test set in SI Section~7 confirm that the outcome for RRINs 
varies for each protein and can also be dependent on the choice of cut-off radii~\cite{ribeiro2015energy}.
We would like to emphasise, however, that our propensity measure is principally agnostic to
the protein network model under analysis, thus allowing for the evaluation 
of distinct graph-construction techniques (e.g., atomistic vs coarse-grained) or the
use of different force-fields.  Again, this would open another interesting avenue for future work.

Finally, it is important to remark that our method is
computationally efficient.  To obtain the bond-to-bond propensities,
we only need to solve a sparse linear system 
(Eq. \ref{eq:M}) involving the (weighted) Laplacian of the protein
graph. As discussed in Section~\ref{sec:comp_cost}, 
recent algorithmic advances allow us to solve such linear systems in \textit{almost linear time}
\cite{spielman2004nearly, kelner2013simple}.  Hence protein complexes of
$\sim100,000$ atoms can be run in minutes on a standard
desktop computer.  We can thus maintain atomistic detail, yet
analyse large biomolecular complexes that are intractable for
traditional computational methods.

\section{Materials and Methods}
\label{sec:Methods}

\subsection{Mathematical derivation of the bond-to-bond propensity}\label{sec:bond-bond}
\subsubsection{Fluctuations and the edge-to-edge transfer matrix of a graph}
\label{sec:M-matrix}
The edge-to-edge transfer matrix $M$ was introduced in
Ref.~\cite{schaub2014structure} as a non-local edge-coupling matrix
for the analysis of weighted undirected graphs, based on the concept
of flow redistribution.  In that work, it was shown that the element
$M_{ji}$ reflects the effect that an injected flux on edge $i$ has on
the flux along edge $j$ after the fluxes are redistributed over the
whole graph when at equilibrium.  Alternatively, $M$ can be understood
as a discrete Green's function in the \textit{edge space} of the
graph.  See~\cite{schaub2014structure} for detailed derivations and
applications.

In this paper, we derive a complementary interpretation of the matrix
$M$.  As shown below, the edge-to-edge transfer matrix can be
understood as describing how the fluctuations of the edge weights 
propagate through the graph. This new re-interpretation
underpins the work in this paper, as it highlights the importance of $M$ 
for the analysis of bond fluctuations in biomolecules.

As our starting point, consider the well-known Langevin equation, sometimes
denoted the \emph{heat kernel equation}
\cite{chung2000discrete, reuveni2010anomalies}:
\begin{equation}\label{eq:langevin}
  \dot{\mathbf{x}} = -L\mathbf{x} + \boldsymbol{\epsilon}.
\end{equation}
Formally, Eq.~\eqref{eq:langevin} has the same structure as the
canonical model for scalar vibrations with nearest neighbour
interactions encoded by the matrix $L$~\cite{nakayama1994dynamical,
  leitner2008energy}.  Alternatively, Eq.~\eqref{eq:langevin} may be
considered as a model of a diffusing particle transitioning like a
random walker on the underlying graph structure represented by $L$. 
In contrast to residue level methods~\cite{chennubhotla2007signal},
the variable $\mathbf{x}$ is associated with \textit{atomic}
fluctuations, i.e., our graph model reflects an atomic
description that incorporates physico-chemical interactions derived
from the three dimensional structure of the protein recorded in a PDB file.  
The resulting graph contains energy-weighted interactions representing
bonds in the protein, including both covalent bonds and weak
interactions such as hydrogen bonds, salt bridges, hydrophobic
tethers, and electrostatic interactions.  For details of
the graph construction see Section~\ref{sec:ntwk_cnstr} and SI.

The matrix $L$ is the graph Laplacian~\cite{biggs1993algebraic, delmotte2011protein}:
\begin{equation}
  L_{ij} = \begin{cases}
    -w_{ij},& i \neq j \\
    \sum_{j}w_{ij}, & i = j,
  \end{cases}
\end{equation}
where $w_{ij}$ is the weight of the edge between nodes (atoms)
$i,j$. In this case, $w_{ij}$ is the energy of the bond between both
atoms. Thermal background fluctuations are modelled by
$\boldsymbol{\epsilon}$, a zero mean white Gaussian noise input
vector, i.e., a (simple) heat bath acting independently on all atomic
sites with covariance matrix
\begin{equation}
  \langle \epsilon_i(t)\epsilon_j(s)\rangle = \delta(t-s)\delta_{ij},
\end{equation}
where $\delta$ stands for the Dirac delta function.

Instead of focusing on the \textit{atomic} (node) variables
$\mathbf{x}$, we wish to study the coupling between \textit{bonds}, 
and thus concentrate on the bond (edge) variables of the
graph:
\begin{equation}
  y_b = x_{\text{head}(b)} - x_{\text{tail}(b)}.  
\end{equation} 
Clearly, $y_b$ describes the difference of the node variables at the
endpoints of the associated bond $b$, i.e., a \textit{fluctuation}
associated with the bond between two atoms.  The vector of bond
fluctuations can be compactly represented in vector notation as:
$$ {\mathbf{y}=B^T \mathbf{x}},$$
where $B$ is the incidence matrix of the graph relating each
edge variable to its corresponding node variables, i.e., $B_{bi} = 1$
if node $i$ is the head of bond $b$; $B_{bi}=-1$  if node $i$ is the
tail of bond $b$; and $B_{bi}=0$ otherwise.

We can now calculate the cross-correlations between edge fluctuations
as:
\begin{flalign}
  \label{eq:edge_corr}
  \mathcal{R}(\tau) :=
  \mathbb E[\mathbf{y}(t)\mathbf{y}^T(t+\tau)] =
  \dfrac{1}{2}B^T\exp(-\tau L)L^\dag B,
\end{flalign}
where $L^\dag$ is the (Moore-Penrose) pseudoinverse of the Laplacian
matrix. 
Each entry $[\mathcal{R}(\tau)]_{b_1b_2}$ describes how a fluctuation
at bond $b_2$ is correlated with a fluctuation at bond~$b_1$ at time
$\tau$. See SI for a full derivation of Eq.~\eqref{eq:edge_corr}.

Biophysically, we are ultimately interested in the
\textit{energy fluctuations} induced by bonds on other
bonds. Therefore, we multiply the correlation matrix
$\mathcal{R}(\tau)$ by the diagonal matrix of bond energies, $G =
\text{diag}(w_b)$:
$$ M(\tau):= G \, \mathcal{R}(\tau),$$
to obtain the matrix of \textit{bond-to-bond energy correlations}
with delay $\tau$.
Our measure of bond-to-bond propensity is obtained from the
\textit{instantaneous} correlations (i.e., $\tau = 0$) leading to the
edge-to-edge transfer matrix:
\begin{equation}
  \label{eq:M}
  M :=  M(0) = \dfrac{1}{2}GB^TL^\dag B.
\end{equation} 
Note that the diagonal entries of $M$ are indeed related to the
average energy stored in the bond fluctuations: $M_{bb} = \frac{1}{2}\langle w_b
y_b y_b\rangle = \frac{1}{2}\langle w_b
(x_{\text{head}(b)}\!-\!x_{\text{tail}(b)})^2\rangle$.  Likewise, the
off-diagonal entries $M_{b_1b_2}$ reflect how a perturbation at bond
$b_2$ affects another bond $b_1$ weighted by the strength of bond
$b_1$. Hence the influence on a stronger bond is considered to be more
important.
Although we have not considered here time-delayed
correlations (i.e., as a function of $\tau$), this is an interesting direction
for future research.

\subsubsection{Definition of the bond-to-bond propensity}\label{sec:bb_propensity}

To construct our measure of propensity, we only assume knowledge of
the active site and proceed as follows. Let us consider all the
ligand-protein interactions formed at the active site and compute
their combined effect on each bond $b$ outside of the active site:
\begin{equation}
  \Pi^{\text{raw}}_b = \sum_{\substack{b' \in \text{ ligand}
    }} |M_{bb'}|.
\end{equation}
This raw propensity reflects how closely the active-site is coupled to
each individual bond.  Note that the computations include \textit{all} 
the bonds in the protein (covalent and non-covalent). However, in the paper 
we only report the effect on weak bonds, as it is changes in
weak-bonding patterns that usually drive allosteric response in proteins.
Since different proteins have different numbers of bonds, we make the
measure consistent by normalising the score:
\begin{equation}
  \label{eq:pp_bond}
  \Pi_b = \frac{\Pi^{\text{raw}}_b}{\sum_{b} \Pi^{\text{raw}}_{b}}.
\end{equation}
Throughout the manuscript, the quantity $\Pi_b$ is referred to as the
\textit{propensity} of bond $b$; a measure of how much edge $b$ is 
affected by the interactions at the active site. The
propensity of a \textit{residue} is defined as the
sum of the (normalised) propensities of its bonds:
\begin{equation}
  \label{eq:pp_residue}
  \Pi_{R} = \sum_{b \in R} \Pi_b.
\end{equation}

\subsubsection{Computational cost of bond-to-bond propensity}
\label{sec:comp_cost}
The computation of the propensities is efficient.  Note
that~Eq.~\eqref{eq:pp_bond} requires the summation over columns of the
$M$ matrix corresponding to protein-ligand interactions. 
Crucially, we do not need to compute the full pseudo-inverse $L^\dag$
in Eq.~\eqref{eq:M}; we can instead solve a sparse linear system
involving the graph Laplacian.  Recent algorithmic
developments~\cite{spielman2004nearly, kelner2013simple} have made
this possible in almost linear time, $\mathcal{O}(E\log^2(N_a))$, where
$E$ is the number of bonds (edges) and $N_a$ is the number of atoms
(nodes).  Our method therefore is scalable to extremely large
systems. Using the Combinatorial Multigrid toolbox written by
Y.~Koutis~\cite{koutis2011nearly} (available at
\url{http://www.cs.cmu.edu/~jkoutis/cmg.html}) propensities for all
the bonds in proteins with $\sim100,000$ atoms can be run in minutes
on a standard desktop computer.

\subsection{Significance of propensities through quantile scores}
\label{sec:qr}

To identify bonds (and residues) with high propensities relative to
others at a similar distance from the active site, we use
\textit{quantile regression}~\cite{koenker2005quantile}, a technique
of wide use in econometrics, ecology, and medical statistics.  In
contrast to standard least squares regression, which focusses on
estimating a model for the conditional mean of the samples, quantile
regression (QR) provides a method to estimate models for conditional
quantile functions.  This is important for two reasons: (i) the
conditional distributions of propensities are highly non-normal; and
(ii) we are interested not in the `average' bond, but in those bonds
with particularly high propensities lying in the tails of
the distribution. Once the fitted models are obtained, the quantile score of a bond $p_b$
is a measure of how high the propensity $\Pi_b$ is relative to other bonds 
in the sample which are at a similar distance from the active site.

Although QR goes back more than 200 years, it has only become widely
used recently, due to the availability of computational resources.
The mathematical basis of the method stems from the fact the
$p^\text{th}$ quantile, $Q_p$, of a distribution is given by the
solution of the following optimisation problem:  given a sample $\{y_i\}_{i=1}^n$
parametrically dependent on $m$ variables $\mathbf{x}_i \in
\mathbb{R}^m$ with parameters $\boldsymbol{\beta}$, the estimate of
the conditional $p^\text{th}$ quantile of the sample distribution is
obtained by solving
\begin{equation}
  \label{eq:quantile}
  \min_{\boldsymbol{\beta}} \sum_{i=1}^n \rho_p(y_i - Q(\mathbf{x}_i, \boldsymbol{\beta})),  \quad p \in [0,1],
\end{equation}
where $\rho_p(\cdot)$ is the tilted absolute value function
\begin{equation}
  \rho_{p}(y) = \left |y \, \left(p - \mathbbm{I}(y < 0)\right) \right |,
\end{equation}
and $\mathbbm{I}(\cdot)$ is the indicator function. If the dependence
is assumed to be linear, $Q(\mathbf{x}_i, \boldsymbol{\beta}) =
\beta_0 + \boldsymbol{\beta}^T \mathbf{x}_i$, the optimisation
can be formulated as a linear program and solved efficiently through
the simplex method to obtain $\hat{\boldsymbol{\beta}} \in
\mathbb{R}^{m+1}$, the estimated parameters defining the
model~\cite{koenker2005quantile}.

In Sections~\ref{sec:caspase-1}--\ref{sec:h-Ras}, we have applied QR
to the propensities $\Pi_b$ of bonds within each protein so as to take into
account their dependence with respect to $d_b$, the minimum distance
between bond $b$ and any bond in the active site:
\begin{equation}
d_b = \min_{b' \in \text{active}} |\mathbf{v}_b - \mathbf{v}_{b'}|,
\end{equation}
where the vector $\mathbf{v}_b$ contains the coordinates of the midpoint of bond $b$.
Based on the observed exponential decay of $\Pi$
with $d$, we adopt a linear model for the logarithm of the
propensities and estimate the conditional quantile functions by
solving the minimisation problem
\begin{equation}
  \label{eq:beta}
  \hat{\boldsymbol{\beta}}^{\text{\tiny{prot}}}(p) = \argmin_{(\beta_0,\beta_1)} \sum_b^{\text{\tiny{protein}}} \rho_p (\log(\Pi_b) - (\beta_0 + \beta_1d)),
\end{equation}
where the sum runs over the weak bonds of the corresponding protein.
From the estimated model for the protein, we then calculate the
quantile score of bond $b$ at distance $d_b$ from the
active site and with propensity $\Pi_b$, by finding the
quantile $p_b$ such that
\begin{equation}
  \label{eq:log_pb_fit}
  p_b = \argmin_{p \in [0,1]} \left| \beta_0^{\text{\tiny{prot}}}(p) + \beta_1^{\text{\tiny{prot}}}(p) d_b - \log(\Pi_b) \right|.
\end{equation}

Similarly, in Section~\ref{sec:scop}, we use QR to obtain \textit{absolute}
quantile scores of bonds and residues with respect to a
reference set of 100 proteins from the SCOP database.  
In this case, the propensities are regressed against both the distance to the active site $d$, 
and the number of non-covalent bonds in the protein, $E$.  Since the mean
propensity scales as $E^{-1}$, we also assume a power-law dependency 
of the quantiles.  Hence, we solve
\begin{equation}
  \label{eq:qr_multiple}
  \hat{\boldsymbol{\beta}}^{\text{\tiny{ref}}}(p) = \argmin_{(\beta_0,\beta_1, \beta_2)} \sum_b^{\text{\tiny{SCOP}}} \rho_p (\log(\Pi_b) - (\beta_0 + \beta_1d + \beta_2 \log(E))), 
\end{equation}
where the sum runs over all the weak bonds of all 
the proteins in the SCOP reference set.
For each quantile $p$, the model is defined by the equation of a plane 
$\beta^{\text{\tiny{ref}}}_0 (p) +
\beta^{\text{\tiny{ref}}}_1(p) d + \beta^{\text{\tiny{ref}}}_2(p) E$ (Fig.~\ref{fig:scop}b).  
The global quantile score $p^{\text{\tiny{ref}}}_b$ for bond $b$ at a distance $d_b$ from the
active site in a protein with $E_b$ non-covalent bonds is found by solving
\begin{equation}
  \label{eq:p_multiple}
  p^{\text{\tiny{ref}}}_b = \argmin_{p \in [0,1]} \left| \beta_0^{\text{\tiny{ref}}}(p) + \beta_1^{\text{\tiny{ref}}}(p) d_b + \beta_2^{\text{\tiny{ref}}}(p) E_b - \log(\Pi_b) \right| .
\end{equation}
Quantile scores for residues are obtained by applying the same process
to the propensities $\Pi_R$.  

The QR computations have been carried out using the R toolbox \textit{quantreg}
(\url{http://cran.r-project.org/web/packages/quantreg/index.html})
developed by R.~Koenker~\cite{koenker2015quantreg}.

\subsection{The SCOP reference set of generic proteins}
\label{sec:scop-mm}

The Structural Classification of Proteins (SCOP) database is a
manually curated database which uses a hierarchical classification
scheme collecting protein domains into structurally similar groups
\cite{murzin1995scop}.  The major classes of cytoplasmic proteins in
the database are $\alpha$, $\beta$, $\alpha/\beta$, $\alpha + \beta$,
and multidomain, covering all the major fold-types for
cytosolic proteins.  To obtain a representative set of proteins from
the database, we randomly selected 20 proteins from each of the
five classes.  Note that we only include proteins for which there is a structure with a
ligand bound to the active site.  Our reference set thus covers a
broad region of protein structure space.  A table containing details
of the 100 proteins selected can be found in the electronic \textit{SI}.

For each protein in the dataset, 
we compute the distance from the active site, $d_b$, 
and we calculate the propensity, $\Pi_b$, for all its $E$ weak bonds.
Across the 100 proteins,
we obtain a total of $465409$ $(d, E, \Pi_b)$ 3-tuples corresponding
to all the weak bonds in the proteins of the reference set  (Fig.~\ref{fig:scop}a).  
We then use QR to fit quantiles to this
reference set, as given by Eq.~\eqref{eq:qr_multiple}.
Note that the estimated quantile models, which are conditional on $d$ and $E$,
are now referred to the whole SCOP reference set and are not specific
to any one particular protein.  We then use the quantiles of the reference
set to compare the bond propensities of any protein of interest and
compute the \textit{absolute} quantile score $p_b^{\text{\tiny{ref}}}$
for each bond, as given by Eq.~\eqref{eq:p_multiple}. 
This score measures how high the bond propensity is, given its distance 
from the active site and the number of weak bonds in the protein of interest, 
as compared to all the bonds contained
in the wide range of proteins represented in the SCOP
reference set.

\subsection{Statistical evaluation of allosteric site quantile scores}
\label{sec:metrics}

To validate our findings on the allosteric protein test set, we evaluated the significance 
of the computed quantile scores according to four statistical measures, based on the
following metrics:
\begin{enumerate}[(i)]
\item The average bond quantile score:
  \begin{equation}\label{eq:testset:site_score}
    \overline{p_{b,\text{site}}} = \frac{1}{N_{b,\text{site}}}\sum_{b \in \text{site}} p_b,
  \end{equation}
  where $N_{b,\text{site}}$ is the number of bonds in the site.  
\item The average residue quantile score:
\begin{equation}
 \overline{p_{R,\text{site}}} = \frac{1}{N_{R,\text{site}}}\sum_{R \in \text{site}} p_R,
\end{equation}
  where $N_{R,\text{site}}$ is the number of bonds in the site.
\item The proportion of allosteric bonds with $p_b >0.95$, denoted $\text{P}(p_{b,\text{allo}} >0.95)$.
Since the quantile scores are uniformly distributed, 0.05 is the expected proportion of bonds with quantile scores above 0.95.
  \item The average reference bond quantile score:
\begin{equation}
 \overline{p_{b,\text{site}}^{\text{ref}}} = \frac{1}{N_{b,\text{site}}}\sum_{b \in \text{site}} p^{\text{ref}}_b,
\end{equation}
  where $N_{b,\text{site}}$ is the number of bonds in the site.
\end{enumerate}

These four measures are introduced to check robustly for the 
significance of the bonds in the allosteric site from distinct perspectives. %
If the functional coupling between active and allosteric sites is
due to a cumulative effect of the entire allosteric site, then average quantile scores over all
bonds in the allosteric site should be an accurate measure of its allosteric
propensity.  Measures (i), (ii) and (iv) capture this property at the level
of bonds and residues for both intrinsic and absolute propensities.
It is also possible that functional coupling to
the active site is concentrated on a small number of high quantile score 
bonds, with most others only being involved in structural or
energetic aspects of binding to the allosteric ligand and having low
quantile scores. Our metric (iii), which measures the number of high quantile score bonds 
in the site, can capture this behaviour based on the tail of the distribution. 
Reassuringly, the four measures provide complementary, yet largely consistent outcomes.

\subsubsection{Structural bootstrapping}
\label{sec:bootstrap}
To establish the significance of the average quantile scores 
$\overline{p_{b,\text{allo}}}$ and $\overline{p_{R,\text{allo}}}$,
we assess them against random surrogate sites sampled from the same protein, 
used as a structural bootstrap.  The surrogate sites generated
satisfy two structural constraints: 1) they have the same number of residues as the
allosteric site;  2) their diameter (i.e., the maximum distance between
any two atoms in the site) is not larger than that of the allosteric
site.  The algorithm for generating these sites is described in
Section S5 of the \textit{SI}.  For each protein, we generate 
1000 surrogate sites and calculate their quantile scores
$\overline{p_{b,\text{site}}}$ and $\overline{p_{R,\text{site}}}$.  
The average scores over the ensemble of 1000 surrogate
sites $\langle\overline{p_{b,\text{site}}}\rangle_{\text{surr}}$
and $\langle\overline{p_{R,\text{site}}}\rangle_{\text{surr}}$, where the
angle brackets denote the ensemble average, are then compared
against the average residue quantile score of the allosteric site (Figure~\ref{fig:test_set}a, b).  
A bootstrap with 10000 resamples with replacement~\cite{efron1994introduction}
was used to obtain 95\%~confidence intervals providing statistical signficance.

\subsubsection{Validation on the allosteric test set}
Figure~\ref{fig:test_set}~(a)--(d) reports these four statistical measures
for all 20 proteins analysed (see SI, Table S3 for the corresponding numerical data). 
Our results indicate robust identification of the allosteric sites in the test set. 
The quantile score of the allosteric site is
higher than that of the surrogate sites and above the 95\% bootstrapped confidence interval  
in 14 out of 20 proteins for the residue score, $\overline{p_{R,\text{allo}}}$, 
and for 16 out of 20 proteins for the bond score, $\overline{p_{b,\text{allo}}}$ 
(Figure~\ref{fig:test_set}a, b). 
The proteins identified by both measures are almost coincident, with few differences:
Glutamate DH (1HWZ) is significant according to the
bond score and marginally below significance according to the
residue score, whereas the opposite applies to Thrombin (1SFQ).
The reason for these differences lies with the distribution of
bond scores: in some cases, allosteric sites have only a few bonds with 
high quantile scores and many other less important bonds.  When considered
at the level of residues, this can lead to high $p_R$ scores; yet when
bonds are considered individually through their $p_b$ scores, 
the high quantile scores are averaged out over the whole allosteric site. 

To evaluate the presence of high scoring bonds, we 
compute the proportion of bonds with high quantile score 
$\text{P}(p_{b,\text{allo}} > 0.95)$ in the allosteric site, as compared to 
the expected proportion (0.05) above this quantile.
The proportion of high quantile score bonds in
the allosteric site is greater than expected in 17 of the 20 proteins
(Fig.~\ref{fig:test_set}c).  Of these 17 proteins, 16 coincide with those 
identified using the average scores reported above,
and we additionally identify 
h-Ras (3K8Y).  This finding confirms that allosteric sites consistently exhibit 
a larger than expected number of bonds with a strong coupling to the active site.

Finally, we compute the average \textit{absolute} quantile score of the allosteric site 
$\overline{p_{b, \text{allo}}^{\text{ref}}}$ against the SCOP reference set (Figure~\ref{fig:test_set}d).  
The results are largely consistent with the intrinsic measure $\overline{p_{b,\text{allo}}}$: 
in 14/20 proteins, the absolute quantile score is greater than the expected
0.5, i.e. $\overline{p_{b, \text{allo}}^{\text{ref}}} > 0.5$.
Yet some proteins (e.g., glutamate dehyrogenase (1HWZ), fructose 1,6-bisphosphatase (1EYI),
and glycogen phosphorylase (7GPB)) 
have high intrinsic quantile scores, as compared to other bonds in the same protein, but do not
score highly in absolute value, as compared to the reference SCOP ensemble.  
This result highlights the fact that a site need not have a high absolute propensity, 
as long as its propensity is high in comparison with the rest of the protein it belongs
to, so that the `signal' from the site outweighs the `noise' from the rest of the protein. 
Interestingly, the lac repressor (1EFA) has an allosteric site 
with large absolute propensity
($\overline{p_{b,\text{allo}}^{\text{ref}}} = 0.60 > 0.5$)
but non-significant intrinsic propensity.

\subsection{Construction of the atomistic graph}
\label{sec:ntwk_cnstr}

An in-depth discussion of the construction of the graph can be found
in Refs.~\cite{delmotte2011protein,amor2014uncovering}, and further
details are given in the SI, Section~2.  Briefly, we use an atomistic graph 
representation of a protein, where each node
corresponds to an atom and the edges represent both covalent and
non-covalent interactions, weighted by bond energies derived from
detailed atomic potentials.  The covalent bond energies are taken from
standard bond dissociation energy tables.  Non-covalent interactions
include hydrogen bonds, salt bridges, hydrophobic tethers and electrostatic
interactions. Hydrogen bond energies are obtained from the DREIDING
force-field~\cite{mayo1990dreiding}.  Attractive hydrophobic
interaction energies are defined between carbon and sulfur atoms,
according to a hydrophobic potential of mean force introduced by Lin
\textit{et al} \cite{lin2007hydrophobic}.  Electrostatic interactions
with coordination ions and ligands are identified from the LINK
entries in the PDB file, with bond energies assigned using a Coulomb
potential.

To compare the results between our atomistic model and residue-level
RRINs~\cite{chennubhotla2007signal}, we use coarse-grained
network models obtained from the oGNM server~\cite{yang2006ognm}.
A detailed comparison of results obtained with atomistic networks 
and RRINs is given in the SI Section~7.

We note that the main methodology (i.e., the propensity measure and methods developed in Sections \ref{sec:bond-bond}--\ref{sec:qr}) is independent of the construction of the graph. Users are free to construct the network using alternative potentials  (e.g., AMBER~\cite{case2015amber} or CHARMM~\cite{brooks2009charmm}) or using coarse-grained networks.

\section{Acknowledgments}

BRCA was supported by a studentship of the EPSRC~Centre for Doctoral
Training under the Institute of Chemical Biology, Imperial College
London.  SNY and MB acknowledge support through EPSRC grant
EP/I017267/1.  
We thank Keith Willison for suggesting h-Ras as an example
and for helpful discussions.

\section{Author Contributions}
BRCA, SNY and MB conceived the study.
BRCA performed the numerical analysis and created the Figures.
SNY and MB supervised the study.
All authors contributed to developing the theoretical tools.
All authors wrote and reviewed the manuscript.

\bibliography{rsc}

\begin{thebibliography}{85}%
\makeatletter
\providecommand \@ifxundefined [1]{%
 \@ifx{#1\undefined}
}%
\providecommand \@ifnum [1]{%
 \ifnum #1\expandafter \@firstoftwo
 \else \expandafter \@secondoftwo
 \fi
}%
\providecommand \@ifx [1]{%
 \ifx #1\expandafter \@firstoftwo
 \else \expandafter \@secondoftwo
 \fi
}%
\providecommand \natexlab [1]{#1}%
\providecommand \enquote  [1]{``#1''}%
\providecommand \bibnamefont  [1]{#1}%
\providecommand \bibfnamefont [1]{#1}%
\providecommand \citenamefont [1]{#1}%
\providecommand \href@noop [0]{\@secondoftwo}%
\providecommand \href [0]{\begingroup \@sanitize@url \@href}%
\providecommand \@href[1]{\@@startlink{#1}\@@href}%
\providecommand \@@href[1]{\endgroup#1\@@endlink}%
\providecommand \@sanitize@url [0]{\catcode `\\12\catcode `\$12\catcode
  `\&12\catcode `\#12\catcode `\^12\catcode `\_12\catcode `\%12\relax}%
\providecommand \@@startlink[1]{}%
\providecommand \@@endlink[0]{}%
\providecommand \url  [0]{\begingroup\@sanitize@url \@url }%
\providecommand \@url [1]{\endgroup\@href {#1}{\urlprefix }}%
\providecommand \urlprefix  [0]{URL }%
\providecommand \Eprint [0]{\href }%
\providecommand \doibase [0]{http://dx.doi.org/}%
\providecommand \selectlanguage [0]{\@gobble}%
\providecommand \bibinfo  [0]{\@secondoftwo}%
\providecommand \bibfield  [0]{\@secondoftwo}%
\providecommand \translation [1]{[#1]}%
\providecommand \BibitemOpen [0]{}%
\providecommand \bibitemStop [0]{}%
\providecommand \bibitemNoStop [0]{.\EOS\space}%
\providecommand \EOS [0]{\spacefactor3000\relax}%
\providecommand \BibitemShut  [1]{\csname bibitem#1\endcsname}%
\let\auto@bib@innerbib\@empty
\bibitem [{\citenamefont {Monod}\ \emph {et~al.}(1963)\citenamefont {Monod},
  \citenamefont {Changeux},\ and\ \citenamefont {Jacob}}]{monod1963allosteric}%
  \BibitemOpen
  \bibfield  {author} {\bibinfo {author} {\bibfnamefont {J.}~\bibnamefont
  {Monod}}, \bibinfo {author} {\bibfnamefont {J.-P.}\ \bibnamefont {Changeux}},
  \ and\ \bibinfo {author} {\bibfnamefont {F.}~\bibnamefont {Jacob}},\
  }\href@noop {} {\bibfield  {journal} {\bibinfo  {journal} {Journal of
  molecular biology}\ }\textbf {\bibinfo {volume} {6}},\ \bibinfo {pages} {306}
  (\bibinfo {year} {1963})}\BibitemShut {NoStop}%
\bibitem [{\citenamefont {Perutz}(1989)}]{perutz1989mechanisms}%
  \BibitemOpen
  \bibfield  {author} {\bibinfo {author} {\bibfnamefont {M.~F.}\ \bibnamefont
  {Perutz}},\ }\href@noop {} {\bibfield  {journal} {\bibinfo  {journal}
  {Quarterly reviews of biophysics}\ }\textbf {\bibinfo {volume} {22}},\
  \bibinfo {pages} {139} (\bibinfo {year} {1989})}\BibitemShut {NoStop}%
\bibitem [{\citenamefont {Nussinov}\ and\ \citenamefont
  {Tsai}(2013)}]{nussinov2013allostery}%
  \BibitemOpen
  \bibfield  {author} {\bibinfo {author} {\bibfnamefont {R.}~\bibnamefont
  {Nussinov}}\ and\ \bibinfo {author} {\bibfnamefont {C.-J.}\ \bibnamefont
  {Tsai}},\ }\href@noop {} {\bibfield  {journal} {\bibinfo  {journal} {Cell}\
  }\textbf {\bibinfo {volume} {153}},\ \bibinfo {pages} {293} (\bibinfo {year}
  {2013})}\BibitemShut {NoStop}%
\bibitem [{\citenamefont {Frauenfelder}\ \emph {et~al.}(1991)\citenamefont
  {Frauenfelder}, \citenamefont {Sligar},\ and\ \citenamefont
  {Wolynes}}]{frauenfelder1991energy}%
  \BibitemOpen
  \bibfield  {author} {\bibinfo {author} {\bibfnamefont {H.}~\bibnamefont
  {Frauenfelder}}, \bibinfo {author} {\bibfnamefont {S.~G.}\ \bibnamefont
  {Sligar}}, \ and\ \bibinfo {author} {\bibfnamefont {P.~G.}\ \bibnamefont
  {Wolynes}},\ }\href@noop {} {\bibfield  {journal} {\bibinfo  {journal}
  {Science}\ }\textbf {\bibinfo {volume} {254}},\ \bibinfo {pages} {1598}
  (\bibinfo {year} {1991})}\BibitemShut {NoStop}%
\bibitem [{\citenamefont {Henzler-Wildman}\ and\ \citenamefont
  {Kern}(2007)}]{henzlerwildman2007}%
  \BibitemOpen
  \bibfield  {author} {\bibinfo {author} {\bibfnamefont {K.}~\bibnamefont
  {Henzler-Wildman}}\ and\ \bibinfo {author} {\bibfnamefont {D.}~\bibnamefont
  {Kern}},\ }\href {\doibase 10.1038/nature06522} {\bibfield  {journal}
  {\bibinfo  {journal} {Nature}\ }\textbf {\bibinfo {volume} {450}},\ \bibinfo
  {pages} {964} (\bibinfo {year} {2007})}\BibitemShut {NoStop}%
\bibitem [{\citenamefont {Volkman}\ \emph {et~al.}(2001)\citenamefont
  {Volkman}, \citenamefont {Lipson}, \citenamefont {Wemmer},\ and\
  \citenamefont {Kern}}]{volkman2001two}%
  \BibitemOpen
  \bibfield  {author} {\bibinfo {author} {\bibfnamefont {B.~F.}\ \bibnamefont
  {Volkman}}, \bibinfo {author} {\bibfnamefont {D.}~\bibnamefont {Lipson}},
  \bibinfo {author} {\bibfnamefont {D.~E.}\ \bibnamefont {Wemmer}}, \ and\
  \bibinfo {author} {\bibfnamefont {D.}~\bibnamefont {Kern}},\ }\href@noop {}
  {\bibfield  {journal} {\bibinfo  {journal} {Science Signaling}\ }\textbf
  {\bibinfo {volume} {291}},\ \bibinfo {pages} {2429} (\bibinfo {year}
  {2001})}\BibitemShut {NoStop}%
\bibitem [{\citenamefont {Gunasekaran}\ \emph {et~al.}(2004)\citenamefont
  {Gunasekaran}, \citenamefont {Ma},\ and\ \citenamefont
  {Nussinov}}]{Gunasekaran2004}%
  \BibitemOpen
  \bibfield  {author} {\bibinfo {author} {\bibfnamefont {K.}~\bibnamefont
  {Gunasekaran}}, \bibinfo {author} {\bibfnamefont {B.}~\bibnamefont {Ma}}, \
  and\ \bibinfo {author} {\bibfnamefont {R.}~\bibnamefont {Nussinov}},\ }\href
  {\doibase 10.1002/prot.20232} {\bibfield  {journal} {\bibinfo  {journal}
  {PROTEINS: Structure, Function and Bioinformatics}\ }\textbf {\bibinfo
  {volume} {57}},\ \bibinfo {pages} {433} (\bibinfo {year} {2004})}\BibitemShut
  {NoStop}%
\bibitem [{\citenamefont {Hardy}\ and\ \citenamefont
  {Wells}(2004)}]{hardy2004searching}%
  \BibitemOpen
  \bibfield  {author} {\bibinfo {author} {\bibfnamefont {J.~A.}\ \bibnamefont
  {Hardy}}\ and\ \bibinfo {author} {\bibfnamefont {J.~A.}\ \bibnamefont
  {Wells}},\ }\href@noop {} {\bibfield  {journal} {\bibinfo  {journal} {Current
  opinion in structural biology}\ }\textbf {\bibinfo {volume} {14}},\ \bibinfo
  {pages} {706} (\bibinfo {year} {2004})}\BibitemShut {NoStop}%
\bibitem [{\citenamefont {Lockless}\ and\ \citenamefont
  {Ranganathan}(1999)}]{lockless1999evolutionarily}%
  \BibitemOpen
  \bibfield  {author} {\bibinfo {author} {\bibfnamefont {S.}~\bibnamefont
  {Lockless}}\ and\ \bibinfo {author} {\bibfnamefont {R.}~\bibnamefont
  {Ranganathan}},\ }\href@noop {} {\bibfield  {journal} {\bibinfo  {journal}
  {Science}\ }\textbf {\bibinfo {volume} {286}},\ \bibinfo {pages} {295}
  (\bibinfo {year} {1999})}\BibitemShut {NoStop}%
\bibitem [{\citenamefont {Grant}\ \emph {et~al.}(2011)\citenamefont {Grant},
  \citenamefont {Lukman}, \citenamefont {Hocker}, \citenamefont {Sayyah},
  \citenamefont {Brown}, \citenamefont {McCammon},\ and\ \citenamefont
  {Gorfe}}]{grant2011novel}%
  \BibitemOpen
  \bibfield  {author} {\bibinfo {author} {\bibfnamefont {B.~J.}\ \bibnamefont
  {Grant}}, \bibinfo {author} {\bibfnamefont {S.}~\bibnamefont {Lukman}},
  \bibinfo {author} {\bibfnamefont {H.~J.}\ \bibnamefont {Hocker}}, \bibinfo
  {author} {\bibfnamefont {J.}~\bibnamefont {Sayyah}}, \bibinfo {author}
  {\bibfnamefont {J.~H.}\ \bibnamefont {Brown}}, \bibinfo {author}
  {\bibfnamefont {J.~A.}\ \bibnamefont {McCammon}}, \ and\ \bibinfo {author}
  {\bibfnamefont {A.~A.}\ \bibnamefont {Gorfe}},\ }\href@noop {} {\bibfield
  {journal} {\bibinfo  {journal} {PLoS One}\ }\textbf {\bibinfo {volume} {6}},\
  \bibinfo {pages} {e25711} (\bibinfo {year} {2011})}\BibitemShut {NoStop}%
\bibitem [{\citenamefont {Weinkam}\ \emph {et~al.}(2012)\citenamefont
  {Weinkam}, \citenamefont {Pons},\ and\ \citenamefont
  {Sali}}]{weinkam2012structure}%
  \BibitemOpen
  \bibfield  {author} {\bibinfo {author} {\bibfnamefont {P.}~\bibnamefont
  {Weinkam}}, \bibinfo {author} {\bibfnamefont {J.}~\bibnamefont {Pons}}, \
  and\ \bibinfo {author} {\bibfnamefont {A.}~\bibnamefont {Sali}},\ }\href@noop
  {} {\bibfield  {journal} {\bibinfo  {journal} {Proceedings of the National
  Academy of Sciences}\ }\textbf {\bibinfo {volume} {109}},\ \bibinfo {pages}
  {4875} (\bibinfo {year} {2012})}\BibitemShut {NoStop}%
\bibitem [{\citenamefont {Ota}\ and\ \citenamefont
  {Agard}(2005)}]{ota2005intramolecular}%
  \BibitemOpen
  \bibfield  {author} {\bibinfo {author} {\bibfnamefont {N.}~\bibnamefont
  {Ota}}\ and\ \bibinfo {author} {\bibfnamefont {D.}~\bibnamefont {Agard}},\
  }\href@noop {} {\bibfield  {journal} {\bibinfo  {journal} {Journal of
  molecular biology}\ }\textbf {\bibinfo {volume} {351}},\ \bibinfo {pages}
  {345} (\bibinfo {year} {2005})}\BibitemShut {NoStop}%
\bibitem [{\citenamefont {Demerdash}\ \emph {et~al.}(2009)\citenamefont
  {Demerdash}, \citenamefont {Daily},\ and\ \citenamefont
  {Mitchell}}]{demerdash2009structure}%
  \BibitemOpen
  \bibfield  {author} {\bibinfo {author} {\bibfnamefont {O.~N.}\ \bibnamefont
  {Demerdash}}, \bibinfo {author} {\bibfnamefont {M.~D.}\ \bibnamefont
  {Daily}}, \ and\ \bibinfo {author} {\bibfnamefont {J.~C.}\ \bibnamefont
  {Mitchell}},\ }\href@noop {} {\bibfield  {journal} {\bibinfo  {journal} {PLoS
  computational biology}\ }\textbf {\bibinfo {volume} {5}},\ \bibinfo {pages}
  {e1000531} (\bibinfo {year} {2009})}\BibitemShut {NoStop}%
\bibitem [{\citenamefont {Panjkovich}\ and\ \citenamefont
  {Daura}(2012)}]{panjkovich2012exploiting}%
  \BibitemOpen
  \bibfield  {author} {\bibinfo {author} {\bibfnamefont {A.}~\bibnamefont
  {Panjkovich}}\ and\ \bibinfo {author} {\bibfnamefont {X.}~\bibnamefont
  {Daura}},\ }\href@noop {} {\bibfield  {journal} {\bibinfo  {journal} {BMC
  bioinformatics}\ }\textbf {\bibinfo {volume} {13}},\ \bibinfo {pages} {273}
  (\bibinfo {year} {2012})}\BibitemShut {NoStop}%
\bibitem [{\citenamefont {Collier}\ and\ \citenamefont
  {Ortiz}(2013)}]{collier2013emerging}%
  \BibitemOpen
  \bibfield  {author} {\bibinfo {author} {\bibfnamefont {G.}~\bibnamefont
  {Collier}}\ and\ \bibinfo {author} {\bibfnamefont {V.}~\bibnamefont
  {Ortiz}},\ }\href@noop {} {\bibfield  {journal} {\bibinfo  {journal}
  {Archives of biochemistry and biophysics}\ }\textbf {\bibinfo {volume}
  {538}},\ \bibinfo {pages} {6} (\bibinfo {year} {2013})}\BibitemShut {NoStop}%
\bibitem [{\citenamefont {Monod}\ \emph {et~al.}(1965)\citenamefont {Monod},
  \citenamefont {Wyman},\ and\ \citenamefont {Changeux}}]{monod1965nature}%
  \BibitemOpen
  \bibfield  {author} {\bibinfo {author} {\bibfnamefont {J.}~\bibnamefont
  {Monod}}, \bibinfo {author} {\bibfnamefont {J.}~\bibnamefont {Wyman}}, \ and\
  \bibinfo {author} {\bibfnamefont {J.}~\bibnamefont {Changeux}},\ }\href@noop
  {} {\bibfield  {journal} {\bibinfo  {journal} {Journal of molecular biology}\
  }\textbf {\bibinfo {volume} {12}},\ \bibinfo {pages} {88} (\bibinfo {year}
  {1965})}\BibitemShut {NoStop}%
\bibitem [{\citenamefont {Koshland~Jr}\ \emph {et~al.}(1966)\citenamefont
  {Koshland~Jr}, \citenamefont {Nemethy},\ and\ \citenamefont
  {Filmer}}]{koshland1966comparison}%
  \BibitemOpen
  \bibfield  {author} {\bibinfo {author} {\bibfnamefont {D.}~\bibnamefont
  {Koshland~Jr}}, \bibinfo {author} {\bibfnamefont {G.}~\bibnamefont
  {Nemethy}}, \ and\ \bibinfo {author} {\bibfnamefont {D.}~\bibnamefont
  {Filmer}},\ }\href@noop {} {\bibfield  {journal} {\bibinfo  {journal}
  {Biochemistry}\ }\textbf {\bibinfo {volume} {5}},\ \bibinfo {pages} {365}
  (\bibinfo {year} {1966})}\BibitemShut {NoStop}%
\bibitem [{\citenamefont {Hilser}\ \emph {et~al.}(2012)\citenamefont {Hilser},
  \citenamefont {Wrabl},\ and\ \citenamefont {Motlagh}}]{hilser2012structural}%
  \BibitemOpen
  \bibfield  {author} {\bibinfo {author} {\bibfnamefont {V.~J.}\ \bibnamefont
  {Hilser}}, \bibinfo {author} {\bibfnamefont {J.~O.}\ \bibnamefont {Wrabl}}, \
  and\ \bibinfo {author} {\bibfnamefont {H.~N.}\ \bibnamefont {Motlagh}},\
  }\href@noop {} {\bibfield  {journal} {\bibinfo  {journal} {Annual review of
  biophysics}\ }\textbf {\bibinfo {volume} {41}},\ \bibinfo {pages} {585}
  (\bibinfo {year} {2012})}\BibitemShut {NoStop}%
\bibitem [{\citenamefont {del Sol}\ \emph {et~al.}(2009)\citenamefont {del
  Sol}, \citenamefont {Tsai}, \citenamefont {Ma},\ and\ \citenamefont
  {Nussinov}}]{del2009origin}%
  \BibitemOpen
  \bibfield  {author} {\bibinfo {author} {\bibfnamefont {A.}~\bibnamefont {del
  Sol}}, \bibinfo {author} {\bibfnamefont {C.}~\bibnamefont {Tsai}}, \bibinfo
  {author} {\bibfnamefont {B.}~\bibnamefont {Ma}}, \ and\ \bibinfo {author}
  {\bibfnamefont {R.}~\bibnamefont {Nussinov}},\ }\href@noop {} {\bibfield
  {journal} {\bibinfo  {journal} {Structure}\ }\textbf {\bibinfo {volume}
  {17}},\ \bibinfo {pages} {1042} (\bibinfo {year} {2009})}\BibitemShut
  {NoStop}%
\bibitem [{\citenamefont {Zhuravlev}\ \emph {et~al.}(2010)\citenamefont
  {Zhuravlev}, \citenamefont {Papoian} \emph {et~al.}}]{zhuravlev2010protein}%
  \BibitemOpen
  \bibfield  {author} {\bibinfo {author} {\bibfnamefont {P.}~\bibnamefont
  {Zhuravlev}}, \bibinfo {author} {\bibfnamefont {G.}~\bibnamefont {Papoian}},
  \emph {et~al.},\ }\href@noop {} {\bibfield  {journal} {\bibinfo  {journal}
  {Quarterly reviews of biophysics}\ }\textbf {\bibinfo {volume} {43}},\
  \bibinfo {pages} {295} (\bibinfo {year} {2010})}\BibitemShut {NoStop}%
\bibitem [{\citenamefont {M{\"u}ller-Werkmeister}\ and\ \citenamefont
  {Bredenbeck}(2014)}]{muller2014donor}%
  \BibitemOpen
  \bibfield  {author} {\bibinfo {author} {\bibfnamefont {H.~M.}\ \bibnamefont
  {M{\"u}ller-Werkmeister}}\ and\ \bibinfo {author} {\bibfnamefont
  {J.}~\bibnamefont {Bredenbeck}},\ }\href@noop {} {\bibfield  {journal}
  {\bibinfo  {journal} {Physical Chemistry Chemical Physics}\ }\textbf
  {\bibinfo {volume} {16}},\ \bibinfo {pages} {3261} (\bibinfo {year}
  {2014})}\BibitemShut {NoStop}%
\bibitem [{\citenamefont {Li}\ \emph {et~al.}(2014)\citenamefont {Li},
  \citenamefont {Magana},\ and\ \citenamefont {Dyer}}]{li2014anisotropic}%
  \BibitemOpen
  \bibfield  {author} {\bibinfo {author} {\bibfnamefont {G.}~\bibnamefont
  {Li}}, \bibinfo {author} {\bibfnamefont {D.}~\bibnamefont {Magana}}, \ and\
  \bibinfo {author} {\bibfnamefont {R.~B.}\ \bibnamefont {Dyer}},\ }\href@noop
  {} {\bibfield  {journal} {\bibinfo  {journal} {Nature communications}\
  }\textbf {\bibinfo {volume} {5}} (\bibinfo {year} {2014})}\BibitemShut
  {NoStop}%
\bibitem [{\citenamefont {Mart{\'\i}nez}\ \emph {et~al.}(2011)\citenamefont
  {Mart{\'\i}nez}, \citenamefont {Figueira}, \citenamefont {Webb},
  \citenamefont {Polikarpov},\ and\ \citenamefont
  {Skaf}}]{martinez2011mapping}%
  \BibitemOpen
  \bibfield  {author} {\bibinfo {author} {\bibfnamefont {L.}~\bibnamefont
  {Mart{\'\i}nez}}, \bibinfo {author} {\bibfnamefont {A.~C.}\ \bibnamefont
  {Figueira}}, \bibinfo {author} {\bibfnamefont {P.}~\bibnamefont {Webb}},
  \bibinfo {author} {\bibfnamefont {I.}~\bibnamefont {Polikarpov}}, \ and\
  \bibinfo {author} {\bibfnamefont {M.~S.}\ \bibnamefont {Skaf}},\ }\href@noop
  {} {\bibfield  {journal} {\bibinfo  {journal} {The Journal of Physical
  Chemistry Letters}\ }\textbf {\bibinfo {volume} {2}},\ \bibinfo {pages}
  {2073} (\bibinfo {year} {2011})}\BibitemShut {NoStop}%
\bibitem [{\citenamefont {Fujii}\ \emph {et~al.}(2014)\citenamefont {Fujii},
  \citenamefont {Mizuno}, \citenamefont {Ishikawa},\ and\ \citenamefont
  {Mizutani}}]{fujii2014observing}%
  \BibitemOpen
  \bibfield  {author} {\bibinfo {author} {\bibfnamefont {N.}~\bibnamefont
  {Fujii}}, \bibinfo {author} {\bibfnamefont {M.}~\bibnamefont {Mizuno}},
  \bibinfo {author} {\bibfnamefont {H.}~\bibnamefont {Ishikawa}}, \ and\
  \bibinfo {author} {\bibfnamefont {Y.}~\bibnamefont {Mizutani}},\ }\href@noop
  {} {\bibfield  {journal} {\bibinfo  {journal} {The Journal of Physical
  Chemistry Letters}\ }\textbf {\bibinfo {volume} {5}},\ \bibinfo {pages}
  {3269} (\bibinfo {year} {2014})}\BibitemShut {NoStop}%
\bibitem [{\citenamefont {Nguyen}\ \emph {et~al.}(2009)\citenamefont {Nguyen},
  \citenamefont {Derreumaux},\ and\ \citenamefont {Stock}}]{nguyen2009energy}%
  \BibitemOpen
  \bibfield  {author} {\bibinfo {author} {\bibfnamefont {P.~H.}\ \bibnamefont
  {Nguyen}}, \bibinfo {author} {\bibfnamefont {P.}~\bibnamefont {Derreumaux}},
  \ and\ \bibinfo {author} {\bibfnamefont {G.}~\bibnamefont {Stock}},\
  }\href@noop {} {\bibfield  {journal} {\bibinfo  {journal} {The Journal of
  Physical Chemistry B}\ }\textbf {\bibinfo {volume} {113}},\ \bibinfo {pages}
  {9340} (\bibinfo {year} {2009})}\BibitemShut {NoStop}%
\bibitem [{\citenamefont {Gnanasekaran}\ \emph {et~al.}(2011)\citenamefont
  {Gnanasekaran}, \citenamefont {Agbo},\ and\ \citenamefont
  {Leitner}}]{gnanasekaran2011communication}%
  \BibitemOpen
  \bibfield  {author} {\bibinfo {author} {\bibfnamefont {R.}~\bibnamefont
  {Gnanasekaran}}, \bibinfo {author} {\bibfnamefont {J.~K.}\ \bibnamefont
  {Agbo}}, \ and\ \bibinfo {author} {\bibfnamefont {D.~M.}\ \bibnamefont
  {Leitner}},\ }\href@noop {} {\bibfield  {journal} {\bibinfo  {journal} {The
  Journal of chemical physics}\ }\textbf {\bibinfo {volume} {135}},\ \bibinfo
  {pages} {065103} (\bibinfo {year} {2011})}\BibitemShut {NoStop}%
\bibitem [{\citenamefont {Gerek}\ and\ \citenamefont
  {Ozkan}(2011)}]{gerek2011change}%
  \BibitemOpen
  \bibfield  {author} {\bibinfo {author} {\bibfnamefont {Z.~N.}\ \bibnamefont
  {Gerek}}\ and\ \bibinfo {author} {\bibfnamefont {S.~B.}\ \bibnamefont
  {Ozkan}},\ }\href@noop {} {\bibfield  {journal} {\bibinfo  {journal} {PLoS
  computational biology}\ }\textbf {\bibinfo {volume} {7}},\ \bibinfo {pages}
  {e1002154} (\bibinfo {year} {2011})}\BibitemShut {NoStop}%
\bibitem [{\citenamefont {Kaya}\ \emph {et~al.}(2013)\citenamefont {Kaya},
  \citenamefont {Armutlulu}, \citenamefont {Ekesan},\ and\ \citenamefont
  {Haliloglu}}]{kaya2013mcpath}%
  \BibitemOpen
  \bibfield  {author} {\bibinfo {author} {\bibfnamefont {C.}~\bibnamefont
  {Kaya}}, \bibinfo {author} {\bibfnamefont {A.}~\bibnamefont {Armutlulu}},
  \bibinfo {author} {\bibfnamefont {S.}~\bibnamefont {Ekesan}}, \ and\ \bibinfo
  {author} {\bibfnamefont {T.}~\bibnamefont {Haliloglu}},\ }\href@noop {}
  {\bibfield  {journal} {\bibinfo  {journal} {Nucleic acids research}\ }\textbf
  {\bibinfo {volume} {41}},\ \bibinfo {pages} {W249} (\bibinfo {year}
  {2013})}\BibitemShut {NoStop}%
\bibitem [{\citenamefont {Nakayama}\ \emph {et~al.}(1994)\citenamefont
  {Nakayama}, \citenamefont {Yakubo},\ and\ \citenamefont
  {Orbach}}]{nakayama1994dynamical}%
  \BibitemOpen
  \bibfield  {author} {\bibinfo {author} {\bibfnamefont {T.}~\bibnamefont
  {Nakayama}}, \bibinfo {author} {\bibfnamefont {K.}~\bibnamefont {Yakubo}}, \
  and\ \bibinfo {author} {\bibfnamefont {R.~L.}\ \bibnamefont {Orbach}},\
  }\href@noop {} {\bibfield  {journal} {\bibinfo  {journal} {Reviews of modern
  physics}\ }\textbf {\bibinfo {volume} {66}},\ \bibinfo {pages} {381}
  (\bibinfo {year} {1994})}\BibitemShut {NoStop}%
\bibitem [{\citenamefont {Leitner}(2008)}]{leitner2008energy}%
  \BibitemOpen
  \bibfield  {author} {\bibinfo {author} {\bibfnamefont {D.~M.}\ \bibnamefont
  {Leitner}},\ }\href@noop {} {\bibfield  {journal} {\bibinfo  {journal} {Annu.
  Rev. Phys. Chem.}\ }\textbf {\bibinfo {volume} {59}},\ \bibinfo {pages} {233}
  (\bibinfo {year} {2008})}\BibitemShut {NoStop}%
\bibitem [{\citenamefont {Del~Sol}\ \emph {et~al.}(2006)\citenamefont
  {Del~Sol}, \citenamefont {Fujihashi}, \citenamefont {Amoros},\ and\
  \citenamefont {Nussinov}}]{del2006residues}%
  \BibitemOpen
  \bibfield  {author} {\bibinfo {author} {\bibfnamefont {A.}~\bibnamefont
  {Del~Sol}}, \bibinfo {author} {\bibfnamefont {H.}~\bibnamefont {Fujihashi}},
  \bibinfo {author} {\bibfnamefont {D.}~\bibnamefont {Amoros}}, \ and\ \bibinfo
  {author} {\bibfnamefont {R.}~\bibnamefont {Nussinov}},\ }\href@noop {}
  {\bibfield  {journal} {\bibinfo  {journal} {Molecular systems biology}\
  }\textbf {\bibinfo {volume} {2}} (\bibinfo {year} {2006})}\BibitemShut
  {NoStop}%
\bibitem [{\citenamefont {Del~Sol}\ \emph {et~al.}(2007)\citenamefont
  {Del~Sol}, \citenamefont {Ara{\'u}zo-Bravo}, \citenamefont {Amoros},
  \citenamefont {Nussinov} \emph {et~al.}}]{del2007modular}%
  \BibitemOpen
  \bibfield  {author} {\bibinfo {author} {\bibfnamefont {A.}~\bibnamefont
  {Del~Sol}}, \bibinfo {author} {\bibfnamefont {M.}~\bibnamefont
  {Ara{\'u}zo-Bravo}}, \bibinfo {author} {\bibfnamefont {D.}~\bibnamefont
  {Amoros}}, \bibinfo {author} {\bibfnamefont {R.}~\bibnamefont {Nussinov}},
  \emph {et~al.},\ }\href@noop {} {\bibfield  {journal} {\bibinfo  {journal}
  {Genome Biol}\ }\textbf {\bibinfo {volume} {8}},\ \bibinfo {pages} {R92}
  (\bibinfo {year} {2007})}\BibitemShut {NoStop}%
\bibitem [{\citenamefont {Chennubhotla}\ and\ \citenamefont
  {Bahar}(2007)}]{chennubhotla2007signal}%
  \BibitemOpen
  \bibfield  {author} {\bibinfo {author} {\bibfnamefont {C.}~\bibnamefont
  {Chennubhotla}}\ and\ \bibinfo {author} {\bibfnamefont {I.}~\bibnamefont
  {Bahar}},\ }\href@noop {} {\bibfield  {journal} {\bibinfo  {journal} {PLoS
  computational biology}\ }\textbf {\bibinfo {volume} {3}},\ \bibinfo {pages}
  {e172} (\bibinfo {year} {2007})}\BibitemShut {NoStop}%
\bibitem [{\citenamefont {Amitai}\ \emph {et~al.}(2004)\citenamefont {Amitai},
  \citenamefont {Shemesh}, \citenamefont {Sitbon}, \citenamefont {Shklar},
  \citenamefont {Netanely}, \citenamefont {Venger},\ and\ \citenamefont
  {Pietrokovski}}]{amitai2004network}%
  \BibitemOpen
  \bibfield  {author} {\bibinfo {author} {\bibfnamefont {G.}~\bibnamefont
  {Amitai}}, \bibinfo {author} {\bibfnamefont {A.}~\bibnamefont {Shemesh}},
  \bibinfo {author} {\bibfnamefont {E.}~\bibnamefont {Sitbon}}, \bibinfo
  {author} {\bibfnamefont {M.}~\bibnamefont {Shklar}}, \bibinfo {author}
  {\bibfnamefont {D.}~\bibnamefont {Netanely}}, \bibinfo {author}
  {\bibfnamefont {I.}~\bibnamefont {Venger}}, \ and\ \bibinfo {author}
  {\bibfnamefont {S.}~\bibnamefont {Pietrokovski}},\ }\href@noop {} {\bibfield
  {journal} {\bibinfo  {journal} {Journal of molecular biology}\ }\textbf
  {\bibinfo {volume} {344}},\ \bibinfo {pages} {1135} (\bibinfo {year}
  {2004})}\BibitemShut {NoStop}%
\bibitem [{\citenamefont {Ghosh}\ and\ \citenamefont
  {Vishveshwara}(2007)}]{ghosh2007study}%
  \BibitemOpen
  \bibfield  {author} {\bibinfo {author} {\bibfnamefont {A.}~\bibnamefont
  {Ghosh}}\ and\ \bibinfo {author} {\bibfnamefont {S.}~\bibnamefont
  {Vishveshwara}},\ }\href@noop {} {\bibfield  {journal} {\bibinfo  {journal}
  {Proceedings of the National Academy of Sciences}\ }\textbf {\bibinfo
  {volume} {104}},\ \bibinfo {pages} {15711} (\bibinfo {year}
  {2007})}\BibitemShut {NoStop}%
\bibitem [{\citenamefont {Sethi}\ \emph {et~al.}(2009)\citenamefont {Sethi},
  \citenamefont {Eargle}, \citenamefont {Black},\ and\ \citenamefont
  {Luthey-Schulten}}]{sethi2009dynamical}%
  \BibitemOpen
  \bibfield  {author} {\bibinfo {author} {\bibfnamefont {A.}~\bibnamefont
  {Sethi}}, \bibinfo {author} {\bibfnamefont {J.}~\bibnamefont {Eargle}},
  \bibinfo {author} {\bibfnamefont {A.}~\bibnamefont {Black}}, \ and\ \bibinfo
  {author} {\bibfnamefont {Z.}~\bibnamefont {Luthey-Schulten}},\ }\href@noop {}
  {\bibfield  {journal} {\bibinfo  {journal} {Proceedings of the National
  Academy of Sciences}\ }\textbf {\bibinfo {volume} {106}},\ \bibinfo {pages}
  {6620} (\bibinfo {year} {2009})}\BibitemShut {NoStop}%
\bibitem [{\citenamefont {Ribeiro}\ and\ \citenamefont
  {Ortiz}(2014)}]{ribeiro2014determination}%
  \BibitemOpen
  \bibfield  {author} {\bibinfo {author} {\bibfnamefont {A.~A.}\ \bibnamefont
  {Ribeiro}}\ and\ \bibinfo {author} {\bibfnamefont {V.}~\bibnamefont
  {Ortiz}},\ }\href@noop {} {\bibfield  {journal} {\bibinfo  {journal} {Journal
  of Chemical Theory and Computation}\ }\textbf {\bibinfo {volume} {10}},\
  \bibinfo {pages} {1762} (\bibinfo {year} {2014})}\BibitemShut {NoStop}%
\bibitem [{\citenamefont {Ribeiro}\ and\ \citenamefont
  {Ortiz}(2015)}]{ribeiro2015energy}%
  \BibitemOpen
  \bibfield  {author} {\bibinfo {author} {\bibfnamefont {A.~A.}\ \bibnamefont
  {Ribeiro}}\ and\ \bibinfo {author} {\bibfnamefont {V.}~\bibnamefont
  {Ortiz}},\ }\href@noop {} {\bibfield  {journal} {\bibinfo  {journal} {The
  Journal of Physical Chemistry B}\ }\textbf {\bibinfo {volume} {119}},\
  \bibinfo {pages} {1835} (\bibinfo {year} {2015})}\BibitemShut {NoStop}%
\bibitem [{\citenamefont {Delmotte}\ \emph {et~al.}(2011)\citenamefont
  {Delmotte}, \citenamefont {Tate}, \citenamefont {Yaliraki},\ and\
  \citenamefont {Barahona}}]{delmotte2011protein}%
  \BibitemOpen
  \bibfield  {author} {\bibinfo {author} {\bibfnamefont {A.}~\bibnamefont
  {Delmotte}}, \bibinfo {author} {\bibfnamefont {E.}~\bibnamefont {Tate}},
  \bibinfo {author} {\bibfnamefont {S.}~\bibnamefont {Yaliraki}}, \ and\
  \bibinfo {author} {\bibfnamefont {M.}~\bibnamefont {Barahona}},\ }\href@noop
  {} {\bibfield  {journal} {\bibinfo  {journal} {Physical Biology}\ }\textbf
  {\bibinfo {volume} {8}},\ \bibinfo {pages} {055010} (\bibinfo {year}
  {2011})}\BibitemShut {NoStop}%
\bibitem [{\citenamefont {Amor}\ \emph {et~al.}(2014)\citenamefont {Amor},
  \citenamefont {Yaliraki}, \citenamefont {Woscholski},\ and\ \citenamefont
  {Barahona}}]{amor2014uncovering}%
  \BibitemOpen
  \bibfield  {author} {\bibinfo {author} {\bibfnamefont {B.}~\bibnamefont
  {Amor}}, \bibinfo {author} {\bibfnamefont {S.}~\bibnamefont {Yaliraki}},
  \bibinfo {author} {\bibfnamefont {R.}~\bibnamefont {Woscholski}}, \ and\
  \bibinfo {author} {\bibfnamefont {M.}~\bibnamefont {Barahona}},\ }\href@noop
  {} {\bibfield  {journal} {\bibinfo  {journal} {Molecular BioSystems}\
  }\textbf {\bibinfo {volume} {10}},\ \bibinfo {pages} {2247} (\bibinfo {year}
  {2014})}\BibitemShut {NoStop}%
\bibitem [{\citenamefont {Schaub}\ \emph {et~al.}(2014)\citenamefont {Schaub},
  \citenamefont {Lehmann}, \citenamefont {Yaliraki},\ and\ \citenamefont
  {Barahona}}]{schaub2014structure}%
  \BibitemOpen
  \bibfield  {author} {\bibinfo {author} {\bibfnamefont {M.~T.}\ \bibnamefont
  {Schaub}}, \bibinfo {author} {\bibfnamefont {J.}~\bibnamefont {Lehmann}},
  \bibinfo {author} {\bibfnamefont {S.~N.}\ \bibnamefont {Yaliraki}}, \ and\
  \bibinfo {author} {\bibfnamefont {M.}~\bibnamefont {Barahona}},\ }\href@noop
  {} {\bibfield  {journal} {\bibinfo  {journal} {Network Science}\ }\textbf
  {\bibinfo {volume} {2}},\ \bibinfo {pages} {66} (\bibinfo {year}
  {2014})}\BibitemShut {NoStop}%
\bibitem [{\citenamefont {Spielman}\ and\ \citenamefont
  {Teng}(2004)}]{spielman2004nearly}%
  \BibitemOpen
  \bibfield  {author} {\bibinfo {author} {\bibfnamefont {D.~A.}\ \bibnamefont
  {Spielman}}\ and\ \bibinfo {author} {\bibfnamefont {S.-H.}\ \bibnamefont
  {Teng}},\ }in\ \href@noop {} {\emph {\bibinfo {booktitle} {Proceedings of the
  thirty-sixth annual ACM symposium on Theory of computing}}}\ (\bibinfo
  {organization} {ACM},\ \bibinfo {year} {2004})\ pp.\ \bibinfo {pages}
  {81--90}\BibitemShut {NoStop}%
\bibitem [{\citenamefont {Kelner}\ \emph {et~al.}(2013)\citenamefont {Kelner},
  \citenamefont {Orecchia}, \citenamefont {Sidford},\ and\ \citenamefont
  {Zhu}}]{kelner2013simple}%
  \BibitemOpen
  \bibfield  {author} {\bibinfo {author} {\bibfnamefont {J.~A.}\ \bibnamefont
  {Kelner}}, \bibinfo {author} {\bibfnamefont {L.}~\bibnamefont {Orecchia}},
  \bibinfo {author} {\bibfnamefont {A.}~\bibnamefont {Sidford}}, \ and\
  \bibinfo {author} {\bibfnamefont {Z.~A.}\ \bibnamefont {Zhu}},\ }in\
  \href@noop {} {\emph {\bibinfo {booktitle} {Proceedings of the forty-fifth
  annual ACM symposium on Theory of computing}}}\ (\bibinfo {organization}
  {ACM},\ \bibinfo {year} {2013})\ pp.\ \bibinfo {pages} {911--920}\BibitemShut
  {NoStop}%
\bibitem [{\citenamefont {Koenker}(2005)}]{koenker2005quantile}%
  \BibitemOpen
  \bibfield  {author} {\bibinfo {author} {\bibfnamefont {R.}~\bibnamefont
  {Koenker}},\ }\href@noop {} {\emph {\bibinfo {title} {{Q}uantile
  regression}}},\ \bibinfo {number} {38}\ (\bibinfo  {publisher} {Cambridge
  university press},\ \bibinfo {year} {2005})\BibitemShut {NoStop}%
\bibitem [{\citenamefont {Wei}\ \emph {et~al.}(2006)\citenamefont {Wei},
  \citenamefont {Pere}, \citenamefont {Koenker},\ and\ \citenamefont
  {He}}]{wei2006quantile}%
  \BibitemOpen
  \bibfield  {author} {\bibinfo {author} {\bibfnamefont {Y.}~\bibnamefont
  {Wei}}, \bibinfo {author} {\bibfnamefont {A.}~\bibnamefont {Pere}}, \bibinfo
  {author} {\bibfnamefont {R.}~\bibnamefont {Koenker}}, \ and\ \bibinfo
  {author} {\bibfnamefont {X.}~\bibnamefont {He}},\ }\href@noop {} {\bibfield
  {journal} {\bibinfo  {journal} {Statistics in medicine}\ }\textbf {\bibinfo
  {volume} {25}},\ \bibinfo {pages} {1369} (\bibinfo {year}
  {2006})}\BibitemShut {NoStop}%
\bibitem [{\citenamefont {Knight}\ and\ \citenamefont
  {Ackerly}(2002)}]{knight2002variation}%
  \BibitemOpen
  \bibfield  {author} {\bibinfo {author} {\bibfnamefont {C.~A.}\ \bibnamefont
  {Knight}}\ and\ \bibinfo {author} {\bibfnamefont {D.~D.}\ \bibnamefont
  {Ackerly}},\ }\href@noop {} {\bibfield  {journal} {\bibinfo  {journal}
  {Ecology Letters}\ }\textbf {\bibinfo {volume} {5}},\ \bibinfo {pages} {66}
  (\bibinfo {year} {2002})}\BibitemShut {NoStop}%
\bibitem [{\citenamefont {Buchinsky}(1994)}]{buchinsky1994changes}%
  \BibitemOpen
  \bibfield  {author} {\bibinfo {author} {\bibfnamefont {M.}~\bibnamefont
  {Buchinsky}},\ }\href@noop {} {\bibfield  {journal} {\bibinfo  {journal}
  {Econometrica: Journal of the Econometric Society}\ ,\ \bibinfo {pages}
  {405}} (\bibinfo {year} {1994})}\BibitemShut {NoStop}%
\bibitem [{\citenamefont {Datta}\ \emph {et~al.}(2008)\citenamefont {Datta},
  \citenamefont {Scheer}, \citenamefont {Romanowski},\ and\ \citenamefont
  {Wells}}]{datta2008allosteric}%
  \BibitemOpen
  \bibfield  {author} {\bibinfo {author} {\bibfnamefont {D.}~\bibnamefont
  {Datta}}, \bibinfo {author} {\bibfnamefont {J.}~\bibnamefont {Scheer}},
  \bibinfo {author} {\bibfnamefont {M.}~\bibnamefont {Romanowski}}, \ and\
  \bibinfo {author} {\bibfnamefont {J.}~\bibnamefont {Wells}},\ }\href@noop {}
  {\bibfield  {journal} {\bibinfo  {journal} {Journal of molecular biology}\
  }\textbf {\bibinfo {volume} {381}},\ \bibinfo {pages} {1157} (\bibinfo {year}
  {2008})}\BibitemShut {NoStop}%
\bibitem [{\citenamefont {Cook}(1979)}]{cook1979influential}%
  \BibitemOpen
  \bibfield  {author} {\bibinfo {author} {\bibfnamefont {R.~D.}\ \bibnamefont
  {Cook}},\ }\href@noop {} {\bibfield  {journal} {\bibinfo  {journal} {Journal
  of the American Statistical Association}\ }\textbf {\bibinfo {volume} {74}},\
  \bibinfo {pages} {169} (\bibinfo {year} {1979})}\BibitemShut {NoStop}%
\bibitem [{\citenamefont {Dyer}\ and\ \citenamefont
  {Dahlquist}(2006)}]{dyer2006switched}%
  \BibitemOpen
  \bibfield  {author} {\bibinfo {author} {\bibfnamefont {C.~M.}\ \bibnamefont
  {Dyer}}\ and\ \bibinfo {author} {\bibfnamefont {F.~W.}\ \bibnamefont
  {Dahlquist}},\ }\href@noop {} {\bibfield  {journal} {\bibinfo  {journal}
  {Journal of bacteriology}\ }\textbf {\bibinfo {volume} {188}},\ \bibinfo
  {pages} {7354} (\bibinfo {year} {2006})}\BibitemShut {NoStop}%
\bibitem [{\citenamefont {Lee}\ \emph {et~al.}(2001)\citenamefont {Lee},
  \citenamefont {Cho}, \citenamefont {Pelton}, \citenamefont {Yan},
  \citenamefont {Berry},\ and\ \citenamefont {Wemmer}}]{lee2001crystal_2}%
  \BibitemOpen
  \bibfield  {author} {\bibinfo {author} {\bibfnamefont {S.-Y.}\ \bibnamefont
  {Lee}}, \bibinfo {author} {\bibfnamefont {H.~S.}\ \bibnamefont {Cho}},
  \bibinfo {author} {\bibfnamefont {J.~G.}\ \bibnamefont {Pelton}}, \bibinfo
  {author} {\bibfnamefont {D.}~\bibnamefont {Yan}}, \bibinfo {author}
  {\bibfnamefont {E.~A.}\ \bibnamefont {Berry}}, \ and\ \bibinfo {author}
  {\bibfnamefont {D.~E.}\ \bibnamefont {Wemmer}},\ }\href@noop {} {\bibfield
  {journal} {\bibinfo  {journal} {Journal of Biological Chemistry}\ }\textbf
  {\bibinfo {volume} {276}},\ \bibinfo {pages} {16425} (\bibinfo {year}
  {2001})}\BibitemShut {NoStop}%
\bibitem [{\citenamefont {McDonald}\ \emph {et~al.}(2012)\citenamefont
  {McDonald}, \citenamefont {Boyer},\ and\ \citenamefont
  {Lee}}]{mcdonald2012segmental}%
  \BibitemOpen
  \bibfield  {author} {\bibinfo {author} {\bibfnamefont {L.~R.}\ \bibnamefont
  {McDonald}}, \bibinfo {author} {\bibfnamefont {J.~A.}\ \bibnamefont {Boyer}},
  \ and\ \bibinfo {author} {\bibfnamefont {A.~L.}\ \bibnamefont {Lee}},\
  }\href@noop {} {\bibfield  {journal} {\bibinfo  {journal} {Structure}\
  }\textbf {\bibinfo {volume} {20}},\ \bibinfo {pages} {1363} (\bibinfo {year}
  {2012})}\BibitemShut {NoStop}%
\bibitem [{\citenamefont {Bourret}\ \emph {et~al.}(1993)\citenamefont
  {Bourret}, \citenamefont {Drake}, \citenamefont {Chervitz}, \citenamefont
  {Simon},\ and\ \citenamefont {Falke}}]{bourret1993activation}%
  \BibitemOpen
  \bibfield  {author} {\bibinfo {author} {\bibfnamefont {R.~B.}\ \bibnamefont
  {Bourret}}, \bibinfo {author} {\bibfnamefont {S.~K.}\ \bibnamefont {Drake}},
  \bibinfo {author} {\bibfnamefont {S.~A.}\ \bibnamefont {Chervitz}}, \bibinfo
  {author} {\bibfnamefont {M.~I.}\ \bibnamefont {Simon}}, \ and\ \bibinfo
  {author} {\bibfnamefont {J.~J.}\ \bibnamefont {Falke}},\ }\href@noop {}
  {\bibfield  {journal} {\bibinfo  {journal} {Journal of Biological Chemistry}\
  }\textbf {\bibinfo {volume} {268}},\ \bibinfo {pages} {13089} (\bibinfo
  {year} {1993})}\BibitemShut {NoStop}%
\bibitem [{\citenamefont {Smith}\ \emph {et~al.}(2003)\citenamefont {Smith},
  \citenamefont {Latiolais}, \citenamefont {Guanga}, \citenamefont {Citineni},
  \citenamefont {Silversmith},\ and\ \citenamefont
  {Bourret}}]{smith2003investigation}%
  \BibitemOpen
  \bibfield  {author} {\bibinfo {author} {\bibfnamefont {J.~G.}\ \bibnamefont
  {Smith}}, \bibinfo {author} {\bibfnamefont {J.~A.}\ \bibnamefont
  {Latiolais}}, \bibinfo {author} {\bibfnamefont {G.~P.}\ \bibnamefont
  {Guanga}}, \bibinfo {author} {\bibfnamefont {S.}~\bibnamefont {Citineni}},
  \bibinfo {author} {\bibfnamefont {R.~E.}\ \bibnamefont {Silversmith}}, \ and\
  \bibinfo {author} {\bibfnamefont {R.~B.}\ \bibnamefont {Bourret}},\
  }\href@noop {} {\bibfield  {journal} {\bibinfo  {journal} {Journal of
  bacteriology}\ }\textbf {\bibinfo {volume} {185}},\ \bibinfo {pages} {6385}
  (\bibinfo {year} {2003})}\BibitemShut {NoStop}%
\bibitem [{\citenamefont {McDonald}\ \emph {et~al.}(2013)\citenamefont
  {McDonald}, \citenamefont {Whitley}, \citenamefont {Boyer},\ and\
  \citenamefont {Lee}}]{mcdonald2013colocalization}%
  \BibitemOpen
  \bibfield  {author} {\bibinfo {author} {\bibfnamefont {L.~R.}\ \bibnamefont
  {McDonald}}, \bibinfo {author} {\bibfnamefont {M.~J.}\ \bibnamefont
  {Whitley}}, \bibinfo {author} {\bibfnamefont {J.~A.}\ \bibnamefont {Boyer}},
  \ and\ \bibinfo {author} {\bibfnamefont {A.~L.}\ \bibnamefont {Lee}},\
  }\href@noop {} {\bibfield  {journal} {\bibinfo  {journal} {Journal of
  molecular biology}\ } (\bibinfo {year} {2013})}\BibitemShut {NoStop}%
\bibitem [{\citenamefont {McCormick}(1995)}]{mccormick1995ras}%
  \BibitemOpen
  \bibfield  {author} {\bibinfo {author} {\bibfnamefont {F.}~\bibnamefont
  {McCormick}},\ }\href@noop {} {\bibfield  {journal} {\bibinfo  {journal}
  {Molecular reproduction and development}\ }\textbf {\bibinfo {volume} {42}},\
  \bibinfo {pages} {500} (\bibinfo {year} {1995})}\BibitemShut {NoStop}%
\bibitem [{\citenamefont {Buhrman}\ \emph {et~al.}(2010)\citenamefont
  {Buhrman}, \citenamefont {Holzapfel}, \citenamefont {Fetics},\ and\
  \citenamefont {Mattos}}]{buhrman2010allosteric}%
  \BibitemOpen
  \bibfield  {author} {\bibinfo {author} {\bibfnamefont {G.}~\bibnamefont
  {Buhrman}}, \bibinfo {author} {\bibfnamefont {G.}~\bibnamefont {Holzapfel}},
  \bibinfo {author} {\bibfnamefont {S.}~\bibnamefont {Fetics}}, \ and\ \bibinfo
  {author} {\bibfnamefont {C.}~\bibnamefont {Mattos}},\ }\href@noop {}
  {\bibfield  {journal} {\bibinfo  {journal} {Proceedings of the National
  Academy of Sciences}\ }\textbf {\bibinfo {volume} {107}},\ \bibinfo {pages}
  {4931} (\bibinfo {year} {2010})}\BibitemShut {NoStop}%
\bibitem [{\citenamefont {Murzin}\ \emph {et~al.}(1995)\citenamefont {Murzin},
  \citenamefont {Brenner}, \citenamefont {Hubbard},\ and\ \citenamefont
  {Chothia}}]{murzin1995scop}%
  \BibitemOpen
  \bibfield  {author} {\bibinfo {author} {\bibfnamefont {A.~G.}\ \bibnamefont
  {Murzin}}, \bibinfo {author} {\bibfnamefont {S.~E.}\ \bibnamefont {Brenner}},
  \bibinfo {author} {\bibfnamefont {T.}~\bibnamefont {Hubbard}}, \ and\
  \bibinfo {author} {\bibfnamefont {C.}~\bibnamefont {Chothia}},\ }\href@noop
  {} {\bibfield  {journal} {\bibinfo  {journal} {Journal of molecular biology}\
  }\textbf {\bibinfo {volume} {247}},\ \bibinfo {pages} {536} (\bibinfo {year}
  {1995})}\BibitemShut {NoStop}%
\bibitem [{\citenamefont {Daily}\ and\ \citenamefont
  {Gray}(2009)}]{daily2009allosteric}%
  \BibitemOpen
  \bibfield  {author} {\bibinfo {author} {\bibfnamefont {M.}~\bibnamefont
  {Daily}}\ and\ \bibinfo {author} {\bibfnamefont {J.}~\bibnamefont {Gray}},\
  }\href@noop {} {\bibfield  {journal} {\bibinfo  {journal} {PLoS computational
  biology}\ }\textbf {\bibinfo {volume} {5}},\ \bibinfo {pages} {e1000293}
  (\bibinfo {year} {2009})}\BibitemShut {NoStop}%
\bibitem [{\citenamefont {Zhu}\ \emph {et~al.}(1996)\citenamefont {Zhu},
  \citenamefont {Amsler}, \citenamefont {Volz},\ and\ \citenamefont
  {Matsumura}}]{zhu1996tyrosine}%
  \BibitemOpen
  \bibfield  {author} {\bibinfo {author} {\bibfnamefont {X.}~\bibnamefont
  {Zhu}}, \bibinfo {author} {\bibfnamefont {C.~D.}\ \bibnamefont {Amsler}},
  \bibinfo {author} {\bibfnamefont {K.}~\bibnamefont {Volz}}, \ and\ \bibinfo
  {author} {\bibfnamefont {P.}~\bibnamefont {Matsumura}},\ }\href@noop {}
  {\bibfield  {journal} {\bibinfo  {journal} {Journal of bacteriology}\
  }\textbf {\bibinfo {volume} {178}},\ \bibinfo {pages} {4208} (\bibinfo {year}
  {1996})}\BibitemShut {NoStop}%
\bibitem [{\citenamefont {Bellsolell}\ \emph {et~al.}(1996)\citenamefont
  {Bellsolell}, \citenamefont {Cronet}, \citenamefont {Majolero}, \citenamefont
  {Serrano},\ and\ \citenamefont {Coll}}]{bellsolell1996three}%
  \BibitemOpen
  \bibfield  {author} {\bibinfo {author} {\bibfnamefont {L.}~\bibnamefont
  {Bellsolell}}, \bibinfo {author} {\bibfnamefont {P.}~\bibnamefont {Cronet}},
  \bibinfo {author} {\bibfnamefont {M.}~\bibnamefont {Majolero}}, \bibinfo
  {author} {\bibfnamefont {L.}~\bibnamefont {Serrano}}, \ and\ \bibinfo
  {author} {\bibfnamefont {M.}~\bibnamefont {Coll}},\ }\href@noop {} {\bibfield
   {journal} {\bibinfo  {journal} {Journal of molecular biology}\ }\textbf
  {\bibinfo {volume} {257}},\ \bibinfo {pages} {116} (\bibinfo {year}
  {1996})}\BibitemShut {NoStop}%
\bibitem [{\citenamefont {Buchli}\ \emph {et~al.}(2013)\citenamefont {Buchli},
  \citenamefont {Waldauer}, \citenamefont {Walser}, \citenamefont {Donten},
  \citenamefont {Pfister}, \citenamefont {Bl{\"o}chliger}, \citenamefont
  {Steiner}, \citenamefont {Caflisch}, \citenamefont {Zerbe},\ and\
  \citenamefont {Hamm}}]{buchli2013kinetic}%
  \BibitemOpen
  \bibfield  {author} {\bibinfo {author} {\bibfnamefont {B.}~\bibnamefont
  {Buchli}}, \bibinfo {author} {\bibfnamefont {S.~A.}\ \bibnamefont
  {Waldauer}}, \bibinfo {author} {\bibfnamefont {R.}~\bibnamefont {Walser}},
  \bibinfo {author} {\bibfnamefont {M.~L.}\ \bibnamefont {Donten}}, \bibinfo
  {author} {\bibfnamefont {R.}~\bibnamefont {Pfister}}, \bibinfo {author}
  {\bibfnamefont {N.}~\bibnamefont {Bl{\"o}chliger}}, \bibinfo {author}
  {\bibfnamefont {S.}~\bibnamefont {Steiner}}, \bibinfo {author} {\bibfnamefont
  {A.}~\bibnamefont {Caflisch}}, \bibinfo {author} {\bibfnamefont
  {O.}~\bibnamefont {Zerbe}}, \ and\ \bibinfo {author} {\bibfnamefont
  {P.}~\bibnamefont {Hamm}},\ }\href@noop {} {\bibfield  {journal} {\bibinfo
  {journal} {Proceedings of the National Academy of Sciences}\ }\textbf
  {\bibinfo {volume} {110}},\ \bibinfo {pages} {11725} (\bibinfo {year}
  {2013})}\BibitemShut {NoStop}%
\bibitem [{\citenamefont {Chung}\ and\ \citenamefont
  {Yau}(2000)}]{chung2000discrete}%
  \BibitemOpen
  \bibfield  {author} {\bibinfo {author} {\bibfnamefont {F.}~\bibnamefont
  {Chung}}\ and\ \bibinfo {author} {\bibfnamefont {S.-T.}\ \bibnamefont
  {Yau}},\ }\href@noop {} {\bibfield  {journal} {\bibinfo  {journal} {Journal
  of Combinatorial Theory, Series A}\ }\textbf {\bibinfo {volume} {91}},\
  \bibinfo {pages} {191} (\bibinfo {year} {2000})}\BibitemShut {NoStop}%
\bibitem [{\citenamefont {Reuveni}\ \emph {et~al.}(2010)\citenamefont
  {Reuveni}, \citenamefont {Granek},\ and\ \citenamefont
  {Klafter}}]{reuveni2010anomalies}%
  \BibitemOpen
  \bibfield  {author} {\bibinfo {author} {\bibfnamefont {S.}~\bibnamefont
  {Reuveni}}, \bibinfo {author} {\bibfnamefont {R.}~\bibnamefont {Granek}}, \
  and\ \bibinfo {author} {\bibfnamefont {J.}~\bibnamefont {Klafter}},\
  }\href@noop {} {\bibfield  {journal} {\bibinfo  {journal} {Proceedings of the
  National Academy of Sciences}\ }\textbf {\bibinfo {volume} {107}},\ \bibinfo
  {pages} {13696} (\bibinfo {year} {2010})}\BibitemShut {NoStop}%
\bibitem [{\citenamefont {Biggs}(1993)}]{biggs1993algebraic}%
  \BibitemOpen
  \bibfield  {author} {\bibinfo {author} {\bibfnamefont {N.}~\bibnamefont
  {Biggs}},\ }\href@noop {} {\emph {\bibinfo {title} {{A}lgebraic graph
  theory}}}\ (\bibinfo  {publisher} {Cambridge university press},\ \bibinfo
  {year} {1993})\BibitemShut {NoStop}%
\bibitem [{\citenamefont {Koutis}\ \emph {et~al.}(2011)\citenamefont {Koutis},
  \citenamefont {Miller},\ and\ \citenamefont {Peng}}]{koutis2011nearly}%
  \BibitemOpen
  \bibfield  {author} {\bibinfo {author} {\bibfnamefont {I.}~\bibnamefont
  {Koutis}}, \bibinfo {author} {\bibfnamefont {G.~L.}\ \bibnamefont {Miller}},
  \ and\ \bibinfo {author} {\bibfnamefont {R.}~\bibnamefont {Peng}},\ }in\
  \href@noop {} {\emph {\bibinfo {booktitle} {Foundations of Computer Science
  (FOCS), 2011 IEEE 52\textsuperscript{nd} Annual Symposium on}}}\ (\bibinfo
  {organization} {IEEE},\ \bibinfo {year} {2011})\ pp.\ \bibinfo {pages}
  {590--598}\BibitemShut {NoStop}%
\bibitem [{\citenamefont {Koenker}(2015)}]{koenker2015quantreg}%
  \BibitemOpen
  \bibfield  {author} {\bibinfo {author} {\bibfnamefont {R.}~\bibnamefont
  {Koenker}},\ }\href {http://CRAN.R-project.org/package=quantreg} {\emph
  {\bibinfo {title} {quantreg: {Q}uantile {R}egression}}} (\bibinfo {year}
  {2015}),\ \bibinfo {note} {{R} package version 5.19}\BibitemShut {NoStop}%
\bibitem [{\citenamefont {Efron}\ and\ \citenamefont
  {Tibshirani}(1994)}]{efron1994introduction}%
  \BibitemOpen
  \bibfield  {author} {\bibinfo {author} {\bibfnamefont {B.}~\bibnamefont
  {Efron}}\ and\ \bibinfo {author} {\bibfnamefont {R.~J.}\ \bibnamefont
  {Tibshirani}},\ }\href@noop {} {\emph {\bibinfo {title} {{A}n introduction to
  the bootstrap}}}\ (\bibinfo  {publisher} {CRC press},\ \bibinfo {year}
  {1994})\BibitemShut {NoStop}%
\bibitem [{\citenamefont {Mayo}\ \emph {et~al.}(1990)\citenamefont {Mayo},
  \citenamefont {Olafson},\ and\ \citenamefont {Goddard}}]{mayo1990dreiding}%
  \BibitemOpen
  \bibfield  {author} {\bibinfo {author} {\bibfnamefont {S.}~\bibnamefont
  {Mayo}}, \bibinfo {author} {\bibfnamefont {B.}~\bibnamefont {Olafson}}, \
  and\ \bibinfo {author} {\bibfnamefont {W.}~\bibnamefont {Goddard}},\
  }\href@noop {} {\bibfield  {journal} {\bibinfo  {journal} {Journal of
  Physical Chemistry}\ }\textbf {\bibinfo {volume} {94}},\ \bibinfo {pages}
  {8897} (\bibinfo {year} {1990})}\BibitemShut {NoStop}%
\bibitem [{\citenamefont {Lin}\ \emph {et~al.}(2007)\citenamefont {Lin},
  \citenamefont {Fawzi},\ and\ \citenamefont
  {Head-Gordon}}]{lin2007hydrophobic}%
  \BibitemOpen
  \bibfield  {author} {\bibinfo {author} {\bibfnamefont {M.~S.}\ \bibnamefont
  {Lin}}, \bibinfo {author} {\bibfnamefont {N.~L.}\ \bibnamefont {Fawzi}}, \
  and\ \bibinfo {author} {\bibfnamefont {T.}~\bibnamefont {Head-Gordon}},\
  }\href@noop {} {\bibfield  {journal} {\bibinfo  {journal} {Structure}\
  }\textbf {\bibinfo {volume} {15}},\ \bibinfo {pages} {727} (\bibinfo {year}
  {2007})}\BibitemShut {NoStop}%
\bibitem [{\citenamefont {Yang}\ \emph {et~al.}(2006)\citenamefont {Yang},
  \citenamefont {Rader}, \citenamefont {Liu}, \citenamefont {Jursa},
  \citenamefont {Chen}, \citenamefont {Karimi},\ and\ \citenamefont
  {Bahar}}]{yang2006ognm}%
  \BibitemOpen
  \bibfield  {author} {\bibinfo {author} {\bibfnamefont {L.-W.}\ \bibnamefont
  {Yang}}, \bibinfo {author} {\bibfnamefont {A.}~\bibnamefont {Rader}},
  \bibinfo {author} {\bibfnamefont {X.}~\bibnamefont {Liu}}, \bibinfo {author}
  {\bibfnamefont {C.~J.}\ \bibnamefont {Jursa}}, \bibinfo {author}
  {\bibfnamefont {S.~C.}\ \bibnamefont {Chen}}, \bibinfo {author}
  {\bibfnamefont {H.~A.}\ \bibnamefont {Karimi}}, \ and\ \bibinfo {author}
  {\bibfnamefont {I.}~\bibnamefont {Bahar}},\ }\href@noop {} {\bibfield
  {journal} {\bibinfo  {journal} {Nucleic acids research}\ }\textbf {\bibinfo
  {volume} {34}},\ \bibinfo {pages} {W24} (\bibinfo {year} {2006})}\BibitemShut
  {NoStop}%
\bibitem [{\citenamefont {Case}\ \emph {et~al.}(2015)\citenamefont {Case},
  \citenamefont {Berryman}, \citenamefont {Betz}, \citenamefont {Cerutti},
  \citenamefont {Cheatham~III}, \citenamefont {Darden}, \citenamefont {Duke},
  \citenamefont {Giese}, \citenamefont {Gohlke}, \citenamefont {Goetz} \emph
  {et~al.}}]{case2015amber}%
  \BibitemOpen
  \bibfield  {author} {\bibinfo {author} {\bibfnamefont {D.}~\bibnamefont
  {Case}}, \bibinfo {author} {\bibfnamefont {J.}~\bibnamefont {Berryman}},
  \bibinfo {author} {\bibfnamefont {R.}~\bibnamefont {Betz}}, \bibinfo {author}
  {\bibfnamefont {D.}~\bibnamefont {Cerutti}}, \bibinfo {author} {\bibfnamefont
  {T.}~\bibnamefont {Cheatham~III}}, \bibinfo {author} {\bibfnamefont
  {T.}~\bibnamefont {Darden}}, \bibinfo {author} {\bibfnamefont
  {R.}~\bibnamefont {Duke}}, \bibinfo {author} {\bibfnamefont {T.}~\bibnamefont
  {Giese}}, \bibinfo {author} {\bibfnamefont {H.}~\bibnamefont {Gohlke}},
  \bibinfo {author} {\bibfnamefont {A.}~\bibnamefont {Goetz}},  \emph
  {et~al.},\ }\href@noop {} {\bibfield  {journal} {\bibinfo  {journal}
  {University of California, San Francisco}\ } (\bibinfo {year}
  {2015})}\BibitemShut {NoStop}%
\bibitem [{\citenamefont {Brooks}\ \emph {et~al.}(2009)\citenamefont {Brooks},
  \citenamefont {Brooks}, \citenamefont {MacKerell}, \citenamefont {Nilsson},
  \citenamefont {Petrella}, \citenamefont {Roux}, \citenamefont {Won},
  \citenamefont {Archontis}, \citenamefont {Bartels}, \citenamefont {Boresch}
  \emph {et~al.}}]{brooks2009charmm}%
  \BibitemOpen
  \bibfield  {author} {\bibinfo {author} {\bibfnamefont {B.~R.}\ \bibnamefont
  {Brooks}}, \bibinfo {author} {\bibfnamefont {C.~L.}\ \bibnamefont {Brooks}},
  \bibinfo {author} {\bibfnamefont {A.~D.}\ \bibnamefont {MacKerell}}, \bibinfo
  {author} {\bibfnamefont {L.}~\bibnamefont {Nilsson}}, \bibinfo {author}
  {\bibfnamefont {R.~J.}\ \bibnamefont {Petrella}}, \bibinfo {author}
  {\bibfnamefont {B.}~\bibnamefont {Roux}}, \bibinfo {author} {\bibfnamefont
  {Y.}~\bibnamefont {Won}}, \bibinfo {author} {\bibfnamefont {G.}~\bibnamefont
  {Archontis}}, \bibinfo {author} {\bibfnamefont {C.}~\bibnamefont {Bartels}},
  \bibinfo {author} {\bibfnamefont {S.}~\bibnamefont {Boresch}},  \emph
  {et~al.},\ }\href@noop {} {\bibfield  {journal} {\bibinfo  {journal} {Journal
  of computational chemistry}\ }\textbf {\bibinfo {volume} {30}},\ \bibinfo
  {pages} {1545} (\bibinfo {year} {2009})}\BibitemShut {NoStop}%
\bibitem [{\citenamefont {Jacobs}\ and\ \citenamefont
  {Thorpe}(1995)}]{Jacobs1995}%
  \BibitemOpen
  \bibfield  {author} {\bibinfo {author} {\bibfnamefont {D.~J.}\ \bibnamefont
  {Jacobs}}\ and\ \bibinfo {author} {\bibfnamefont {M.~F.}\ \bibnamefont
  {Thorpe}},\ }\href@noop {} {\bibfield  {journal} {\bibinfo  {journal}
  {Physical review letters}\ }\textbf {\bibinfo {volume} {75}},\ \bibinfo
  {pages} {4051} (\bibinfo {year} {1995})}\BibitemShut {NoStop}%
\bibitem [{\citenamefont {Jacobs}\ and\ \citenamefont
  {Thorpe}(1999)}]{Jacobs1999}%
  \BibitemOpen
  \bibfield  {author} {\bibinfo {author} {\bibfnamefont {D.~J.}\ \bibnamefont
  {Jacobs}}\ and\ \bibinfo {author} {\bibfnamefont {M.~F.}\ \bibnamefont
  {Thorpe}},\ }\href@noop {} {\enquote {\bibinfo {title}
  {{C}omputer-implemented system for analyzing rigidity of substructures within
  a macromolecule},}\ } (\bibinfo {year} {1999}),\ \bibinfo {note} {uS Patent
  6,014,449}\BibitemShut {NoStop}%
\bibitem [{\citenamefont {Huheey}\ \emph {et~al.}(1993)\citenamefont {Huheey},
  \citenamefont {Keitler},\ and\ \citenamefont
  {Keitler}}]{huheey1993inorganic}%
  \BibitemOpen
  \bibfield  {author} {\bibinfo {author} {\bibfnamefont {J.}~\bibnamefont
  {Huheey}}, \bibinfo {author} {\bibfnamefont {E.}~\bibnamefont {Keitler}}, \
  and\ \bibinfo {author} {\bibfnamefont {R.}~\bibnamefont {Keitler}},\
  }\href@noop {} {\emph {\bibinfo {title} {{I}norganic {C}hemistry,
  {P}rinciples of {S}tructure and {B}onding}}}\ (\bibinfo  {publisher} {Harper
  Collins College Publishers, New York},\ \bibinfo {year} {1993})\BibitemShut
  {NoStop}%
\bibitem [{\citenamefont {Jacobs}\ \emph {et~al.}(2001)\citenamefont {Jacobs},
  \citenamefont {Rader}, \citenamefont {Kuhn},\ and\ \citenamefont
  {Thorpe}}]{jacobs2001protein}%
  \BibitemOpen
  \bibfield  {author} {\bibinfo {author} {\bibfnamefont {D.}~\bibnamefont
  {Jacobs}}, \bibinfo {author} {\bibfnamefont {A.}~\bibnamefont {Rader}},
  \bibinfo {author} {\bibfnamefont {L.}~\bibnamefont {Kuhn}}, \ and\ \bibinfo
  {author} {\bibfnamefont {M.}~\bibnamefont {Thorpe}},\ }\href@noop {}
  {\bibfield  {journal} {\bibinfo  {journal} {Proteins: Structure, Function,
  and Bioinformatics}\ }\textbf {\bibinfo {volume} {44}},\ \bibinfo {pages}
  {150} (\bibinfo {year} {2001})}\BibitemShut {NoStop}%
\bibitem [{\citenamefont {Dahiyat}\ \emph {et~al.}(1997)\citenamefont
  {Dahiyat}, \citenamefont {Benjamin~Gordon},\ and\ \citenamefont
  {Mayo}}]{dahiyat1997automated}%
  \BibitemOpen
  \bibfield  {author} {\bibinfo {author} {\bibfnamefont {B.}~\bibnamefont
  {Dahiyat}}, \bibinfo {author} {\bibfnamefont {D.}~\bibnamefont
  {Benjamin~Gordon}}, \ and\ \bibinfo {author} {\bibfnamefont {S.}~\bibnamefont
  {Mayo}},\ }\href@noop {} {\bibfield  {journal} {\bibinfo  {journal} {Protein
  Science}\ }\textbf {\bibinfo {volume} {6}},\ \bibinfo {pages} {1333}
  (\bibinfo {year} {1997})}\BibitemShut {NoStop}%
\bibitem [{\citenamefont {Gilson}\ and\ \citenamefont
  {Honig}(1986)}]{gilson1986dielectric}%
  \BibitemOpen
  \bibfield  {author} {\bibinfo {author} {\bibfnamefont {M.~K.}\ \bibnamefont
  {Gilson}}\ and\ \bibinfo {author} {\bibfnamefont {B.~H.}\ \bibnamefont
  {Honig}},\ }\href@noop {} {\bibfield  {journal} {\bibinfo  {journal}
  {Biopolymers}\ }\textbf {\bibinfo {volume} {25}},\ \bibinfo {pages} {2097}
  (\bibinfo {year} {1986})}\BibitemShut {NoStop}%
\bibitem [{\citenamefont {Jorgensen}\ and\ \citenamefont
  {Tirado-Rives}(1988)}]{jorgensen1988opls}%
  \BibitemOpen
  \bibfield  {author} {\bibinfo {author} {\bibfnamefont {W.~L.}\ \bibnamefont
  {Jorgensen}}\ and\ \bibinfo {author} {\bibfnamefont {J.}~\bibnamefont
  {Tirado-Rives}},\ }\href@noop {} {\bibfield  {journal} {\bibinfo  {journal}
  {Journal of the American Chemical Society}\ }\textbf {\bibinfo {volume}
  {110}},\ \bibinfo {pages} {1657} (\bibinfo {year} {1988})}\BibitemShut
  {NoStop}%
\bibitem [{\citenamefont {Schu{\`e}ttelkopf}\ and\ \citenamefont
  {Van~Aalten}(2004)}]{schuettelkopf2004prodrg}%
  \BibitemOpen
  \bibfield  {author} {\bibinfo {author} {\bibfnamefont {A.~W.}\ \bibnamefont
  {Schu{\`e}ttelkopf}}\ and\ \bibinfo {author} {\bibfnamefont {D.~M.}\
  \bibnamefont {Van~Aalten}},\ }\href@noop {} {\bibfield  {journal} {\bibinfo
  {journal} {Acta Crystallographica Section D: Biological Crystallography}\
  }\textbf {\bibinfo {volume} {60}},\ \bibinfo {pages} {1355} (\bibinfo {year}
  {2004})}\BibitemShut {NoStop}%
\bibitem [{\citenamefont {Delmotte}(2014)}]{delmotte2014allscale}%
  \BibitemOpen
  \bibfield  {author} {\bibinfo {author} {\bibfnamefont {A.}~\bibnamefont
  {Delmotte}},\ }\emph {\bibinfo {title} {{A}ll-scale structural analysis of
  biomolecules through dynamical graph partitioning}},\ \href@noop {} {Ph.D.
  thesis},\ \bibinfo  {school} {Imperial College London} (\bibinfo {year}
  {2014})\BibitemShut {NoStop}%
\bibitem [{\citenamefont {Conte}\ \emph {et~al.}(2000)\citenamefont {Conte},
  \citenamefont {Ailey}, \citenamefont {Hubbard}, \citenamefont {Brenner},
  \citenamefont {Murzin},\ and\ \citenamefont {Chothia}}]{conte2000scop}%
  \BibitemOpen
  \bibfield  {author} {\bibinfo {author} {\bibfnamefont {L.~L.}\ \bibnamefont
  {Conte}}, \bibinfo {author} {\bibfnamefont {B.}~\bibnamefont {Ailey}},
  \bibinfo {author} {\bibfnamefont {T.~J.}\ \bibnamefont {Hubbard}}, \bibinfo
  {author} {\bibfnamefont {S.~E.}\ \bibnamefont {Brenner}}, \bibinfo {author}
  {\bibfnamefont {A.~G.}\ \bibnamefont {Murzin}}, \ and\ \bibinfo {author}
  {\bibfnamefont {C.}~\bibnamefont {Chothia}},\ }\href@noop {} {\bibfield
  {journal} {\bibinfo  {journal} {Nucleic acids research}\ }\textbf {\bibinfo
  {volume} {28}},\ \bibinfo {pages} {257} (\bibinfo {year} {2000})}\BibitemShut
  {NoStop}%
\bibitem [{\citenamefont {Daily}\ \emph {et~al.}(2008)\citenamefont {Daily},
  \citenamefont {Upadhyaya},\ and\ \citenamefont {Gray}}]{daily2008contact}%
  \BibitemOpen
  \bibfield  {author} {\bibinfo {author} {\bibfnamefont {M.}~\bibnamefont
  {Daily}}, \bibinfo {author} {\bibfnamefont {T.}~\bibnamefont {Upadhyaya}}, \
  and\ \bibinfo {author} {\bibfnamefont {J.}~\bibnamefont {Gray}},\ }\href@noop
  {} {\bibfield  {journal} {\bibinfo  {journal} {Proteins: Structure, Function,
  and Bioinformatics}\ }\textbf {\bibinfo {volume} {71}},\ \bibinfo {pages}
  {455} (\bibinfo {year} {2008})}\BibitemShut {NoStop}%
\bibitem [{\citenamefont {Atilgan}\ \emph {et~al.}(2004)\citenamefont
  {Atilgan}, \citenamefont {Akan},\ and\ \citenamefont
  {Baysal}}]{atilgan2004small}%
  \BibitemOpen
  \bibfield  {author} {\bibinfo {author} {\bibfnamefont {A.~R.}\ \bibnamefont
  {Atilgan}}, \bibinfo {author} {\bibfnamefont {P.}~\bibnamefont {Akan}}, \
  and\ \bibinfo {author} {\bibfnamefont {C.}~\bibnamefont {Baysal}},\
  }\href@noop {} {\bibfield  {journal} {\bibinfo  {journal} {Biophysical
  journal}\ }\textbf {\bibinfo {volume} {86}},\ \bibinfo {pages} {85} (\bibinfo
  {year} {2004})}\BibitemShut {NoStop}%
\end{thebibliography}%

\clearpage
\newpage
\onecolumngrid

\setcounter{figure}{0}
\setcounter{table}{0}
\setcounter{section}{0}
\setcounter{equation}{0}
\renewcommand\thesection{S\arabic{section}}
\renewcommand\thefigure{S\arabic{figure}}
\renewcommand\thetable{S\Roman{table}}
\renewcommand\theequation{S\arabic{equation}}

\begin{center}\Large\textbf{Supplementary Information}\end{center}

\section{Derivation of the graph-theoretical formula for edge fluctuations}

We now derive in more detail Eq.(5), presented in Materials and Methods (Section~IVA) in the main text.
Let us consider the Langevin equation, Eq.(1) in the main text:
\begin{equation}\label{eq:lang_node}
  \mathbf{\dot x} =-L\mathbf{x} + \boldsymbol \epsilon,
\end{equation}
where $\boldsymbol \epsilon$ is white Gaussian noise.  Without loss of
generality, we may assume that the system started initially from a condition $\mathbf{x}(-\infty) = 0$.
A standard result from linear system theory is that the solution of equation~\eqref{eq:lang_node} is given
by:
\begin{equation}
  \mathbf{X}(t) = \int_{-\infty}^t \exp[-L(t-s)] \boldsymbol{\epsilon}(s)ds.
\end{equation}
Since our input $\boldsymbol \epsilon$ is random, $\mathbf{X}(t)$ is a
random process, which we indicate by the upper-case notation. Likewise, the
edge variables will be described by the random process:
\begin{equation}
  \mathbf Y(t) = B^T \int_{-\infty}^t \exp[-L(t-s)] \boldsymbol{\epsilon}(s)ds,
\end{equation}
where $B$ is the incidence matrix of the graph of the protein.  

The autocorrelation of the process $\mathbf{Y}(t)$ for $\tau>0$ is then
  \begin{align}
    \mathcal R(\tau) &= \mathbb E[\mathbf{Y}(t)\mathbf{Y}^T(t+\tau)] 
    = \mathbb E \left[ \int_{-\infty}^{t+\tau} \int_{-\infty}^t B^T
      \exp [-L(t-s)] \boldsymbol\epsilon(s) \boldsymbol\epsilon(\xi)^T
      \exp [-L(t + \tau - \xi)]^T B \, ds \, d\xi
    \right] \nonumber \\
    &= \int_{-\infty}^{t+\tau} \int_{-\infty}^t B^T \exp [-L(t-s)]
    \mathbb E \left[ \boldsymbol\epsilon(s) \boldsymbol\epsilon(\xi)^T \right]
    \exp [-L(t + \tau - \xi)]^T B \, ds \, d\xi \nonumber \\
    &= \int_{-\infty}^{t+\tau} \int_{-\infty}^t B^T \exp [-L(t-s)]
    \left[ \delta(s - \xi) \, I  \right] \exp [-L(t + \tau - \xi)]^T B \, ds \, d\xi \nonumber \\
    &=\int_{- \infty}^t B^T \exp[-L(t -\xi)] \exp[-L(t+\tau-\xi)]^T B \, d\xi =
    \int_{-\infty}^t B^T \exp[-L(2t-2\xi +\tau)]  B \, d\xi, 
  \end{align}
where we have used the fact that the noise vector $\boldsymbol
\epsilon$ is delta-correlated in time and across nodes (i.e., $I = \delta_{ij}$ is the identity matrix).
The last equality follows from fact that $L=L^T$; hence it commutes and this implies that 
$\exp(Lt)\exp(Lt)^T=\exp(2Lt)$.

This integral can be computed using the eigendecomposition of the matrix 
exponential as follows:
\begin{align}
  \mathcal R(\tau) &=  \int_{-\infty}^t B^T \exp[-L(2t-2\xi +\tau)]  B \, d\xi 
  = \sum_{i=1}^N \int_{-\infty}^t B^T e^{-\lambda_i (2t-2\xi +\tau)}\mathbf{v}_i\mathbf{v}_i^T B\: d\xi   \nonumber \\
  &= \dfrac{1}{N}\int_{-\infty}^t  B^T \mathbf{1}\mathbf{1}^TB \, d\xi + \sum_{i=2}^N \int_{-\infty}^t  B^T e^{-\lambda_i (2t-2\xi +\tau)}\mathbf{v}_i\mathbf{v}_i^TB \, d\xi
  = \sum_{i=2}^N \int_{-\infty}^t  B^T e^{-\lambda_i (2t-2\xi +\tau)}\mathbf{v}_i\mathbf{v}_i^TB \, d\xi \nonumber \\
  &= B^T \left [\sum_{i=2}^N \left .\dfrac{e^{-\lambda_i(\tau+2t-2\xi})}{2\lambda_i} \right|_{\xi=-\infty}^t\mathbf{v}_i\mathbf{v}_i^T \right] B
  = \dfrac{1}{2} B^T \left [\sum_{i=2}^N \dfrac{1}{\lambda_i}  e^{-\lambda_i \tau} \mathbf{v}_i\mathbf{v}_i^T \right]B  \nonumber \\ 
  &= \dfrac{1}{2} B^T \left [\sum_{i=2}^N \dfrac{1}{\lambda_i} \mathbf{v}_i\mathbf{v}_i^T \sum_{j=1}^N e^{-\lambda_j\tau} \mathbf{v}_j\mathbf{v}_j^T \right]B
  = \dfrac{1}{2}B^TL^\dagger \exp(-\tau L) B.
\end{align}
Here we have used the fact that the leading eigenvector of $L$ associated with $\lambda_1 =0$ 
is the vector of ones ($\mathbf{v}_1 = \mathbf{1}$), which is in the null space of $B^T$, i.e., 
$B^T\mathbf{1} = 0$.
In the last two equations we have made use of the orthonormality of the eigenvectors ($\mathbf{v}_i^T\mathbf{v}_j^{} =\delta_{ij}$), 
which implies that $\mathbf{v}_i \mathbf{v}^T_i = \mathbf{v}_i \mathbf{v}_i^T\sum_{j=1}^N\mathbf{v}_j\mathbf{v}_j^T$.

\newpage

\section{Construction of the atomistic protein network}
As discussed in Materials and methods (Section~IVE),
the protein network is constructed by assigning edges between atoms which
interact covalently and non-covalently.  Each edge is weighted by the
strength of the interaction.  Covalent bond strengths are obtained
from tables assuming standard bond lengths.  We include three types of
non-covalent interactions: hydrophobic interactions, hydrogen bonds,
and electrostatic interactions.   The assignment of bonds in the graph follows from 
the well established FIRST framework~\cite{Jacobs1995,Jacobs1999}.
More in detail:
\begin{itemize}
\item \textbf{Covalent bonds:} Covalent bonds are weighted according
  to standard bond dissociation energies given in
  Ref.~\cite{huheey1993inorganic}.
\item \textbf{Hydrophobic tethers:} Hydrophobic tethers are assigned
  between C-C or C-S pairs based on proximity: two atoms have a
  hydrophobic tether if their Van der Waals' radii are within 2~\AA.
  The hydrophobic tethers are identified using
  FIRST~\cite{jacobs2001protein}, which does not assign them an
  energy, and the energy is then determined based on the double-well
  potential of mean force introduced by Lin \textit{et
    al}~\cite{lin2007hydrophobic}, which gives an energy of $\approx$
  -0.8kcal/mol for atoms within 2~\AA.
\item \textbf{Hydrogen bonds:} The energies of hydrogen bonds were
  calculated using the same formula used by the program
  FIRST~\cite{jacobs2001protein} and is based on the potential
  introduced by Mayo \textit{et al}~\cite{dahiyat1997automated}.
\item \textbf{Electrostatic interactions:} Important electrostatic
  interactions between ions and ligands, as defined in the LINK
  entries of the PDB file, are added with energies derived from a
  Coulomb potential
  \begin{equation}
    E_{\text{Coul}} = \frac{332}{\epsilon}\frac{q_1q_2}{r},
  \end{equation}
  where $q_1$ and $q_2$ are the atom charges, $r$ is the distance
  between them, and $\epsilon$ is the dielectric constant, which is
  set to $\epsilon = 4$ as in Ref. ~\cite{gilson1986dielectric}.
  Atom charges for standard residues are obtained from the
  OPLS-AA force field~\cite{jorgensen1988opls}, whereas charges for
  ligands and non-standard residues are found using the PRODRG
  web-server~\cite{schuettelkopf2004prodrg}.
\end{itemize}

An extended discussion of the construction of the atomistic graph can
be found in
Refs.~\cite{delmotte2011protein,delmotte2014allscale,amor2014uncovering}

\newpage
\section{Propensities of CheY conformations:  different activation states and NMR ensemble}

In the following Table and Figures we give additional information about the active and inactive conformations and NMR data of CheY used in Section~IIB of the main text.  

\subsection{Active and inactive conformations of CheY}
\begin{figure*}[!h]
  \centering
  \includegraphics[width = 0.6\textwidth]{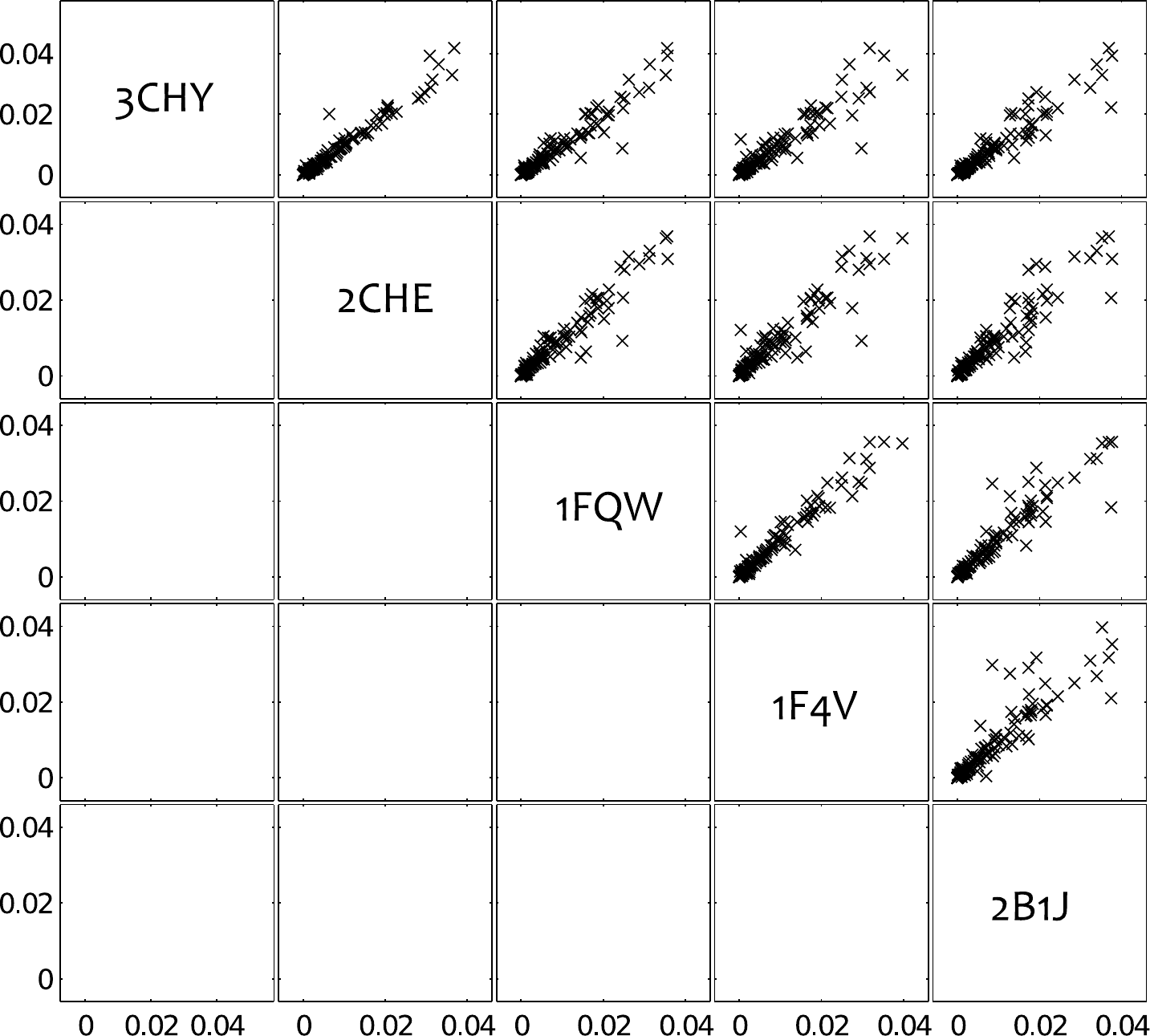}
  \caption{\textbf{Propensities in different conformations of CheY.} Comparison of propensities of residues in
    across different structures of CheY: unbound (3CHY);
    bound to Mg$^{2+}$ (2CHE); bound to Mn$^{2+}$ and phosphate mimic BeFx
    (1FQW); bound to Mn$^{2+}$, BeFx and FliM (1F4V); and bound only to
    FliM (2B1J).  The propensities of the residues 
    are strongly correlated across states.}
  \label{fig:chey_struct_comparison}
\end{figure*}

We calculated the propensities of residues for several CheY structures representing different activation states.  
Details of the different structures are given in Table~\ref{tbl:chey_structs} and a comparison of
the perturbation propensities across the different structures is shown in Figure~\ref{fig:chey_struct_comparison}.
As discussed in Section~IIB.2, the propensities of the residues are strongly correlated across states.
In the main text (Section~IIB.2 and Figure~3), we concentrate on the comparison of 1F4V (active) against 3CHY (inactive). 
\begin{table}[!h]
  \centering
  \caption{\textbf{Details of X-ray structures of CheY analysed.} The conformations correspond to different stages of activation.}
  \label{tbl:chey_structs}
  \begin{tabular}{ccc}
    PDB ID & Structural state & Resolution \\
    \hline
    3CHY & Unbound & 1.7~\AA\\
    2CHE & Bound to Mg$^{2+}$ & 1.8~\AA\\
    1FQW & Bound to Mn$^{2+}$ and BeFx & 2.37~\AA\\
    2B1J & Bound to FliM & 2.4~\AA\\
    1F4V & Bound to Mn$^{2+}$, BeFx and FliM & 2.22~\AA\\
  \end{tabular}
\end{table}

\subsection{CheY structures from NMR experiments}
\begin{figure*}[!htb]
  \centering
  \includegraphics[width = .9\textwidth]{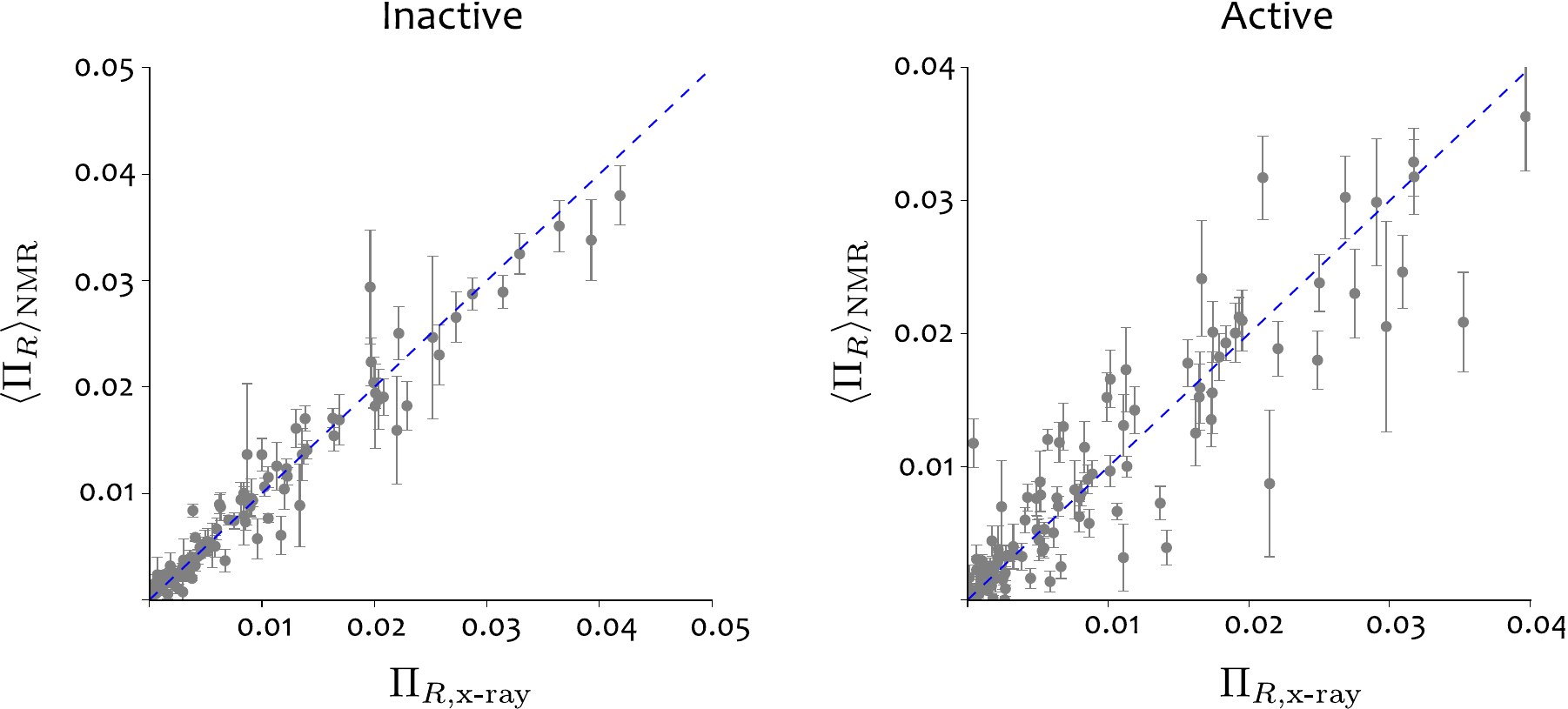}
  \caption{\textbf{Propensities computed from CheY NMR ensembles.} 
  Average propensity obtained from all structures in an NMR ensemble of CheY against the
  propensity obtained from the corresponding X-ray structure for inactive
    (left) and active (right).  The inactive ensemble contains 20 structures and the active ensemble contains 27 structures.  The error bars show the standard deviation of the propensities $\Pi_R$ over the NMR ensemble. Both the variance and the deviation from the X-ray structure is greater for the active conformation.}
  \label{fig:chey_xray_nmr_comparison}
\end{figure*}

We also calculated the perturbation propensities of residues
across two ensembles of NMR structures for active CheY (PDB ID: 1DJM; 27 structures) and
inactive CheY (PDB ID: 1CYE; 20 structures).  A comparison of the average
propensity of each residue (averaged across the NMR ensemble) versus its
propensity in the X-ray structure is shown in Figure~\ref{fig:chey_xray_nmr_comparison} for both
the active ensemble (1DJM) and the inactive ensemble (1CYE). This data is discussed in
the main text (Section~IIB.3) and summarised in Figure~4.

\clearpage

\section{The protein reference set from the SCOP database and absolute quantile scores}
As discussed in Section~IID, we have collected a random reference set of 100 proteins drawn from
the Structural Classifiation of Proteins (SCOP) database~\cite{conte2000scop}.  This reference protein set is used to obtain absolute quantile scores for the propensities, as detailed in Materials and Methods (Section~IVC).
Here we give further details on the reference set and the comparison of absolute and intrinsic quantile scores.

\paragraph*{SCOP database:}
Protein domains in the SCOP database are classified according to a hierarchy based on structural similarity.
Although proteins are additionally divided into superfamilies and
subfamilies according to structural and sequence similarity, 
the major classes are:
\begin{enumerate}
\item All $\alpha$: protein domains containing only alpha-helices
\item All $\beta$: protein domains containing only beta-sheets
\item Alpha and beta ($\alpha/\beta$): protein domains containing both
  $\alpha$-helices and $\beta$-sheets, with mainly parallel
  $\beta$-sheets.
\item Alpha and beta ($\alpha + \beta$): protein domains containing
  both $\alpha$-helices and $\beta$-sheets, with mainly anti-parallel
  $\beta$-sheets.
\item Multi-domain: folds of two or more domains from different
  classes.
\end{enumerate}
We chose 20 proteins from each of these five classes uniformly at random from all proteins in each class,
yet choosing only from structures where there is a ligand bound to the active site.  

\paragraph*{Absolute quantile scores:}
On this set of 100 proteins, we then identified the active site in each protein and
computed the propensity for all its bonds relative to the
active site.  Across the set of 100 proteins in the reference set, 
we have a total of 465,409 non-covalent bonds, on which we
apply quantile regression to obtain absolute quantile scores $p^{\text{\tiny{ref}}}$.
In Figure~\ref{fig:p_pref_comparison} below, the quantile scores $p_b$ for all the bonds of the 
three proteins studied in detail in the main text (caspase-1, Che-Y, h-Ras) are plotted against 
their absolute quantile score $p_b^{\text{\tiny{ref}}}$, showing a good correlation overall.  
In general, we observe a tighter correlation for larger proteins (e.g., caspase-1), 
as a result of the QR fit being based on the number of bonds, $E$, which is related to the size of the protein. 

\begin{figure*}[!htb]
  \centering
  \includegraphics[width = .6\textwidth]{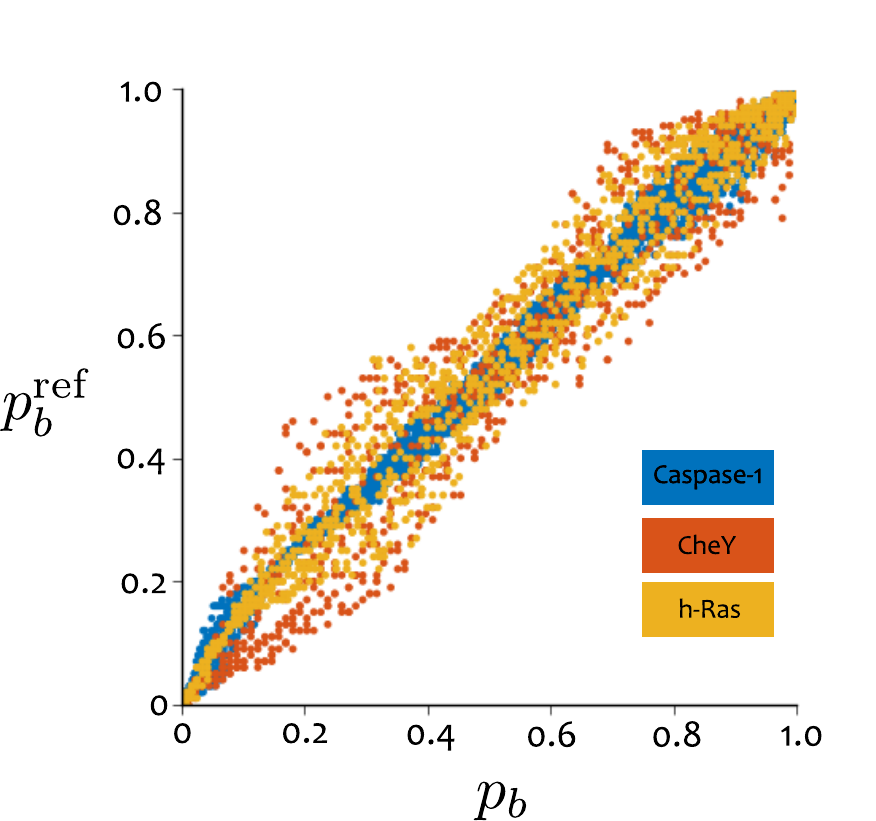}
  \caption{\textbf{Absolute quantile scores versus intrinsic quantile scores.} The absolute quantile scores calculated from the reference set ($p_b^{\text{\tiny{ref}}}$) are plotted against the 
      intrinsic quantile scores ($p_b$) for caspase-1 (blue), CheY (red), and h-Ras (yellow).}
  \label{fig:p_pref_comparison}
\end{figure*}

\clearpage

\section{Bond-to-bond propensities of the allosteric test set}\label{sec:testset:test_set}
\begin{figure*}[!h]
  \centering \includegraphics[width=0.8\textwidth]{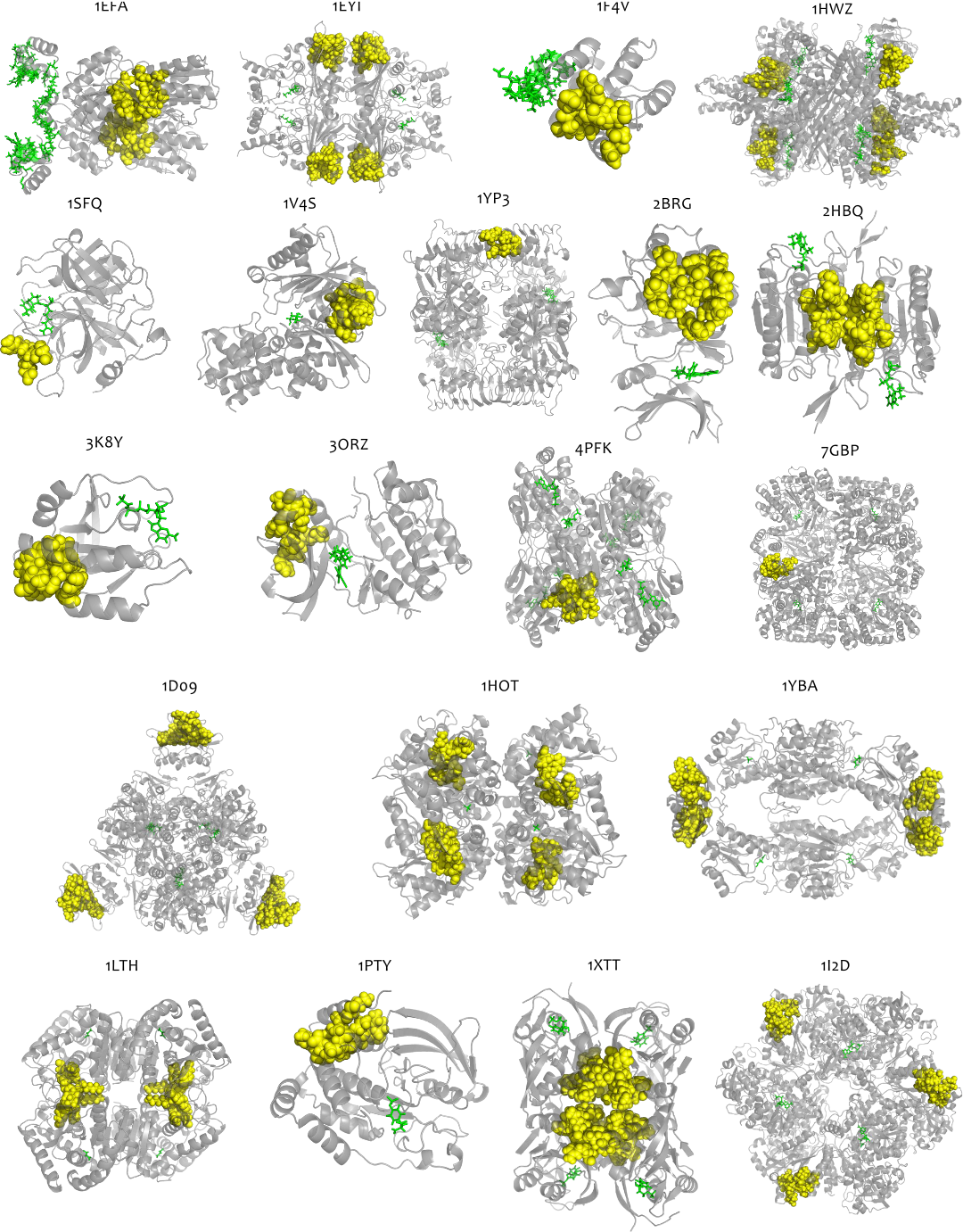}
  \caption[Test set proteins]{\textbf{Allosteric test set.}  The
    structures of the 20 proteins in the allosteric test set are shown
    with the active site ligand (green sticks) and allosteric site residues
    (yellow spheres).}
  \label{fig:test_set_structures}
\end{figure*}

\subsection{Description of the allosteric test set}
As discussed in the main text (Section~IIE), we have constructed a test set of 20 allosteric proteins 
on which to benchmark our algorithm. Each protein in our test set has a structure with a bound active site
ligand and a structure with a bound allosteric ligand.  If the protein
is allosterically activated then we use a single structure in which
the protein is complexed with both the activator and the active site
ligand. Ref. \cite{daily2008contact} collected a test set of 15 allosteric
proteins for which both active site bound and allosteric site bound
structures are available. We have used 10 of these
proteins (the other five were found to be unsuitable for our
analysis due to the presence of many non-standard
  amino-acids, mismatch between the oligomeric state of the active and
  inactive structures, or the absence of an allosteric ligand). 
  We have enlarged the set with a further 10 proteins from an extensive search of the
literature. The structures of the 20 proteins are shown in Figure
\ref{fig:test_set_structures}, with the active site indicated by the green ligand, and the
allosteric site indicated by the yellow spheres.  
The allosteric site is defined as any residue
containing an atom within 4\AA~of the allosteric ligand;
allosteric site bonds are defined as any weak interactions formed by
an allosteric residue.  Full details of the proteins and allosteric
site residues are shown in Table~\ref{tbl:test_set}.

\begin{table}[!h]
  \begin{small}
  \caption[Test set]{\textbf{Proteins in the allosteric test set.}  The active site and allosteric site bound structures for each of the 20 test set proteins.
    If the protein is allosterically activated then the PDB ID for both states will be the same.  The ligand identifier is that used in the PDB file. 
    Exceptions to this are CheY and caspase-1.  As the ligand in these proteins is a peptide, the name and chain ID of the peptide is given instead.}
  \label{tbl:test_set}
  \begin{center}
    \begin{tabular}{l l l l l l}
      &  & \multicolumn{2}{l}{\quad \quad Active} & \multicolumn{2}{l}{\quad Allosteric} \\ \vspace*{-.3cm} \\
      \cline{3-4} \cline{5-6}  \\ \vspace*{-.6cm} \\
      Protein  & Residues & PDB & Ligand & PDB & Ligand \\ \vspace*{-.3cm} \\
      \hline \\
      ATCase & 2790 & 1D09 & PAL & 1RAC & CTP \\
      Lac repressor  & 658 & 1EFA & NPF & 1TLF & IPT \\
      Fructose-1, 6-Bisphosphatase  & 1344 & 1EYI  & F6P & 1EYJ & AMP \\
      CheY & 144 & 1F4V & FliM (D) & 1F4V & BEF \\
      Glutamate DH  & 3018 & 1HWZ  & NDP & 1HWZ & GTP \\
      ATP Sulfurylase  & 3444 & 1I2D  & ADX & 1M8P & PPS\\
      PTP1B  & 299 & 1PTY  & PTR & 1T48 & BB3 \\
      Thrombin  & 281 & 1SFQ & O6G & 1SFQ & NA\\
      Glucokinase  & 449 & 1V4S  & GLC & 1V4S & MRK \\
      UPRTase  & 852 & 1XTT & U5P & 1XTU & CTP \\
      Phosphoglycerate DH & 1644 & 1YBA  & AKG & 1PSD & SER451 \\
      ADP-glucose phosphorylase & 1727 & 1YP3  & ATP & 1YP2 & PMB \\
      CHK1 & 258 & 2BRG & DFY & 3JVS & AGY \\
      Caspase-1 & 520 & 2HBQ  & z-VAD-FMK (C/F) & 2FQQ & F1G\\
      PDK1 & 278 & 3ORZ & BI4 & 3ORZ  & 2A2\\
      Phosphofructokinase & 1288 & 4PFK & F6P & 6PFK & PGA\\
      Glycogen Phosphorylase & 3304 & 7GPB & PLP/SO4 & 7GPB  & SO4/AMP\\
      glcN-6-P deaminase & 1604 & 1HOT & PO4 & 1HOT & NAG/PHS\\
      h-Ras & 175 & 3K8Y & GNP & 3K8Y & ACT\\
      lactate DH & 1260 & 1LTH & NAD & 1LTH & FBP\\
      \hline
    \end{tabular}
  \end{center}
  \end{small}
\end{table}

\subsection{Summary of results on the allosteric test set}
As explained in the main text (Section~IIE and Materials and Methods, Section~IVD),
for each of the 20 proteins in the test set, we analyse the propensities of all bonds
with respect to the active site of the bound structure, using the ligands shown in 
Fig.~\ref{fig:test_set_structures} as the source for the bond-to-bond propensity calculations.  
For each protein, we obtain the propensity $\Pi_b$ of every weak bond 
and its associated quantile score ($p_b$).
To establish their statistical significance, the bond quantile scores $p_b$ 
(and residue averaged quantile scores $p_R$) 
of the allosteric site are compared against an ensemble of randomly generated surrogate sites from each protein.
The ensemble of  
surrogate sites is constructed at random by picking sites that satisfy two structural constraints: 
(i) they have the same number of residues as the allosteric site; 
and (ii) their diameter (the maximum distance between any two atoms in the site) is no larger than
that of the allosteric site.  The sites are generated using Algorithm~\ref{alg:surrogate} with pseudocode given below. The propensities averaged over the ensemble of surrogate sites 
are then used for statistical comparison with the allosteric site. 
We also obtain absolute propensity scores for each bond ($p^{\tiny{\text{ref}}}_b$) by comparing against 
the reference SCOP ensemble of 100 proteins.
These quantities are defined in 
the main text (Materials and Methods, Section~IVD).
\begin{algorithm}[H]
  \caption{Pseudocode for surrogate site sampling}
  \begin{algorithmic}[1]
    \State site $\leftarrow \varnothing$ \While{\# residues in site
      $<$ \# residues in allosteric site} \State choose a residue $R$
    at random \If{diameter(site $\cup$ R) $<$ diameter(allosteric
      site)} \State site $\leftarrow$ site $\cup$ R
    \EndIf
    \EndWhile
  \end{algorithmic}
\label{alg:surrogate}
\end{algorithm}

Using all these scores we obtain our four statistical measures 
of significance summarised in Table~\ref{tbl:test_set_results}.  
These numerical results are presented also in the form of a graph in 
Figure 7 of the main text.

\begin{table*}[h!]
  \centering
  \caption[Allosteric site quantile scores in test set
  proteins]{\textbf{Allosteric site quantile scores in test set
  proteins.}
  The four scores described in Figure 7 of the main text for the test set of 20 proteins. 
  The difference between the allosteric site average quantile score and the average surrogate site score for both residues and bonds are shown in bold if they are greater than 0, and starred if they lie above the 95\% confidence interval computed by a bootstrap with 10000 resamples.
  The average reference quantile score  $\overline{p_{b, \text{allo}}^{\text{ref}}}$ is shown in bold if it is greater than 0.5 (the expected value).
    The proportion $p_{b,\text{allo}} > 0.95$
    is shown in bold if it is greater than 0.05.  }
  \label{tbl:test_set_results}
  \begin{small}
  \begin{tabular}{ll | cccc | c}
    Protein & PDB ID 
    & $\overline{p_{R,\text{allo}}} - \langle \overline{p_{R,\text{site}}}\rangle_\text{surr}$  
    & $\overline{p_{b,\text{allo}}} - \langle \overline{p_{b,\text{site}}}\rangle_\text{surr}$ 
    & $\text{P}(p_{b,\text{allo}} > 0.95)$
    & $\overline{p_{b, \text{allo}}^{\text{ref}}}$ 
    & Summary \\
    \hline  
    %
    %
    Glucokinase & 1V4S &     
    \textbf{0.35}$^{*}$ & \textbf{0.14}$^{*}$  & \textbf{0.12} & \textbf{0.66} &    
    \filledcirc\filledcirc\filledcirc\filledcirc \\
    PDK1 & 3ORZ &    
     \textbf{0.30}$^{*}$ & \textbf{0.030}$^{*}$ & \textbf{0.080} & \textbf{0.56} &
     \filledcirc\filledcirc\filledcirc\filledcirc \\    
    ADP-glucose phosphorylase & 1YP3 &  
    \textbf{0.28}$^{*}$ & \textbf{0.074}$^{*}$ & \textbf{0.10} & \textbf{0.59} &
     \filledcirc\filledcirc\filledcirc\filledcirc\\
    ATCase & 1DO9 &     
     \textbf{0.23}$^{*}$ & \textbf{0.036}$^{*}$ & \textbf{0.091} & \textbf{0.68} &
     \filledcirc\filledcirc\filledcirc\filledcirc\\  
    Caspase-1 & 2HBQ &  
    \textbf{0.15}$^{*}$ & \textbf{0.0032}$^{*}$ & \textbf{0.070} & \textbf{0.54} &
    \filledcirc\filledcirc\filledcirc\filledcirc\\ 
    glcN-6-P deaminase & 1HOT &     
    \textbf{0.13}$^{*}$ & \textbf{0.031}$^{*}$ & \textbf{0.079} & \textbf{0.51} &
     \filledcirc\filledcirc\filledcirc\filledcirc\\       
    PTP1B & 1PTY &     
    \textbf{0.11}$^{*}$ & \textbf{0.0088}$^{*}$ & 0.048 & \textbf{0.50} &
     \filledcirc\filledcirc\emptycirc\filledcirc \\
    Fructose-1, 6-Bisphosphatase & 1EYI &   
    \textbf{0.11}$^{*}$ & \textbf{0.033}$^{*}$  & \textbf{0.052} & 0.49 &
     \filledcirc\filledcirc\filledcirc\emptycirc \\
    Glycogen Phosphorylase & 7GPB &   
    \textbf{0.11}$^{*}$ & \textbf{0.0027}$^{*}$ & \textbf{0.058} & 0.47 &
     \filledcirc\filledcirc\filledcirc\emptycirc \\
     Chemotaxis Y & 1F4V &    
    \textbf{0.096}$^{*}$ & \textbf{0.055}$^{*}$ & \textbf{0.074} & \textbf{0.58} &
     \filledcirc\filledcirc\filledcirc\filledcirc\\
   Phosphofructokinase & 4PFK & 
    \textbf{0.092}$^{*}$ & \textbf{0.068}$^{*}$ & \textbf{0.16} & \textbf{0.54} &
     \filledcirc\filledcirc\filledcirc\filledcirc \\    
    ATP Sulfurylase & 1I2D & 
    \textbf{0.091}$^{*}$ & \textbf{0.0313}$^{*}$ & \textbf{0.068} & \textbf{0.52} &
     \filledcirc\filledcirc\filledcirc\filledcirc\\
    Phosphoglycerate DH & 1YBA & 
     \textbf{0.062}$^{*}$ & \textbf{0.076}$^{*}$ & \textbf{0.075} & \textbf{0.59} &
      \filledcirc\filledcirc\filledcirc\filledcirc \\
    Lactate DH & 1LTH & 
    \textbf{0.063}$^{*}$ & \textbf{0.024}$^{*}$ & \textbf{0.063} & \textbf{0.52} &
    \filledcirc\filledcirc\filledcirc\filledcirc\\   
    UPRtase & 1XTT & 
     \textbf{0.0024} & -0.013 & \textbf{0.06} & 0.44 &
     \emptycirc\emptycirc\filledcirc\emptycirc \\
    Glutamate DH & 1HWZ &   
     -0.015 & \textbf{0.039}$^{*}$  & \textbf{0.068} & 0.44 &
     \emptycirc\filledcirc\filledcirc\emptycirc\\
    h-Ras & 3K8Y & 
     -0.043 & -0.016 & \textbf{0.059} & 0.49 &
      \emptycirc\emptycirc\filledcirc\emptycirc\\   
    Lac repressor & 1EFA &   
    -0.066 & -0.016 & 0.014 & \textbf{0.60} &
    \emptycirc\emptycirc\emptycirc\filledcirc \\  
    Thrombin & 1SFQ & 
     -0.081 & \textbf{0.077}$^{*}$  & \textbf{0.16} & \textbf{0.64} &
     \emptycirc\filledcirc\filledcirc\filledcirc\\    
    CHK1 & 2BRG & 
    -0.24  & -0.15 & 0.0052 & 0.36 &
     \emptycirc\emptycirc\emptycirc\emptycirc \\    
    \hline
  \end{tabular}
\end{small}
\end{table*}

\clearpage

\section{Robustness of the bond-to-bond propensities to random perturbations of the weak interactions}

Proteins are dynamic objects undergoing motions and fluctuations under the influence of the environment.
Such dynamic fluctuations induce changes in the bond energies of the protein, potentially 
leading to the breaking of weak bonds (hydrogen bonds, salt bridges, hydrophobic tethers).
As discussed in the main text when studying the NMR ensemble of conformations of CheY (Section~IIB.3),
whilst there is considerable agreement between the results from the NMR structures 
and the X-ray structure (Fig.~\ref{fig:chey_xray_nmr_comparison}),  
the variability in the ensemble can reveal further information.
It is also important to check that the computation of propensities is generally 
robust to the presence of such noise.
To do this, we have developed two schemes to add random perturbations to our protein 
networks. These schemes mimic the effect of small dynamic fluctuations,
without carrying out expensive molecular dynamics simulations.

Firstly, for each of the 20 proteins in our dataset, we add zero mean Gaussian noise to the edge weights (energies) of non-covalent bonds in the graph, so as to mimic the effect of thermal fluctuations. 
Note that we allow the bonds to break if their randomised energy becomes zero. 
We then recompute our quantile scores for the allosteric site for 10 realisations of 
the noisy networks generated after the addition of the Gaussian fluctuations. 
We do this for 3 levels of noise, i.e., we increase the standard deviation of the Gaussian 
from 1kT=0.6 kcal/mol to 4kT=2.4 kcal/mol. 
The average results of these randomisations for all proteins in the allosteric test set are presented in Table~\ref{tbl:gaussian_randomisation}. Our calculations show that the results are generally robust to fluctuations induced in this way: the signal at the allosteric site only drops slightly when introducing relatively high levels of noise. 

\begin{table}[!h]
\begin{small} 
 \caption{\textbf{Robustness of propensity scores to additive randomness.} Mean ($\pm$ standard deviation) of propensity scores $\overline{p_{R,\text{allo}}} - \langle \overline{p_{R,\text{site}}}\rangle_\text{surr}$  computed from randomisations of the protein networks of the allosteric test set obtained by adding Gaussian noise to the edge weights (bond energies).  The noise level varies between 1kT and 4kT (corresponding to the standard deviation of the added Gaussian) and at each noise level the results were calculated from 10 randomised graphs.  The difference between the allosteric site average quantile score and the average surrogate site score for both residues and bonds are shown in bold if they are greater than 0, and starred if they lie above the 95\% confidence interval computed by a bootstrap with 10000 resamples. The unperturbed result is also shown for comparison.}
  \label{tbl:gaussian_randomisation}
  \begin{center}
   \begin{tabular}{ccccc}
    PDB ID&\multicolumn{1}{p{2cm}}{\centering Unperturbed \\ network} &
    \multicolumn{1}{p{3cm}}{\centering Gaussian noise \\ 1kT}&
    \multicolumn{1}{p{3cm}}{\centering Gaussian noise \\ 2kT}&
       \multicolumn{1}{p{3cm}}{\centering Gaussian noise \\ 4kT} \\
    \hline
    1V4S&
    \textbf{0.35}$^{*}$&\textbf{0.32}$\pm$\textbf{0.011}$^{*}$&\textbf{0.31}$\pm$\textbf{0.019}$^{*}$&\textbf{0.27}$\pm$\textbf{0.017}$^{*}$\\
    3ORZ&
    \textbf{0.30}$^{*}$&\textbf{0.28}$\pm$\textbf{0.0087}$^{*}$&\textbf{0.23}$\pm$\textbf{0.0090}$^{*}$&\textbf{0.24}$\pm$\textbf{0.014}$^{*}$\\
    1YP3&
    \textbf{0.28}$^{*}$&\textbf{0.27}$\pm$\textbf{0.0010}$^{*}$&\textbf{0.25}$\pm$\textbf{0.0088}$^{*}$&\textbf{0.17}$\pm$\textbf{0.016}$^{*}$\\
    1D09&
    \textbf{0.23}$^{*}$&\textbf{0.22}$\pm$\textbf{0.0071}$^{*}$&\textbf{0.21}$\pm$\textbf{0.0024}$^{*}$&\textbf{0.20}$\pm$\textbf{0.0035}$^{*}$\\
    2HBQ&
    \textbf{0.15}$^{*}$&\textbf{0.18}$\pm$\textbf{0.0096}$^{*}$&\textbf{0.20}$\pm$\textbf{0.0058}$^{*}$&\textbf{0.13}$\pm$\textbf{0.0097}$^{*}$\\
    1HOT&
    \textbf{0.13}$^{*}$&\textbf{0.13}$\pm$\textbf{0.0061}$^{*}$&\textbf{0.098}$\pm$\textbf{0.024}$^{*}$&\textbf{0.12}$\pm$\textbf{0.021}$^{*}$\\
    1PTY&
    \textbf{0.11}$^{*}$&\textbf{0.096}$\pm$\textbf{0.020}$^{*}$&\textbf{0.11}$\pm$\textbf{0.022}$^{*}$&\textbf{0.088}$\pm$\textbf{0.031}$^{*}$\\
    1EYI&
    \textbf{0.11}$^{*}$&\textbf{0.13}$\pm$\textbf{0.0065}$^{*}$&\textbf{0.13}$\pm$\textbf{0.0022}$^{*}$&\textbf{0.16}$\pm$\textbf{0.0050}$^{*}$\\
    7GPB&
    \textbf{0.11}$^{*}$&\textbf{0.096}$\pm$\textbf{0.018}$^{*}$&\textbf{0.13}$\pm$\textbf{0.010}$^{*}$&\textbf{0.14}$\pm$\textbf{0.015}$^{*}$\\
    1F4V&
    \textbf{0.096}$^{*}$&\textbf{0.093}$\pm$\textbf{0.018}$^{*}$&\textbf{0.14}$\pm$\textbf{0.0097}$^{*}$&\textbf{0.12}$\pm$\textbf{0.027}$^{*}$\\
    4PFK&
    \textbf{0.092}$^{*}$&\textbf{0.091}$\pm$\textbf{0.0052}$^{*}$&\textbf{0.11}$\pm$\textbf{0.022}$^{*}$&\textbf{0.12}$\pm$\textbf{0.0075}$^{*}$\\
    1I2D&
    \textbf{0.091}$^{*}$&\textbf{0.14}$\pm$\textbf{0.029}$^{*}$&\textbf{0.14}$\pm$\textbf{0.030}$^{*}$&\textbf{0.14}$\pm$\textbf{0.037}$^{*}$\\
    1YBA&
    \textbf{0.062}$^{*}$&\textbf{0.076}$\pm$\textbf{0.0034}$^{*}$&\textbf{0.091}$\pm$\textbf{0.0048}$^{*}$&\textbf{0.073}$\pm$\textbf{0.0051}$^{*}$\\
    1LTH&
    \textbf{0.063}$^{*}$&\textbf{0.070}$\pm$\textbf{0.0099}$^{*}$&\textbf{0.063}$\pm$\textbf{0.016}$^{*}$&\textbf{0.039}$\pm$\textbf{0.019}$^{*}$\\
    1XTT&
    \textbf{0.0024}&\textbf{0.0084}$\pm$\textbf{0.0069}&\textbf{0.015}$\pm$\textbf{0.0083}&\textbf{0.0077}$\pm$\textbf{0.0070}\\
    1HWZ&
    -0.015&-0.0090$\pm$0.0071&-0.0028$\pm$0.0043&\textbf{0.011}$\pm$\textbf{0.0065}\\
    3K8Y&
    -0.043&-0.033$\pm$0.012&-0.012$\pm$0.010&-0.025$\pm$0.022\\
    1EFA&
    -0.066&-0.047$\pm$0.0054&-0.019$\pm$0.0078&-0.0027$\pm$0.0077\\
    1SFQ&
    -0.081&-0.090$\pm$0.0089&-0.083$\pm$0.023&-0.10$\pm$0.028\\
    2BRG&
    -0.24&-0.23$\pm$0.010&-0.24$\pm$0.013&-0.23$\pm$0.026\\
    \hline
  \end{tabular}
  \end{center}
  \end{small}
\end{table}

Secondly, to test a different kind of variability introduced by the environment,   
we have considered the effect of breaking all bonds in our network with energy below a threshold.
Starting with the original unperturbed structure, all weak bonds below a given threshold are removed from the graph.  In this way, we mimic the possibility of extended structural changes that could lead to breaking of bonds in a more global fashion. 

For each of the 20 proteins in the test set, we generate two perturbed networks obtained by bond removal of \textit{all} bonds with energy below two different thresholds: 0.5 kT$\simeq$ 0.3 kcal/mol and 1kT $\simeq$ 0.6 kcal/mol. The effect of this thresholding is extensive.  For the 0.5kT threshold, we delete all hydrophobic tethers and electrostatic interactions as well as a percentage of hydrogen bonds that ranges from 31\% in 1SFQ to 44\% in 1HWZ and 1LTH.  For the 1kT threshold, even further hydrogen bonds are removed, corresponding to eliminating 44\% of H-bonds in 1SFQ up to 57\% of the H-bonds in 7GPB, 2BRG, 1LTH (in addition to all hydrophobic interactions).
 
The calculations of the propensity for the thresholded networks for all 20 proteins in our test set are presented in Table~\ref{tbl:energy_sweep}. Our results show that, overall, the propensity of the allosteric site remains largely robust to such changes across all 20 proteins considered, yet with notable differences in the magnitude of the effect across the set. 
In some proteins, the signal at the
allosteric site is mildly affected by bond deletion (e.g. 3ORZ, 1YP3, 2HBQ, 1HOT, 1PTY). In other cases, however,  the deletion of weaker hydrogen bonds has a large effect in destroying the communication between the allosteric site and the active site (e.g. 1V4S, 1D09, 1EYI, 7GPB, 1F4V).
These differences could be a measure of how robust the allosteric signalling is to energetic fluctuations in the local environment of the protein, and also provide clues as to different structural features connected with the distributed nature of allosteric signalling in the different proteins.
The study of such differences will be the object of future work.

\begin{table}[!h]
\begin{small}
  \caption{\textbf{Robustness of propensity scores to deletion of weak bonds.} The propensity score $\overline{p_{R,\text{allo}}} - \langle \overline{p_{R,\text{site}}}\rangle_\text{surr}$ for networks obtained by deleting all bonds below two energy thresholds.  The results are shown in bold when they are greater than 0 and starred if they lie above the 95\% confidence interval computed by a bootstrap with 10000 resamples. The unperturbed score is reported also for comparison.} 
\label{tbl:energy_sweep}
 \begin{center}
   \begin{tabular}{cccc}
    PDB ID&\multicolumn{1}{p{2cm}}{\centering Unperturbed \\ network} &
    \multicolumn{1}{p{2cm}}{\centering Threshold \\ 0.5 kT}&
    \multicolumn{1}{p{2cm}}{\centering Threshold \\ 1kT} \\
    \hline
    1V4S&
    \textbf{0.35}$^{*}$&\textbf{0.061}$^{*}$&\textbf{0.049}$^{*}$\\
    3ORZ&
    \textbf{0.30}$^{*}$&\textbf{0.24}$^{*}$&\textbf{0.25}$^{*}$\\
    1YP3&
    \textbf{0.28}$^{*}$&\textbf{0.24}$^{*}$&\textbf{0.30}$^{*}$\\
    1D09&
    \textbf{0.23}$^{*}$&\textbf{0.088}$^{*}$&\textbf{0.10}$^{*}$\\
    2HBQ&
    \textbf{0.15}$^{*}$&\textbf{0.16}$^{*}$&\textbf{0.18}$^{*}$\\
    1HOT&
    \textbf{0.13}$^{*}$&\textbf{0.14}$^{*}$&\textbf{0.17}$^{*}$\\
    1PTY&
    \textbf{0.11}$^{*}$&\textbf{0.13}$^{*}$&\textbf{0.080}$^{*}$\\
    1EYI&
    \textbf{0.11}$^{*}$&\textbf{0.026}$^{*}$&\textbf{0.022}$^{*}$\\
    7GPB&
    \textbf{0.11}$^{*}$&\textbf{0.056}$^{*}$&\textbf{0.062}$^{*}$\\
    1F4V&
    \textbf{0.096}$^{*}$&-0.0010&\textbf{0.0085}$^{*}$\\
    4PFK&
    \textbf{0.092}$^{*}$&\textbf{0.17}$^{*}$&\textbf{0.20}$^{*}$\\
    1I2D&
    \textbf{0.091}$^{*}$&\textbf{0.0012}&-0.033\\
    1YBA&
    \textbf{0.062}$^{*}$&\textbf{0.079}$^{*}$&\textbf{0.052}$^{*}$\\
    1LTH&
    \textbf{0.063}$^{*}$&\textbf{0.056}$^{*}$&-0.081\\
    1XTT&
    \textbf{0.0024}&-0.016&-0.023\\
    1HWZ&
    -0.075&-0.016&-0.20\\
    3K8Y&
    -0.043&-0.14&-0.16\\
    1EFA&
    -0.066&\textbf{0.052}$^{*}$&\textbf{0.051}$^{*}$\\
    1SFQ&
    -0.081&\textbf{0.073}$^{*}$&\textbf{0.11}$^{*}$\\
    2BRG&
    -0.24&-0.20&-0.17\\
    \hline
  \end{tabular}
  \end{center}
  \end{small}
\end{table}

\clearpage
\section{Propensities from residue-residue interaction networks}

The computational efficiency of our methodology allows us to analyse all-atom networks without many of the restrictions on system size inherent to other methods. Proteins or protein complexes of hundreds of thousands of atoms can be analysed in a few minutes on a standard desktop.  We can thus keep atomistic detail at the single bond level without restricting the scope of the analysis. Hence there is a less acute need to seek computational savings by obtaining coarse-grained representations of proteins at the level of residue interactions.
However, it is still instructive to consider propensity measures computed 
from residue-level networks (RRINs)~\cite{chennubhotla2007signal}.  
We have undertaken this comparison for all 20 proteins in our test set and report the results below.

As discussed in the main text (Sections~IIA and IIB),  in some cases (e.g., caspase-1, Fig 1b) we found that the additional information contained in the atomistic network leads to increased signal in the detection of the allosteric site, whereas in other cases (e.g., CheY), RRINs already capture well the site connectivity that reveals the presence of the allosteric site.  Our analysis of the full test set (Table~\ref{tbl:mean_RRIN}) confirms that the results from RRINs depend on the protein analysed, and also vary substantially depending on the choice of the cut-off distance (a tunable parameter which needs to be chosen when generating the coarse-grained RRINs). 

The coarse-grained RRINs for each of the 20 proteins in the test set 
were obtained by submitting the corresponding PDB files to the
oGNM server~\cite{yang2006ognm}.  We obtained RRINs at four different cut-off radii: 
6 \AA, 7 \AA, 8 \AA~and 10 \AA. The cut-off radius is a tunable parameter necessary 
to generate a RRIN from PDB files, which establishes how close two residues must be in order to be connected in the RRIN. 
A range of different cut-off radii has been used throughout the literature. However, the usual radius is
around  6.7-7.0~\AA, which corresponds to the first coordination shell~\cite{atilgan2004small}.

Table~\ref{tbl:mean_RRIN} shows the propensity score of the allosteric site 
 $\overline{p_{R,\text{allo}}} - \langle \overline{p_{R,\text{site}}}\rangle_\text{surr}$, 
computed from RRINs obtained at four cut-offs (between 6\AA~and 10\AA) for the 20 proteins 
in the allosteric test set.  
For comparison purposes, we also report the same score obtained from
the all-atom network. It is important to note that this is just one of four scores obtained 
from the all-atom network, reflecting only the \emph{averaged} behaviour over the residues.
This score is complemented by the three other bond-based statistics,
which can pick up inhomogeneities in the propensities of the bonds in the allosteric site, 
as given by the All-atom Summary column carried over from Table~\ref{tbl:test_set_results}.

Our results indicate broad consistency between RRINs and the all-atom network. 
However, the RRIN results vary widely depending on the choice of cut-off radius in the generation of the 
network. Moreover this variability with respect to the cut-off behaves differently for each of the proteins.
As an illustration, the allosteric site of caspase-1 (2HBQ) was not found to be significant in
the RRINs  with cut-off radii of 6~\AA, 7 \AA and 8 \AA~and only weakly significant
for 10~\AA, whereas 1LTH and 2BRG are both only detected in RRINS with cut-off radius of 6~\AA~but
not for larger radii.
Our results are consistent with previous studies that found that allosteric pathway
identification in RRINs is dependent on the chosen cut-off~\cite{ribeiro2014determination}. 
For the different cut-offs, the number of proteins with $p_{R,\text{allo}} > p_{R,\text{rest}}$
varies between 11/20 (at 7, 8, and 10\AA) and 13/20 (at 6\AA), and  only 8/20 proteins 
have $p_{R,\text{allo}} > p_{R,\text{rest}}$ for the RRINS at \textit{all} the cut-off radii.  
This is compared to 15/20 proteins for the atomistic network.

Even when the allosteric site is detected in the RRIN, the signal when 
using the atomistic network is considerably higher in a 
number of proteins (e.g., 1V4S, 1YP3, 7GPB, 1I2D, 2HBQ). 
In other cases (e.g., 1EYI, 4PFK), the RRIN directly loses the 
detectability of the allosteric site even if the cut-off is adjusted.
This observation suggests that these are proteins where the specific chemistry of intra-protein
bonds is important for the allosteric communication. 

On the other hand, there are several other cases (e.g., 3ORZ, 1D09, 1HOT, 1PTY, 1LTH) 
where the RRIN can provide similar results to the atomistic network, 
yet still with some variability depending on the choice of appropriate cut-off.
Interestingly,  there are also some proteins (specifically 1F4V, 1YBA, 3K8Y and 2BRG)
in which the propensity score is higher for RRINs than for the atomistic network.  
In these cases, there tends to be a large heterogeneity in the propensities of the bonds in the allosteric site (see Figure~7 in the main text) with some bonds with large negative values as well as other bonds with large positive values. Our bond statistical measures can account for some of this variability. 
Indeed, both 1F4V and 1YBA are detected by all our four bond measures, and 3K8Y is picked by the measure based on the distributions of $p_b$. Intriguingly, only 2BRG (corresponding to CHK1) cannot be detected by our bond measures. This suggests other areas of future research, in which the importance of averaging at the level of pathways could be used to enrich the findings presented here.

\begin{figure}[!h]
  \centering
  \includegraphics[width = \textwidth]{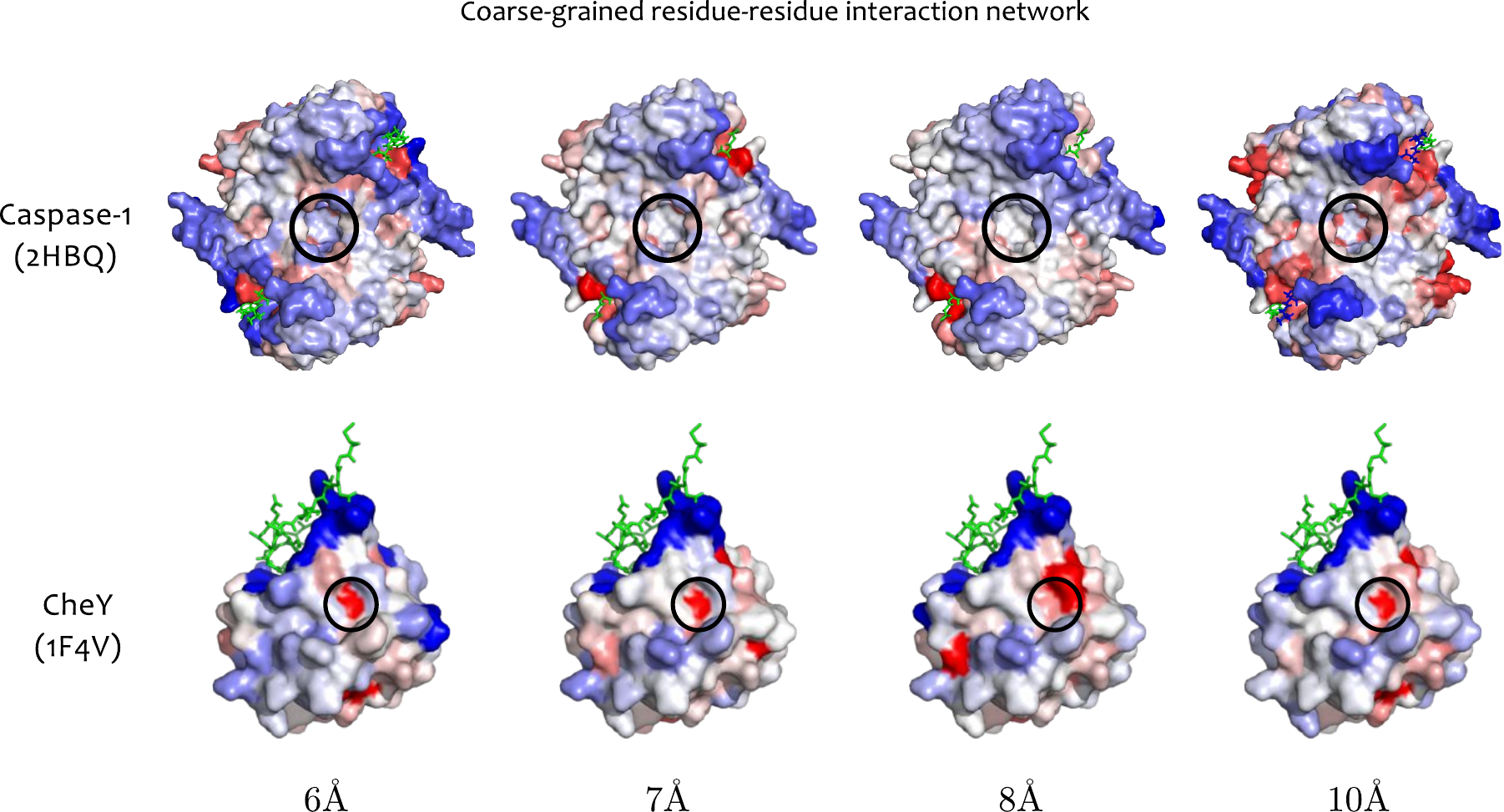}
  \caption{\textbf{Quantile scores computed from RRINs for caspase-1 and CheY at different cut-off radii.} 
 Surface mapping of the residue quantile scores $p_{R}$ of caspase-1 and CheY for RRINs generated 
  with radii cut-offs from 6~\AA~and 10~\AA.  The active-site ligand is shown in green sticks and the allosteric site is circled.  The allosteric site in caspase-1 is not identified for 6, 7, and 8~\AA.  It is identified at 10~\AA, but the signal is weaker than when using an atomistic graph.  In contrast, for CheY the allosteric site is identified as significant across the full range of cut-offs.}
  \label{fig:caspase2}
\end{figure}

\begin{table}[h]
\begin{small}
  \caption{\textbf{Propensities computed from RRINs.} Values of  $\overline{p_{R,\text{allo}}} - \langle \overline{p_{R,\text{site}}}\rangle_\text{surr}$  for residue-residue interaction networks with 
 four cut-off radii from 6\AA--10\AA.  The propensity scores are shown in bold if they are greater than 0, and starred if they lie above the 95\% confidence interval computed by a bootstrap with 10000 resamples.
 The comparable statistic computed from the all-atom network is also presented, as well as the 
 summary of the four bond statistics for each protein from Table~\ref{tbl:test_set_results}.}
  \label{tbl:mean_RRIN}
\begin{center}  
 \begin{tabular}{@{\extracolsep{\fill}}c c  ccccc}
& & \multicolumn{5}{c}{$\overline{p_{R,\text{allo}}} - \langle \overline{p_{R,\text{site}}}\rangle_\text{surr}$ \vspace*{.1cm}} \\ 
\cline{3-7} \vspace*{-.1cm} \\
   \multicolumn{1}{p{1.3cm}}{\centering PDB ID}&
   \multicolumn{1}{p{1.5cm}}{\centering All-atom \\ summary} &
    \multicolumn{1}{p{1.9cm}}{\centering All-atom \\ network}&
    \multicolumn{1}{p{2.1cm}}{\centering RRIN \\ cut-off = 6\AA} &
    \multicolumn{1}{p{2.1cm}}{\centering RRIN \\ cut-off = 7\AA} &
    \multicolumn{1}{p{2.1cm}}{\centering RRIN \\ cut-off = 8\AA} &
    \multicolumn{1}{p{2.1cm}}{\centering RRIN \\ cut-off = 10\AA}\\
    \hline
    1V4S &     \filledcirc\filledcirc\filledcirc\filledcirc&
    \textbf{0.35}$^{*}$&
    \textbf{0.065}$^{*}$&\textbf{0.010}$^{*}$&\textbf{0.047}$^{*}$&\textbf{0.16}$^{*}$\\
    3ORZ&    \filledcirc\filledcirc\filledcirc\filledcirc& 
\textbf{0.30}$^{*}$&
\textbf{0.31}$^{*}$&\textbf{0.34}$^{*}$&\textbf{0.37}$^{*}$&\textbf{0.22}$^{*}$\\

    1YP3&     \filledcirc\filledcirc\filledcirc\filledcirc& 
\textbf{0.28}$^{*}$&
-0.043&\textbf{0.11}$^{*}$&\textbf{0.046}$^{*}$&\textbf{0.13}$^{*}$\\

    1D09&    \filledcirc\filledcirc\filledcirc\filledcirc& 
\textbf{0.23}$^{*}$&
\textbf{0.20}$^{*}$&\textbf{0.17}$^{*}$&\textbf{0.21}$^{*}$&\textbf{0.15}$^{*}$\\

    2HBQ&    \filledcirc\filledcirc\filledcirc\filledcirc& 
\textbf{0.15}$^{*}$&
-0.079&-0.053&-0.062&\textbf{0.098}$^{*}$\\
        1HOT&    \filledcirc\filledcirc\filledcirc\filledcirc& 
\textbf{0.13}$^{*}$&
    -0.065&\textbf{0.13}$^{*}$&\textbf{0.18}$^{*}$&\textbf{0.20}$^{*}$\\
    1PTY&      \filledcirc\filledcirc\emptycirc\filledcirc&
    \textbf{0.11}$^{*}$&
    \textbf{0.11}$^{*}$&\textbf{0.088}$^{*}$&\textbf{0.050}$^{*}$&-0.032\\
    1EYI&      \filledcirc\filledcirc\filledcirc\emptycirc&
\textbf{0.11}$^{*}$&
    -0.036&-0.081&-0.098&-0.029\\
    7GPB&    \filledcirc\filledcirc\filledcirc\emptycirc& 
    \textbf{0.11}$^{*}$&
    \textbf{0.048}$^{*}$&\textbf{0.073}$^{*}$&\textbf{0.047}$^{*}$&-0.095\\
    1F4V&      \filledcirc\filledcirc\filledcirc\filledcirc& 
 \textbf{0.096}$^{*}$&
    \textbf{0.14}$^{*}$&\textbf{0.11}$^{*}$&\textbf{0.23}$^{*}$&\textbf{0.071}$^{*}$\\
    4PFK&    \filledcirc\filledcirc\filledcirc\filledcirc &
 \textbf{0.092}$^{*}$&
    -0.13&-0.24&-0.19&-0.067\\
    1I2D&        \filledcirc\filledcirc\filledcirc\filledcirc& 
\textbf{0.091}$^{*}$&
    \textbf{0.034}$^{*}$&-0.091&\textbf{0.010}$^{*}$&\textbf{0.12}$^{*}$\\
    1YBA&     \filledcirc\filledcirc\filledcirc\filledcirc& 
\textbf{0.062}$^{*}$&
    \textbf{0.18}$^{*}$&\textbf{0.16}$^{*}$&\textbf{0.20}$^{*}$&\textbf{0.29}$^{*}$\\
    1LTH&     \filledcirc\filledcirc\filledcirc\filledcirc&
\textbf{0.063}$^{*}$&
    \textbf{0.080}$^{*}$&-0.11&-0.22&-0.073\\
    1XTT&    \emptycirc\emptycirc\filledcirc\emptycirc& 
\textbf{0.0024}&
    \textbf{0.025}$^{*}$&-0.017&-0.012&\textbf{0.14}$^{*}$\\
    1HWZ&     \emptycirc\filledcirc\filledcirc\emptycirc& 
-0.015&
    \textbf{0.071}$^{*}$&\textbf{0.041}$^{*}$&-0.016&-0.0072\\
    3K8Y&     \emptycirc\emptycirc\filledcirc\emptycirc& 
-0.043&
    \textbf{0.29}$^{*}$&\textbf{0.24}$^{*}$&\textbf{0.17}$^{*}$&\textbf{0.30}$^{*}$\\
    1EFA&     \emptycirc\emptycirc\emptycirc\filledcirc& 
-0.066&
    -0.035&-0.0028&-0.064&-0.075\\
    1SFQ&     \emptycirc\filledcirc\filledcirc\filledcirc&
-0.081&
    -0.18&-0.19&-0.16&-0.19\\
    2BRG&     \emptycirc\emptycirc\emptycirc\emptycirc &
-0.24&
    \textbf{0.13}$^{*}$&-0.043&-0.057&-0.093\\
    \hline
  \end{tabular}
\end{center}
\end{small}
\end{table}

\end{document}